\documentclass[12pt]{article}
\usepackage{graphicx}
\usepackage{color}
\usepackage{fullpage}
\usepackage{bm}

\begin{document}
\title{Quantification of errors in large-eddy simulations of a spatially-evolving mixing layer}
\author{M. Meldi, M.V. Salvetti \& P. Sagaut}
\date{2012}
\maketitle
\begin{abstract}

A stochastic approach based on generalized Polynomial Chaos (gPC) is used to quantify the error in Large-Eddy Simulation (LES) of a spatially-evolving mixing layer flow and its sensitivity to different simulation parameters, viz. the grid stretching in the streamwise and lateral directions and the subgrid scale model constant ($C_S$). The error is evaluated with respect to the results of a highly resolved LES (HRLES) and for different quantities of interest, namely the mean streamwise velocity, the momentum thickness and the shear stress.

A typical feature of the considered spatially evolving flow is the progressive transition from a laminar regime, highly dependent on the inlet conditions, to a fully-developed turbulent one. Therefore the computational domain is divided in two different zones (\textit{inlet dependent} and \textit{fully turbulent}) and the gPC error analysis is carried out for these two zones separately. An optimization of the parameters is also carried out for both these zones.

For all the considered quantities, the results point out that the error is mainly governed by the value of the $C_S$ constant. At the end of the inlet-dependent zone a strong coupling between the normal stretching ratio and the $C_S$ value is observed. The error sensitivity to the parameter values is significantly larger in the inlet-dependent upstream region; however, low error values can be obtained in this region for all the considered physical quantities by an ad-hoc tuning of the parameters. Conversely, in the turbulent regime the error is globally lower and less sensitive to the parameter variations, but it is more difficult to find a set of parameter values leading to optimal results for all the analyzed physical quantities.

{Comparing the databases generated with different subgridscale models, it is possible to observe that the error cost function computed for the streamwise velocity and for the momentum thickness is not significantly sensitive to the used SGS closure. Conversely, the prediction of the shear stress is much more accurate when using a dynamic subgrid scale model and the variance of the considered eCF is lower in magnitude.}
\end{abstract}
\maketitle
\section{Introduction}
\label{IntroCS}
Thanks to the huge growth of available computational resources, nowadays large-eddy simulation (LES) is applied to flow configurations of increasing complexity, such as arise in many technology applications and natural flows. In LES the turbulence scales larger than a given cutoff length scale are directly simulated and flow unsteadiness and three-dimensionality are naturally taken into account. Therefore, LES is apriori more suitable than the Reynolds-averaged Navier-Stokes (RANS) approach for flows in which unsteady and three-dimensional phenomena are important, e.g. in presence of massive separation or large unsteady wakes, or when it is important to capture a significant part of the variable time fluctuations, as, for instance, for mixing prediction or in combustion applications. In this context the assessment of quality and reliability of LES results has become a topic of increasing interest (see e.g. the proceeding books \cite{Meyers2008,Salvetti2011}). To this aim, a quantification of the errors and of their evolution and interaction in LES is certainly needed. This is for LES a much more complicated task than for RANS. Indeed, in RANS the error is usually dominated by closure modeling, while independence of grid resolution and numerical discretization is rather easy to be demonstrated. In LES grid independence can be obtained in the DNS limit or if the cutoff length of the explicit filter is significantly larger than the grid resolution\cite{Sagaut2006Multi}. However, due to the huge computational resources required which are not achievable for realistic Reynolds numbers and complex configurations, this kind of approach is rarely used to recover accurate and robust results. Moreover, if filtering is not explicitly applied to the equations, the separation between resolved and subgrid (SGS) scales is linked in a not completely clear way to the grid resolution and to the numerical scheme: indeed the discretization and modeling errors can be of similar importance \cite{Ghosal1996} and interact in a complex way. These non linear error interactions may lead to counter-intuitive behaviors, such as, for instance, a deterioration of the accuracy of the results when the grid resolution is increased or when a more accurate numerical scheme is used \cite{Meyers2007,Vreman1996}.

Another critical issue is the sensitivity of LES results to the different simulation parameters; the reliability of the results is indeed strictly linked to the \textit{stability} of the simulations, i.e. a small variation of the simulation parameters should not yield a dramatic change in the quality of the results. Benchmarks and comparisons with reference data, either from experiments or direct numerical simulations, have been used in the last decades for appraising the accuracy of LES results and the suitability of numerical methods or physical models. However brute-force statistical methods, like Monte Carlo sampling, lead to a systematic exploration of the space of solution spanned by the optimization parameters that is not realistic in practical cases, requiring an enormous number of deterministic simulations. This obviously leads to an unaffordable computational costs. The main goal to achieve in a sensitivity analysis is to reduce the number of simulations to be performed. An idea is to build a discrete parametrization of the space of solutions through a computationally inexpensive model, and to use this model as input for the the sensitivity analysis and, eventually, for an optimization algorithm. Such a model is often referred to as the \textit{response surface} of the system to be optimized, leading to the definition of a so-called surrogate-model based optimization methodology. 

The development of methodologies aimed at obtaining new insights in the behavior, the interaction and the sensitivity to the parameters of the different errors in LES has recently gained considerable attention.  
One is the error-landscape approach in which a full response surface of the LES error behavior is built from a systematic variation of influencing parameters (e.g., model constants and grid resolution) \cite{mey03,mey06b,Meyers2007,mey10}. This approach provides a framework to characterize the combined effects of modeling and discretization, but at the cost of a large number of simulations, which may become unaffordable for complex cases or when a large number of parameters is involved. 

A possible way to construct error response surfaces at reduced computational costs is to use Polynomial Chaos\cite{Knio2006} or generalized Polynomial Chaos (gPC) theory\cite{Xiu2002}. In the gPC approach the uncertain quantities are modeled by the introduction of input random variables with given statistics. Statistical information on the system response can be built through the prescribed statistical properties of the input random variables. This methodology has been applied in fluid dynamics to investigate the responses to different types of uncertainties, e.g. inflow conditions\cite{Ko2008,Xiu2003,Simon2010,Bruno2009} or boundary conditions\cite{Xiu2003}. A specific application of gPC to LES can be found in the works of Lucor et al.\cite{Lucor2007}, in which a sensitivity analysis to parametric uncertainty in SGS modeling has been carried out for LES of decaying homogeneous turbulence. The gPC approach was shown there to be able to provide an accurate statistical description of the space of possible LES solutions with a limited computational effort.

Using the same numerical approach Meldi et al.\cite{Meldi2011_pof} investigated the sensitivity of the \textit{optimum} model constant of the Smagorinsky model to uncertainties in the shape of the energy spectrum. Indeed, it is well known that the parameter tuning the Smagorinsky model is flow dependent and, thus, should be tuned for each particular application. For many types of flows, if {\em standard} values of $C_S$ are used, the Smagorinsky model has been found to introduce an excessive dissipation. A large amount of work in SGS modeling has been carried out to overcome the negative features of the Smagorinsky model; we cite, for instance, the dynamic calculation of the Smagorinsky constant \cite{Germano1991}, mixed-models \cite{Meneveau2000} or the variational multiscale approach, in which the $SGS$ model to the smallest resolved scales only through variational projection \cite{Hughes2001}. Nonetheless, the classical Smagorinsky model has the advantage of being very simple to be implemented and of implying very limited additional costs. This makes this model still interesting for industrial and engineering applications, possibly with an ad-hoc tuned constant \cite{Lucor2007,mey06b}.

 Another important source of uncertainty in LES is grid refinement. Since the grid itself acts as an implicit filter over the Navier-Stokes equations, a local increase of the Reynolds number theoretically implies a correspondent refinement of the physical resolution. In fact, grid stretching is normally used in complex flow simulations or in canonical test cases in which shear forces are investigated. However, since the Reynolds number is local in space and time, it is not possible to define an priori exact meshing strategy. As already stated previously, the grid resolution and the subgrid scale modeling interact in complex dynamics: complete statistical studies, investigating how grid and modeling uncertainties couple and propagate in the numerical solution, are still not reported in literature.    
The gPC approach is used herein to quantify the errors and their sensitivity to some simulation parameters in large-eddy simulations of a spatially-developing mixing-layer. This flow configuration has been extensively studied due to its significance in many technological applications or environmental flows (see e.g. the review articles\cite{Ho1984,Gutmark1995}). Mixing layers are characterized by the evolution and the interaction of coherent vortex structures forming at the interface between two parallel coflowing streams with different velocities. These structures entrain the fluid into the mixing layer from both sides and play a major role in the bulk mixing of the fluids and in the growth of the mixing layer. The structures then go through successive merging and three-dimensional instabilities; eventually, as the flow undergoes transition to turbulence, progressively finer scale eddies can be observed. Thus, the spatially evolving mixing-layer allows the behavior of LES errors to be investigated both in a transitional regime, characterized by large vortical structures, as well as in the downstream developed turbulent region. This is a situation encountered in several cases of practical interest, to which LES is applied. One of the goal of the present study is indeed to investigate whether the behavior of the errors and the sensitivity to the parameters in the upstream transitional region is qualitatively and quantitatively different from the one in the downstream turbulent zone. Note that the transitional regime is not the classical framework for which LES has been developed and significantly differs from the decaying homogeneous turbulence case previously investigated in Lucor et al. \cite{Lucor2007}.

The reference data for evaluating the errors are obtained through a highly-resolved large-eddy simulation (HRLES). Inlet forcing is used to speed-up the formation and the evolution of the large-scale vortices \cite{Colonius1997} and the SGS scales are modeled using the dynamic Smagorinsky model\cite{Germano1991}.{ Two sets of LES simulations are carried out using a coarser grid resolution: the difference between the two mentioned databases is that in the first one a classical Smagorinsky model is used, while in the second one the subgrid scale effects on the flow are recovered by the dynamic procedure proposed by Germano\cite{Germano1991}. The considered uncertain parameters are the grid stretching ratios in the streamwise and lateral directions and, for the first mentioned database, the Smagorinsky model constant. Starting from a database of $(4^n)$ simulations, being $n$ the number of uncertainty parameter in the considered database,} gPC is used to characterize the statistics and the pdf of the errors evaluated with respect to the HRLES reference data for some quantities characterizing the evolution of the mixing-layer, viz. the mean streamwise velocity, the momentum thickness and the shear stress.
The gPC application is herein used in its non-intrusive approach, i.e. the errors are directly projected over the orthogonal basis spanning the random space, without any modification of the deterministic solver. 

Finally, the error response surface built through gPC is used to carry out an optimization of the of the considered parameters, i.e. the values minimizing the considered errors are sought. An issue for the considered flow is whether global optimal values can be found for all the analyzed quantities and for both the transitional and the developed turbulent regimes.

{Summarizing, the aim of the present sensitivity analysis is twofold: (i) {to quantify through the gPC approach the error behavior and its sensitivity to some simulation parameters in LES of a spatially-evolving  flow characterized by both transitional and fully-turbulent zones.} (ii) The comparison between the error behaviors computed for the classical and the dynamic Smagorinsky  models is useful to highlight whether a satisfactory accuracy can actually be recovered with the Smagorinsky model by an ad-hoc tuning of the constant. This, together with the definition of the $C_S$ value ranges for which the Smagorinsky closure can lead to accurate results  is of significant practical interest, due to the simplicity and low costs of this model and to the fact that it is 
implemented in most of the commercial and open-source codes.

The paper is structured as follows: in Section \ref{numtools} some brief reminders dealing with the mathematical and numerical methodology are given. The characteristics of the test case and the error cost function formulation, i.e. the algebraic relation used to quantify the error, are introduced and discussed. Comparisons of the reference results with DNS and LES reported in literature are furnished in Section \ref{validation}; a division of the physical domain into a transitional and a turbulent zone has been proposed, based on the \textit{memory} of the inlet perturbations. Following this partition, the error cost function recovered in the classical Smagorinsky database has been analyzed through gPC in the two regions separately and the results are reported in Sections \ref{turb} and \ref{transition} respectively. The analysis is successively carried out for the database generated with the use of the dynamic Smagorinsky model, the results reported in Section \ref{dynamic}. discussion about the optimization of the parameters is reported in Section \ref{optimisationHPC}, while a summary of the results and concluding remarks are given in Section \ref{conclusionHPC}.
\section{Methodology and flow configuration}
\label{numtools}
\subsection{Modeling and numerical ingredients}
\label{les}
The scale separation in LES is achieved by the application of a low-pass filter, which is mathematically formulated as a convolution product in the physical space. The filtered Navier-Stokes equations for incompressible flows are given by:
\begin{equation}
\frac{\partial{\bar{u}_i}}{\partial{t}} + \frac{\partial{\bar{u}_i \bar{u}_j}}{\partial{x_j}} + \frac{\partial{\bar{p}}}{\partial{x_i}} - 2 \nu \frac{\partial{\bar{S}_{ij}}}{\partial{x_j}} - \frac{\partial{\tau_{ij}}}{\partial{x_j}} = 0  \hspace{0.5cm} i=1,2,3
\end{equation} 
where $\bar{u}_i$ is the filtered velocity component in the $x_i$ direction, $\bar{p}$ is the filtered pressure and $\bar{S}_{ij}$ is the resolved strain tensor. The filtering of the convective term leads to the addiction of a so called subgrid-scale (SGS) stress tensor: 
\begin{equation}
\tau_{ij}=\overline{u_i u_j}- \bar{u}_i\bar{u}_j
\end{equation}
This term has to be modeled in order to close the problem. A simple and popular approach is represented by $eddy$-$viscosity$ models, characterized by the introduction of a SGS viscosity term in the equations in order to mimic the dissipative behavior of the smallest scales eliminated by filtering:
\begin{equation}
\tau_{ij}= - 2 \nu_{SGS} \bar{S}_{ij}
\end{equation}
where $\nu_{SGS}$ is the so called eddy viscosity. The oldest and most widely used eddy-viscosity SGS model is the Smagorinsky model \cite{Smagorinsky1963}, in which this viscosity is expressed as follows:
\begin{equation}
\nu_t = C_S^2 \Delta^2 (2 \bar{S}_{ij} \bar{S}_{ij})^{1/2} 	
\end{equation}
where $\Delta$ is the LES filter width and $C_S$ is a model parameter, which must be a-priori given and possibly tuned for each particular application. One of the parameters that will be the object of gPC analysis and of optimization in the following is indeed the value of the Smagorinsky constant. For the highly-resolved LES simulation, carried out to obtain a reference solution (see Sec. IIC), the dynamic Smagorinsky model\cite{Germano1991} is instead used.

The filtered Navier-Stokes equations have been solved by OpenFOAM, a code based on C++ libraries. The code is built on libraries grouping several different classes and the ones for LES computations have been extensively used and tested \cite{Tabor2009,Fureby1997-1,DeVillers2006}. A finite-volume face-centered approach is used for space discretization on hexahedral grid elements. Through integration over the computational cells, the Navier-Stokes equations may be rewritten as a function of the fluxes across the cell faces: the fluxes must be discretized through numerical interpolation. Second-order centered schemes have been used herein consistently with many of the LES simulations in the literature. A Poisson equation is then constructed \cite{Rhie1983}, which implements the
solenoidal field incompressibility condition; the equation set is then solved sequentially by the use of the PISO algorithm.  The simulations are advanced in time by the Crank-Nicolson scheme, which is second-order accurate in time.
The resulting linear system is then solved at each time step by the application of an algebraic multigrid (AMG) method.
\subsection{Generalized polynomial chaos}
\label{gpc}
The main features of the generalized polynomial chaos approach are briefly recalled here; for more details we refer to Ghanem \& Spanos\cite{Ghanem1991} and Le Ma\^itre \& Knio\cite{LeMaitre2011}.  In the last years, this stochastic approach has been applied to turbulent flow analysis with satisfying results\cite{Lucor2007,Meldi2011_jfm}. 

Let us define a {\it probability space} ($\Omega$,$\mathcal{A}$,$\mathcal{P}$) where $\Omega$ is the event space, $\mathcal{A} \subset 2^{\Omega}$ its $\sigma$-algebra and $\mathcal{P}$ its probability measure. Being $\omega$ an element of the event space, we define a random field $X(\omega)$ such that it maps the probability space into a function space $V$, $X:\Omega \rightarrow V$. In the following we will consider second-order random fields, i.e. those satisfying the relation:
\begin{equation}
 {\bf E}(X,X) < + \infty
\end{equation} 
where ${\bf E}$ denotes the expectation of a random variable. In this context, gPC is a tool allowing second-order random fields to be represented through a set of random variables ${\bf \xi}(\omega)$.
An approximation of the random field $X(\omega)$ is then recovered through its Galerkin projection onto a polynomial orthogonal basis taking the following form:
\begin{equation}
X(\omega)=a_0B_0 + \sum_{i_1=1}^{\infty}a_{i_1}B_1({ \xi_{i_1}})+ \sum_{i_1=1}^{\infty} \sum_{i_2=1}^{i_1}a_{i_1 i_2}B_2({ \xi_{i_1}, \xi_{i_2}}) + ...
\label{EldredEq}
\end{equation}
where $\xi = (\xi_1, ..., \xi_N)^T$ is a N-dimensional random vector and $B_i$ is a polynomial of order $i$ depending on the $\sigma$ algebra of $\xi$. 
This expression can be easily reformulated using a term-based indexing instead of a order-based indexing. Let $\Phi_k({\bf \xi}(\omega))$ be a single polynomial, the pseudospectral approximation (\ref{EldredEq}) can be written as follows:
\begin{equation}
X(\omega)=\sum_{k=0}^{\infty}a_k\Phi_k({\bf \xi}(\omega))
\label{LucorEq}
\end{equation}
in which there is a one-to-one correspondence between $a_{i_1 \, i_2 \, \cdots \, I_n}$ and $a_k$ and between $B_n({ \xi_{i_1}, \, \xi_{i_2}, \, \cdots \,, \, \xi_{i_n}})$ and $\Phi_k({\bf \xi})$
The polynomial expansion is truncated to a finite limit and the orthogonality of the polynomials is set through the relation
\begin{equation}
<\Phi_i \Phi_j>=<\Phi_i^2> \delta_{ij}
\end{equation}
where $<\cdot,\cdot>$ denotes an ensemble average. This inner product is defined over the measure $W(\xi)$ of the random variables as follows:
\begin{equation}
<f(\xi)g(\xi)>= \int_{\omega \in \Omega} f(\xi)g(\xi)dP(\omega)= \int f(\xi)g(\xi) W(\xi) d \xi
\end{equation}
Thanks to the orthogonality of the polynomial basis, each coefficient of the Galerkin projection (\ref{LucorEq}) can be recovered through the following definition:
\begin{equation}
a_k = \frac{<X,\Phi_k>}{<\Phi_{k}^2>}= \frac{1}{<\Phi_{k}^2>}\int_{\omega \in \Omega} X \Phi_k \rho({\bf \xi}) d({\bf \xi})
\end{equation}
where each inner product involves a multidimensional integral over the support range of the weighting function. The integrals can be computed through different mathematical methods: considering the number of random variables investigated, the coefficients $a_k$ have been computed through Gaussian quadrature in the present work.

The polynomial family to be used must be a priori specified. The choice of the polynomials affects the speed of the convergence of the series: a unsuitable polynomial family may lead to the need of a large number of degrees of freedom to obtain a given level of accuracy (i.e. a higher order of the polynomials), while a suitable polynomials family is able to interpolate both the input and the random variables by means of a few degrees of freedom. In the case of the input, when dealing with Gaussian quadrature, an optimal family has a weight coefficient similar to the function $W$.

The probability density function $pdf$ of the random variables has been considered as \textit{uniform} leading to the choice of the Legendre polynomial family: this is a well suited choice since the inner product weighting function is directly proportional with a factor $0.5$ to the set probability density function. The polynomial expansion has been limited to the third order since the contribution of fourth and higher order polynomials is negligible. This leads to the use of a 20 polynomial basis to generate the error cost function over the uncertainty space. 

The gPC application is herein used in its non-intrusive approach, i.e. the variables, and more precisely the error cost functions defined in Sec. \ref{ecf}, are directly projected over the orthogonal basis spanning the random space, without any modification of the deterministic solver. In the following the analysis has been performed by considering as uncertainty variables the grid stretching ratios in the streamwise and in the lateral directions, defined more in detail in Sec. IIC, and the Smagorinsky model constant (see section \ref{les}).

The number of points to discretize each random variable space is chosen in order to recover converged integrals when computing the polynomial coefficients. The polynomial expansion being truncated to the third order, 4 points for each random variable are sufficient to compute the coefficients $a_k$. The accuracy of the method has been investigated by checking that the contribution of polynomials of order higher than three remains very low for all the considered quantities. An example of such an analysis is reported in the Appendix.
\subsection{Test case}
\label{testcase}
The considered test case is a spatially-evolving mixing layer. Following previous studies in the literature \cite{Wang2007,McMullan2007},the dimensions of the computational domain have been chosen as follows: $25 \Lambda$ x $6 \Lambda$ x $4 \Lambda$, in the streamwise, lateral and spanwise directions respectively, where $\Lambda = A \delta_0$ and $\delta_0$ is the vorticity thickness at the inlet. The value of $A$ has been set equal to $14.132$, because in this way $\Lambda$ represents the most unstable wavelength estimated through the linear instability theory \cite{McMullan2007}. The Reynolds number of the flow is $Re = (U_1-U_2) \delta_0 / \nu = 700$, where $U_1$ and $U_2$ represent the asymptotic velocities and $\nu$ is the kinematic viscosity.\\
A hyperbolic-tangent mean base streamwise velocity profile has been imposed at the inlet ($x=0$) \cite{Wang2007,McMullan2007}:
\begin{eqnarray}
U&=& \frac{U_1 + U_2}{2} + \frac{U_1 - U_2}{2} \tanh(\frac{2y}{\delta_0}) \; \; \; \; \; \; -3 \Lambda < y < 3 \Lambda \nonumber \\
V&=&0 \nonumber \\
W&=&0 \nonumber
\end{eqnarray}
Following the results by Colonius and Lele \cite{Colonius1997}, tridimensional perturbations have been added to the mean profile to trigger transition to turbulence. These modes are related to $\Lambda$ in the following way:
\begin{eqnarray}
f  &=& (2\pi)/ \Lambda \nonumber \\
P(0,z) &=& \sum{2 \, cos(i*f*z)} \; \; \; \; \; \; i=1,2,4,8 \; \; \; \; \; \; 0 < z < 4 \Lambda \nonumber \\
u^{\prime} &=& v^{\prime} = w^{\prime} =0.05* P * \exp(-0.5 \sqrt{y/\delta_0}) \nonumber
\end{eqnarray}
where $u^{\prime}$, $v^{\prime}$, $w^{\prime}$, are the perturbation terms in the streamwise, lateral and spanwise direction to be added to $U$, $V$ and $W$ respectively.
This tridimensional time independent field is then perturbed at each time step by the addition of a white noise of maximum intensity equal to 2 \% of the local inlet velocity value; this random noise is applied as a step function in the lateral direction only in the region $[-\Lambda, \Lambda]$.
The pressure at inlet is assumed to have a zero gradient. In the lateral direction slip conditions are used, while periodic boundary conditions are imposed in the spanwise direction. At the outlet, a mixed fixed value-zero gradient mass conserving boundary condition for velocity and pressure is applied \cite{Tabor2009}. The initial field is uniform and the velocity is everywhere equal to to the inlet mean hyperbolic-tangent profile.

A single highly refined LES simulation (HRLES) has been run as a reference solution, to be used in the following error analysis. The grid used for HRLES has 1024$\times$144$\times$144 cells in the streamwise, lateral and spanwise directions respectively. The nodes are uniformly spaced in the streamwise and spanwise directions, while in the lateral one they are clustered near $y=0$. For this grid and for the coarser ones generated in the following, we may introduce grid stretching either in the streamwise or in the lateral directions, through a stretching ratio defined as follows:
\begin{equation}
d_{i+1} = d_i \sqrt[m-1]{str_Q}
\end{equation}
where $Q$ identifies the direction (either $x$ or $y$), $m$ is the number of elements in which the length of the domain in the considered direction is divided, $d_i$ and $d_{i+1}$ are the size (in the considered direction) of the $i$ and $i+1$ grid elements; $i$ increases with the distance from $x = 0$ or $y = 0$. In the $y$ direction stretching is symmetric with respect to the $y = 0$ plane. 
For HRLES, $strX$=1 and $strY$=4.5. Consequently, the grid resolution in HRLES is $5 \eta $ $\times$ $4 \eta \rightarrow 18 \eta$ $\times$ $6 \eta$, $\eta$ being the Kolmogorov scale of the associated energy spectrum \cite{Wang2007}. The resolution in the normal direction is of $4 \eta$ for $y=0$, increasing up to $18 \eta$ for $y = \pm 3 \Lambda$.
In HRLES the LES equations are closed using the dynamic version of the Smagorinsky model \cite{Germano1991}. The time step has been fixed to $\Delta t = 2.5e^{-6}$: this value has been chosen to keep the CFL number $Co < 0.5$.
A first transient period $t_1 = 25 t_c$ with $t_c =  \Lambda / (0.5*(U_1+U_2))$ has been simulated, followed by a second simulation time $t_2 = 100 t_c$ over which averages have been computed. The time $t_2$ corresponds to $40000$ time steps.

A database of $64$ $(4^3)$ simulations, carried out over $16$ different coarser grids all having $256$$\times$$72$$\times$$72$ cells, has been successively generated. Let us recall that the uncertainty variables are the stretching ratio in the streamwise direction ($strX$), the stretching in the lateral direction ($strY$) and the value of the Smagorinsky model constant $C_S$: each simulation corresponds to a quadrature point over the 3D uncertainty space $[3,10] \times [3,12] \times [0,0.2]$. While the extreme values considered for $C_S$ are the ones found in literature, the maximum value of the other two variables has been chosen as it may be considered a \textit{ limit value}: indeed, at the maximum stretching value the spatial resolution at $x=y=0$ is the same as the one of the HRLES simulation. At the maximum stretching value the resolution in the streamwise direction varies from $5 \eta$ to $50 \eta$ and in the lateral direction from $4 \eta$ to $48 \eta$. The spacing in the spanwise direction is in all cases constant and corresponds to $12 \eta$. {The previously mentioned $4^2$ grid set has been used to generate a second database: for each grid, a single LES simulation closed with the dynamic version of the Smagorinsky model has been carried out. For both the databases,} the boundary conditions and the simulation $t_1$ and $t_2$ time intervals are the same as for HRLES, as well as the time step. It has been verified that results are invariant to further $\Delta t$ reductions: the error due to time integration may then be considered negligible.
\subsection{Error cost function}
\label{ecf}
Let us denote with $S$ a flow quantity of interest, the corresponding error cost function (eCF) is defined as:
\begin{eqnarray}
eCF_{S} &=&  \frac{ \int_{\Omega} \sqrt{(S_{LES}-S_{HRLES})^2} d \Omega}{\|S_{HRLES} \|}  \nonumber \\
\|S_{HRLES} \| &=& \int_{\Omega} \sqrt{(S_{HRLES})^2} d \Omega \nonumber
\end{eqnarray}
The flow quantities of interest for which $eCF$ has been computed are the mean streamwise velocity, the momentum thickness $\Theta$ (see section \ref{validation}) and the shear stress $\tau = \overline{u^{\prime} v^{\prime}}$ respectively (see section \ref{validation}).
The error cost function has been computed over $10$ different planes in the streamwise direction uniformly distributed between $x = 2\Lambda$ and $x = 20 \Lambda$.
At these locations, $eCF$ has been computed by integration in the spanwise and lateral directions, leading to an error definition for each LES simulation and each streamwise section. Starting from the $4^n$ computed discrete $eCF$ values for each streamwise section and flow variable, gPC has been applied as a post processing tool to generate a continuous distribution over the uncertainty variable space, $eCF(\omega)$.
\section{Main flow features}
\label{validation}
The momentum thickness $\Theta$ is often used to evaluate the grow rate of the mixing layer; the following definition is used:
\begin{equation}
\Theta (x) = \int_{-3 \Lambda}^{3 \Lambda} (U_1 - U(x,y))(U(x,y)-U_2) d y
\end{equation}
where $U(x,y)$ is the streamwise velocity averaged in time and in the spanwise direction. In figure \ref{fig001}(a) the momentum thickness $\Theta$ is compared with DNS\cite{Wang2007} and LES\cite{McMullan2007} data obtained for the same flow conditions, and similar computational domain and boundary conditions. Note, however, that the inlet velocity perturbation herein imposed is different from the one used in the DNS and LES studies\cite{Wang2007,McMullan2007} : this is expected to have a noticeable impact on the flow evolution. Following the arguments by Colonius and Lele \cite{Colonius1997}, the imposition of the tridimensional perturbation modes in HRLES is expected to shift upstream the mixing layer growth and indeed  $\Theta$ grows upstream in HRLES, if compared to the other studies, as shown in figure \ref{fig001}(a). However, the slope in the turbulent region, where the growth is almost linear, is very similar: an approximate estimation is $0.094$ for HRLES, $0.091$ for DNS\cite{Wang2007} and $0.098$ for LES\cite{McMullan2007}. Figure \ref{fig001}(b) shows the profile of the normalized average streamwise velocity $U_n = (U - U_a)/(U_1 - U_2)$, with $U_a = (U_1 + U_2)/2$, as a function of a self similar coordinate, $\eta$, defined as in McMullan et al.\cite{McMullan2007}; data from McMullan et al.\cite{McMullan2007} are used for comparison. In the self similar region the scaled velocity profiles computed at different streamwise locations are expected to overlap. For our data this is almost obtained for $x \ge 12\Lambda$ (see figure \ref{fig001}(b)). Although the curve taken from LES by McMullan et al.\cite{McMullan2007} is computed at $x = 12\Lambda$, it is close to HRLES profile at $x = 6\Lambda$. This reflects again the effects of the earlier growth of the mixing layer thickness in HRLES, due to the different inlet conditions.

As previously discussed, a typical feature of the considered spatially evolving flow is the progressive transition from a laminar regime, highly dependent on the inlet conditions, to a fully-developed turbulent one. Only the turbulent regime can be considered as a  canonical context of application of LES.
It is therefore interesting to carry out the gPC error analysis in the laminar-transition region and in the fully-developed turbulent one separately. An accurate separation of these two regions is difficult, though results in the literature\cite{Colonius1997} suggest that the transition to turbulence is almost space fixed for the considered flow configuration.
Figure \ref{fig99} shows the instantaneous isocontours of the second invariant of the velocity gradient tensor obtained in HRLES and LES. It can be seen from the isocontours in the $(y,z)$ planes at fixed $x$ locations (figures \ref{fig99}(b)-(d)) that well organized vortical structures, connected with the form of the imposed inlet perturbation, are initially present and they progressively loose coherence moving downstream. A similar situation can be observed at different time instants. The comparison between LES isocontours (figures \ref{fig99}(e)-(f)) with HRLES isocontours (figure \ref{fig99}(d)) at $x / \Lambda = 14$ shows that the LES simulations do not resolve completely the flow field. This intended lack of resolution, which is common in practical LES applications, has been set to highlight the effects of the subgridscale model over the error cost function. {Similar considerations may be drawn if the subgrid activity parameter \cite{Geurts2002} $s = \varepsilon_t / ( \varepsilon + \varepsilon_t )$ is considered, being $\varepsilon$ and $\varepsilon_t$ the molecular and subgrid dissipation, respectively. The results observed confirm that while the dissipation rate introduced by the model in the LES simulations is of the same order of magnitude of the physical dissipation rate, $\varepsilon_t$ in HRLES is more than one order of magnitude smaller.} A definition of the two zones may therefore be obtained by assuming that the fully-developed turbulent region starts at the location where the signature of the inlet perturbations on the flow is no more apparent.
Therefore, the quantitative division between the inlet-dependent zone and the fully-developed turbulent one has been set through the analysis of the correlation coefficient between the spanwise energy spectra obtained at the different streamwise sections with the one at $x = 2\Lambda$, reported in table \ref{tab:corrCoeff}. The correlation coefficient has a strong drop in correspondence of the section $\Lambda = 12$: thus, the fully-developed turbulent region is assumed in the following to start at $\Lambda = 12$. This assumption is also supported by the self similarity of the streamwise scaled velocity profiles previous observed for $x \ge 12\Lambda$ (see figure \ref{fig001}(b)).

%
\section{Error analysis in the fully-developed turbulent region}
\label{turb}
The error analysis is first carried out in the \textit{turbulent} part of the computational domain ($x / \Lambda=12-20$).
The results of gPC applied to the error cost function values, generated for the already mentioned set of discrete points spanning the uncertainty space in the considered random parameters ( viz. $strX$, $strY$ and $C_S$), are summarized in table \ref{tab:turbStatU}.
It is important to stress that in this region a higher streamwise stretching $strX$ corresponds to a progressively lower local resolution  moving from $\Lambda=12$ to $\Lambda=20$.
In table \ref{tab:turbStatU}, $eCF_{d}$ is the error obtained in a single deterministic simulation carried out for the mean values of the uncertainty parameters in the considered range, i.e. $strX=6.5$, $strY=7.5$ and $C_S=0.1$, while $\overline{eCF}$ is the coefficient of the zero-order polynomial in the gPC expansion, that is the stochastic mean value of the error. The ${c}_{v}$ coefficient is the ratio between the stochastic recovered standard deviation and $\overline{eCF}$: it gives a measure of the variability of the error and values close to zero indicate a quasi-deterministic phenomenon.
In figures \ref{fig:StatPropECF}(b)-(d) the partial variances\cite{Sobol1993} are reported. These quantities, which are normalized over the total variance, quantify the sensitivity of the error to the single parameters and to their interactions.
The gPC analysis has been carried out for different streamwise locations and the results are reported for each single location; the results obtained by applying gPC to the eCF values averaged over all the considered sections for each quadrature point are also shown in table \ref{tab:turbStatU}. These should give an indication of the general $eCF$ behavior in the turbulent region.

Let us analyze first the error relative to the streamwise average velocity prediction (table \ref{tab:turbStatU} and figure \ref{fig:StatPropECF}(a)). At all the considered sections, the error is generally low ( the maxima never exceed $1.5\%$) and slightly increases with the distance from the inlet. The variance coefficient also tends to increase with the distance from the inlet, varying from $20\%$ to $30\%$. This is generally related to an increase of the maximum error, as it will be shown later in the probability density functions (pdf), and it is probably due to the progressive coarsening of the grid in the streamwise direction, which always occurs for $strX \neq 1$ and becomes more significant as $strX$ increases. Nonetheless, it can be concluded that the error in the prediction of the streamwise average velocity remains very low in the considered parameter range. Figure \ref{fig:turbUpdf} shows the pdfs of the errors for the different considered sections. These pdfs are built through the application of a Monte Carlo method to the polynomial basis, using the polynomial coefficients computed by gPC. In this way a high number of samples may be generated with limited computational resources. The pdfs reported have been normalized over the intensity of the peak of the average section and are presented along with the correspondent $\overline{eCF}$ and $eCF^d$. A significant number of occurrences is always present for eCF $= 0.003 - 0.005$, while the largest amount tends to gather towards an  eCF value that increases moving downstream. The distribution of the occurrences around this peak becomes less and less sharp as x increases, consistently with the larger $c_v$ values obtained at the most downstream sections (see table \ref{tab:turbStatU}). Summarizing, the analysis of the parameters in table \ref{tab:turbStatU} and of the pdfs in figure \ref{fig:turbUpdf} shows that, in the turbulent region, at the most downstream sections it is probable to have a larger value of the error than upstream and that the error value is more sensitive to the choice of the parameters.It is interesting to remark that the deterministic error obtained for the mean values of the parameters is generally closer to the most probable stochastic error than to the mean error recovered in the stochastic analysis. This points out how a deterministic simulation carried out with a set of average parameter values may actually give the eCF values corresponding to the most probable error occurrence. However, it can not furnish any information about the eCF probabilistic distribution.

The partial variances shown in figure \ref{fig:StatPropECF}(b) are useful to quantify the sensitivity to the single parameters and to their interactions. As expected, considering the right part of the figure between $12 \leq x/\Lambda \leq 20$, the sensitivity to the value of the Smagorinsky constant $C_S$ is always important and increases moving downstream, while the impact of the grid lateral stretching $strY$ progressively decreases. Conversely, the sensitivity to the streamwise grid stretching becomes significant only at the most downstream sections. These results can be interpreted by recalling that the grid resolution in the streamwise direction becomes coarser moving downstream for all the considered values of $strX$, and this loss of resolution at the most downstream sections is more dramatic for high $strX$. This explains why the sensitivity to $strX$ increases with x. This is also consistent with the convective dynamics of the error, which can be described by a differential monodimensional equation\cite{Hoffman2005,Hoffman2006}. In shear flows, error growth is associated with convective/absolute instabilities. A lack of resolution close to the inlet boundary condition triggers a fast growth of the error and a grid refinement operated downstream is usually not effective to inhibit the error growth. On the other hand, moving downstream, progressively finer flow structures form. The increase of the error and of its sensitivity to $C_S$ may thus be explained by the lack of resolution of the turbulent scales becoming progressively more important moving downstream. Finally, the lateral velocity gradients become less important with increasing x, and this justifies the progressively lower error sensitivity to $strY$. Note how the combined effects of the different parameters remain always low, except for the coupling between $strY$ and $C_S$ at the most upstream sections. This apparent coupling between $C_S$ and $strY$ effects will be investigated more in details in section \ref{transition}.  

Similar considerations to those made for the errors in the prediction of the mean streamwise velocity may be extended to the momentum thickness eCF (see table \ref{tab:turbStatU}). As for the mean streamwise velocity, the error tends to increase with $x$, although the error in the prediction of the momentum thickness is globally larger than in the previous case. This can be clearly observed in figure \ref{fig:StatPropECF}(a), where the magnitude of the $eCF$ computed for the different flow quantities are shown. Also the error sensitivity to the parameters increases by moving downstream resulting in a progressively more homogeneous distribution of the occurrences over larger subsets of $eCF$ values (see $c_v$ in table \ref{tab:turbStatU} and the pdf shapes in figure \ref{fig:turbMpdf}). Also for the error sensitivity to the single parameters and their interactions, the same observations previously made for the mean streamwise velocity hold true.

Before analyzing the error behavior for the shear stress, let us recall that the relevant error function for all the considered LES simulations has been computed with respect to the HRLES values, without filtering the HRLES variables or including the subgrid scales (SGS) model contribution.
Thus, larger errors are a priori expected on this quantity than for the previously analyzed ones, as well as a higher sensitivity to the grid resolution and to the SGS viscosity. This is indeed verified by the results of $eCF_{d}$, $\overline{eCF}$ and $c_v$ reported in table \ref{tab:turbStatU} and observable in figure \ref{fig:StatPropECF}(a), which are much larger than the corresponding ones for the mean streamwise velocity and the momentum thickness. As previously, the error increases with $x$, but the variance remains almost constant and close to $60\%$. Note, however, how the most probable values, roughly corresponding to the peaks of the pdf (see figure \ref{fig:turbSpdf}), are significantly lower than the deterministic and stochastic average ones and they are approximately of the same order of the one computed for the momentum thickness. This indicates the presence of a limited region in the parameter space giving a much larger error in the prediction of this quantity than that obtained for the majority of the possible parameter combinations. As for the sensitivity to each single parameter, the Smagorinsky constant has the strongest effect, as expected on the basis of the previous considerations.

\section{Error analysis in the inlet-dependent region}
\label{transition}
The same error analysis as in Sec. \ref{turb} is extended here to the \textit{inlet-dependent} region ($x / \Lambda=2-10$), in which a correlation with  the inlet conditions is still present and fully-developed turbulent characteristics are not yet reached. An interesting issue is to investigate whether the behavior of the error is also qualitatively different with respect to the one found in the turbulent region.
Table \ref{tab:transStatU} summarizes the main statistical features of the error for the mean streamwise velocity, the momentum thickness and the shear stress respectively. The corresponding error partial variances and error pdf are shown in figures \ref{fig:StatPropECF}(b)-(d), \ref{fig:transUpdf}, \ref{fig:transMpdf} and \ref{fig:transSpdf} respectively.

Generally speaking, the behavior of $eCF$ is comparable to the one discussed in the previous section; a few considerations should however be made. First of all, it looks like that the error in the mean streamwise velocity prediction is very low with an almost negligible probability to be over $1\%$ value (see figure \ref{fig:transUpdf}). The pdf analysis also points out that a huge number of occurrences cluster close to the most probable value at the section $x = 2\Lambda$, while the mean streamwise velocity prediction becomes more and more sensitive to parameters variation as the flow undergoes transition. Conversely, the error recovered for the momentum thickness and the shear stress may be considerably high, in particular if the most upstream sections are considered. Moving towards higher $x$ values, the stochastic mean error decreases, as well as the most probable error value and the error maximum value (see figure \ref{fig:transUpdf}). For all the considered quantities, the error sensitivity to the parameters is generally significantly higher in this region than in the turbulent one, as shown by the values of $c_v$ in table \ref{tab:transStatU} and in figure \ref{fig:VarPlot}. Inside the inlet-dependent region, at the most upstream section the error coefficient of variance is relatively low but it is associated with a large value of the error for the momentum thickness and the shear stress (see table \ref{tab:transStatU} and figures \ref{fig:StatPropECF}(a), \ref{fig:transMpdf} and \ref{fig:transSpdf}). Moving downstream the error variance, and thus its sensitivity to the parameters, progressively increases reaching a maximum around $x / \Lambda=6 - 8$ and then decreases again. Thus, at the most downstream sections there is again a very high probability that the error be almost equal to the most probable value, which is however significantly lower than at the more upstream sections, as shown for instance by the peaks in the pdfs in figures \ref{fig:transUpdf}-\ref{fig:transSpdf}. 

Considering the partial variances, $C_S$ is again the most important parameter, with the only exception of the error on the mean streamwise velocity at the most upstream section. As could be expected, this indicates that the contribution of the SGS model is particularly critical when the flow undergoes transition. The stretching ratio in the lateral direction, $strY$, is the second most important parameter and its effect on the error prediction is considerably larger if compared to the one in the \textit{turbulent} region. This is probably due to the intense lateral velocity gradient present close to $y = 0$, suggesting that a lack of resolution may occur if the lateral stretching is not sufficiently high. The error is not particular sensitive to $strX$ variations, probably because the streamwise grid resolution in this region is sufficiently high to capture the relevant flow scales even at low $strX$ values. 

Finally, it is also possible to observe that at the most downstream section in this region, $x =10\Lambda$, the error in the shear stress prediction exhibits a strong dependence on the coupling between the normal stretching and the the model constant $C_S$. The same trend was found in section \ref{turb} for the mean streamwise velocity and the momentum thickness prediction at $x = 12\Lambda$. In the mentioned sections, the sensitivity to the SGS model constant significantly decreases and a minimum of the global coefficient of variance $c_v$ is also found, as shown in figure \ref{fig:VarPlot}. Summarizing, a significant drop of the flow sensitivity to $C_S$ and a coupling between $strY$ and $C_S$ is observed in proximity of the latest stages of transition: this characteristic may, thus, be seen as an indicator of an incipient fully developed turbulent state.

\section{Error behavior for the dynamic Smagorinsky model}
\label{dynamic}

{The error cost function is now investigated starting from the database of $4^2$ LES simulations applying the dynamic version of the Smagorinsky model. The aim of this analysis is to observe the main differences in the error dynamics using a more complex model.}

{The mean stochastic values and the coefficients of variance are reported in Table \ref{tab:turbStatU_dyn} and \ref{tab:transStatU_dyn} for the fully-turbulent and the inlet-dependent regions respectively. Comparing the results with those obtained for the classical Smagorinsky model reported in the Tables \ref{tab:turbStatU} and \ref{tab:transStatU}, no significant differences are observed in the predicted stochastic mean values for $eCF_U$ of $eCF_{\theta}$. Conversely, $\overline{eCF_{\tau}}$ is significantly smaller, meaning that dynamic SGS model on the average leads to a better prediction for this quantity. Another important conclusion that can be drawn from the coefficients of variation is that the error cost function is less sensitive to the set of random variables. This could have been expected since it has been previously observed that for the Smagorinsky model the error variance is dominated by the value of $C_S$. The drop in the sensitivity is however not as intense as the partial contribution to the variance due to the model constant found for the Smagorinsky model: in fact, the $C$ values computed by the dynamic procedure depend on the local grid topology and, in particular, on the local stretching. Therefore, for the dynamic model the effect of the stretching ratios on the introduced SGS viscosity is larger than for the Smagorinsky one and the sensitivity of the error to those parameters is hence increased.}

{A comparison of the pdfs reported in Figure \ref{fig:turbUpdf_dyn} to \ref{fig:transSpdf_dyn} with the ones recovered for the classical Smagorinsky model database in Figure \ref{fig:turbUpdf} to \ref{fig:transSpdf} shows that the range of the error prediction is approximately the same when $eCF_U$ and $eCF_{\theta}$ are considered. In particular, the pdfs recovered for $eCF_U$ are extremely similar to the ones computed for $eCF_{\theta}$ in the turbulent region $12 \leq x / \Lambda \leq 20$. This feature has not been observed in the analysis reported in Section \ref{turb}. The probabilistic distribution of $eCF_{\tau}$ is bounded to a significantly shorter range if compared to the one computed from the classical Smagorinsky database for each streamwise section considered. As in the case of the classical Smagorinsky model, the most probable event is not significantly close to the stochastic average: when the transition region is considered, the error mean value is usually larger in magnitude.}

\section{Parameter optimization}
\label{optimisationHPC}

The results presented in sections \ref{turb} and \ref{transition} point out significant differences in the error response surfaces when considering different physical reference quantities and different flow regimes. In this section, the gPC representation of the error over the parameter uncertainty space is used to find optimal parameter values, i.e. the values leading to the minimum of the error, for the different considered physical quantities and domain regions. The minimum has been recovered through the comparison of a high number of samples over the parameter space using the same methodology employed to recover pdfs (see section \ref{turb}).  

Let us first consider the results of this optimization process in the inlet-dependent region. The first three rows of table \ref{tab:optPoint} report the parameter values, in the considered range, corresponding to a minimum value of the eCF for each physical quantity of interest. These optimal values are computed from the the already introduced \textit{average} results in the inlet-dependent region, which represent a general trend of the error in the considered zone. As expected, different optimal values are obtained for the different considered quantities. Therefore, the optimum has also been computed for a global eCF function, built as $eCF_G = a \; eCF_U + b \; eCF_{\Theta} + c \; eCF_{\tau}$, where $a$, $b$ and $c$ are weight coefficients that may be tuned if the optimization of one of the physical quantities is considered more important than that of the other ones. The three weight coefficients have been set to $1$ herein, so that the relative importance of a physical quantity is related to the magnitude of the error only. The global optimal parameters obtained in this way, reported in the fourth row of table \ref{tab:optPoint}, correspond to a $strX$ value belonging to the lower part of the considered range, to a high value of $strY$ and to $C_S$ also belonging to the lower part of the considered range. In particular the optimal $C_S$ appears to be significantly lower than the value of 0.1, usually suggested for shear flows. We stress again that this parameter set represents a global optimum for the considered physical quantities, but the values minimizing the error for each single quantity are different. Thus, it is interesting to analyze whether the global optimum combination of the parameters leads to low errors also on each single quantity. Indeed, in practical applications it is usually sufficient to identify combinations of parameters yielding acceptably low errors, also if these combinations do not exactly correspond to the minimum of the error.
To this aim the volumes in the parameters space verifying the relation eCF $< 1.2$ $*$ min(eCF) for each of the considered quantities are reported in figures \ref{fig:optVolumesTrans}(a)-\ref{fig:optVolumesTrans}(c) together with a black point indicating the corresponding optimum configuration parameter set. The volumes relative to the different physical quantities clearly overlap each other, indicating that a zone in the parameter space corresponding to low errors in the prediction of all the considered physical quantities is detectable. Note that the global optimum parameter set, identified by a square, lies inside all the low-error regions in figures \ref{fig:optVolumesTrans}(a)-\ref{fig:optVolumesTrans}(c) and this shows that this parameter set thus actually corresponds to a low error value on all the considered variables. This is confirmed by the magnitude of the minimum error computed for the global eCF function, which is only $10\%$ larger than the sum of the three minimum error values for the single reference quantity eCF. Finally, the volume in the parameter space in which the global eCF is lower than 1.1 the minimum value, reported in figure \ref{fig:optVolumesTrans}(d), shows that it is possible to recover error values close to the minimum in a large part of the parameter space. Thus, small uncertainties over the chosen parameters do not lead to an abrupt increase of the error. 

The results of the same optimization procedure carried out the turbulent region are reported in the second part of table \ref{tab:optPoint}. As a first observation, except for the stretching ratio $strX$, the minimum errors for the different considered quantities are obtained for completely different parameters sets. Note how the opposite extreme values of $strY$ and $C_s$ in the considered range are found as optimal for the prediction of the streamwise velocity and of the momentum thickness respectively. It is also worth noting that the minimum of the error for each single quantity is obtained for the maximum $strX$ value considered in the analysis, while the minimum of global error function previously defined is found for the minimum considered value of $strX$. This behavior can be better understood by looking at the regions of the parameter space in which eCF $< 1.75$ $*$ min(eCF), reported in figures \ref{fig:optVolumesTurb}(a)-\ref{fig:optVolumesTurb}(c). Although we are looking for parameter space regions in which the error can be up to 75\% larger than the minimum one, the zones identified by this criterion for the different physical quantities do not overlap, confirming that there is no combination of the parameters yielding to low errors in the prediction of all the physical quantities. For instance, the global optimum does not lie in the low-error region for the mean streamwise velocity. This point out that, performing a global optimization, poor results for the mean streawise velocity prediction are obtained, with an error close to its maximum value. However, since the error on the streamwise velocity prediction is globally lower than that on the other quantities, this has only a limited impact on the global error cost function. As a consequence of this difficulty in finding parameter values leading to low error values on all the considered quantities, the global eCF minimum error is more than $50\%$ greater in magnitude than the sum of the three minimum errors computed over the single quantity eCFs. 
Finally, figure \ref{fig:optVolumesTurb}(d) shows the part of the parameter space complying with the relation eCF$_G$ $< 1.1*$ min(eCF$_G$). Differently from the results observed for the global optimum zone computed for the inlet-dependent domain, it can be seen that small changes in the parameters may lead to a significant increase of the error, when considering a global optimization over the turbulent region. 

As a final remark, the parameter values leading to a global optimum error in the turbulent region are quite close to the ones computed for the inlet-dependent zone. This suggests that there are combinations of the parameters which lead to acceptably low global errors, independently of the flow local characteristics. This consideration does not hold true if each single physical quantity is considered.

\section{Concluding remarks}
\label{conclusionHPC}
The error in the predictions of different physical quantities obtained in large-eddy simulations of a spatially-developing mixing layer has been quantified and investigated. The considered physical quantities are the mean streamwise velocity, the momentum thickness and the mean shear stress. The errors have been evaluated with respect to the results of a highly-resolved LES, considered here as the reference solution. The error sensitivity to changes in the values of some simulations parameters, viz. the grid stretching in the streamwise and lateral directions and the constant in the SGS Smagorinsky model, has also been assessed. Thanks to the use of the generalized polynomial chaos representation, error response surfaces in the parameter space could be built from a limited number of simulations. This approach allows a probabilistic characterization of the error, as well as a quantification of its sensitivity to each considered parameter and to their combinations. A typical feature of the considered spatially evolving flow is the progressive transition from a laminar regime, highly dependent on the inlet conditions, to a fully-developed turbulent one. Thus, the computational domain has been divided in a transitional upstream region, in which the signature of the inlet perturbation on the flow development is still evident, and in a downstream turbulent zone. The error analysis has been carried out separately in the two zones, in order to investigate in which extent the error behavior is quantitatively and qualitatively affected by the local flow characteristics. 

In all the considered parameter range and in both the inlet-dependent and turbulent zones, the error on the streamwise velocity is generally very low, much lower than for the other two quantities. For all the quantities, the mean, most probable, maximum and minimum error values significantly change with the streamwise location. In the inlet-dependent zone, moving downstream, the stochastic error decreases, as well as the most probable error value and the magnitude of the extreme error values. This indicates that the most critical part of this region is connected with the initial stages of transition. The opposite behavior of the error is observed in the turbulent region; this is probably related to a progressively coarsening of the streamwise resolution when moving dowstream (the grid nodes are clustered near the inlet), together with the presence of progressively smaller turbulent flow scales. 

The error variance, and thus the sensitivity to changes in the parameter values, also noticeably depends on the streamwise location. For all the considered quantities, inside the inlet-dependent region, at the most upstream section the error variance is relatively low but it is associated with a large value of the error for the momentum thickness and the shear stress. Moving downstream the error coefficient of variance progressively increases reaching a maximum around $x=6 - 8 \Lambda$ and then decreases, reaching a minimum around $x=10 - 12 \Lambda$. In the turbulent zone the error variance increases again, but it remains lower than in the inlet-zone. Thus, for all the considered quantities, the sensitivity of the error to the parameters is found to be generally higher in the inlet-dependent zone than in the turbulent one.

As for the sensitivity of the error to the single parameters, the parameter having the largest impact of the error, for all the considered quantities and in both zones, is the Smagorinsky constant. In the inlet-dependent zone the error has been found to be noticeably sensitive to the grid stretching in the lateral direction, probably because of the strong lateral velocity gradient still present in this region. Conversely, in the turbulent zone the stretching in the streamwise direction has been found to have a significant impact on the errors in the prediction of the mean velocity and of the momentum thickness; this is related to the progressive lack of grid resolution when moving downstream, which becomes more pronounced when the stretching ratio $strX$ is increased.

{The application of a dynamic subgridscale model does not significantly alter neither quantitatively nor qualitatively the error behavior for the streawise average velocity and the momentum thickness, while a significant improvement of the accuracy in the shear stress prediction is observed. This indicates that through a proper tuning of the model constant an accuracy comparable to the one of the dynamic model can be recovered with the Smagorinsky closure, except for the shear stress. Moreover, since a similar qualitative error behavior has been obtained with the different considered SGS closures, we may argue  that the conclusions previously drawn might be rather independent of the used SGS closure.}

Finally, the gPC representation of the error over the parameter uncertainty space has been used to find optimal parameter values, i.e. the values leading to the minimum of the error, for the different considered physical quantities. For both regions, these optimal values are computed from  \textit{average} results, which represent a general trend of the error in the considered zones. A set of parameter values minimizing a global error, i.e. the sum of the errors in the predictions of each single quantity, has also been computed. In the inlet-dependent region the parameter values minimizing the global error also yield low errors in the prediction of each physical quantity and this optimum is \textit{robust}, i.e. small changes in the parameter values do not lead to an abrupt increase of the error. Conversely, in the turbulent region, performing a global optimization, does not guarantee low errors on the single quantities; for instance, the globally optimal parameter values bring to poor results for the mean streamwise velocity prediction, with an error close to its maximum value and the global minimum error is more than $50\%$ greater in magnitude than the sum of the three minimum errors computed for the single quantities.   

Summarizing, in the transitional regime LES may produce a maximum error huge in magnitude and that the variance over the uncertainty space is quite high, but a common region at low error for all the physical quantities is recoverable and the resulting error is not extremely sensitive to small parameter variations. In the turbulent regime the error is globally lower and it is less sensitive to the variables analyzed, meaning that an inappropriate  choice of the parameters values leads to a flow prediction closer to the correct values if compared with predictions performed over the transition part of the domain: anyway, the parameters have to be selected with extreme care to reach a satisfying optimization.

HPCEuropa2 (project nr. 267) and GENCI-CINES computation centers are acknowledged for the resources furnished to generate the database used in the sensitivity analysis.

\appendix
\section{Validation of the accuracy of the stochastic approach}
The accuracy of the proposed stochastic approach is assessed considering as an example the coefficients $a_k$ for the dynamic Smagorinsky database, without loss of generality. The validation is performed considering the error cost function for the momentum thickness $\theta$ recovered at the section $\mbox{x / } \theta \mbox{= 12} $. The coefficients $a_k$ reported in Figure \ref{fig:gPCaccuracy}, which are grouped accordingly to the order of the associated polynomial, are renormalized over a basis $[-1, \, 1] \times [-1, \, 1]$: in this way a clear comparison of the coefficients associated to different variables can be performed. The coefficient $a_0$ is the stochastic mean value of the recovered error cost function, while the other coefficient can be used to recover the total variance as $V = \sum_{i=1}^M a_k^2 \, W(i)$. The reader can observe in Figure \ref{fig:gPCaccuracy} that contribution of higher order polynomials is not significant if compared to low order polynomials: it has been observed that the first order polynomials account for the $40\% \, - \, 70\%$ of the total variance while the third order polynomials affect the solution for the $1\% \, - \, 5\%$. Being the effects of higher order polynomials less and less significant in the spectral projection, we can state that an accuracy of $\approx 1\%$ in the results is recovered when representing the error cost functions through a third order polynomial expansion.

\newpage
\section*{tables}
\begin{table*}[h]
\centering
\begin{tabular}{lcccccc}
 & $x = 2\Lambda$ & $x = 6\Lambda$ & $x = 10\Lambda$ & $x = 12\Lambda$ & $x = 16\Lambda$ & $x = 20\Lambda$ \cr \hline
Correlation coefficient & $1$ & $0.47$ & $0.24$ & $0.09$ & $0.19$ & $0.15$ \cr
\end{tabular}
\caption{Correlation coefficient between spanwise energy spectra at different streamwise locations and the spectrum at $\Lambda = 2$.}
\label{tab:corrCoeff}
\end{table*} 
\begin{table*}[h]
\centering
\begin{tabular}{l|ccc|ccc|ccc}
 & $eCF_U^d$ &$\overline{eCF_U}$& ${c}_{vU}$& $eCF_{\Theta}^d$ &$\overline{eCF_{\Theta}}$& ${c}_{v \Theta}$ & $eCF_{\tau}^d$ &$\overline{eCF_{\tau}}$& ${c}_{v \tau}$ \cr \hline
x = 12$\Lambda$ & 4.06e-003 & 3.54e-003 & 21.77\% & 0.055 & 0.053 & 16.89\% & 0.076 & 0.104 & 54.30\% \cr
x = 14$\Lambda$ & 5.87e-003 & 4.40e-003 & 26.12\% & 0.072 & 0.057 & 20.09\% & 0.207 & 0.145 & 60.51\% \cr
x = 16$\Lambda$ & 7.40e-003 & 5.86e-003 & 27.03\% & 0.086 & 0.070 & 21.79\% & 0.275 & 0.177 & 59.90\% \cr
x = 18$\Lambda$ & 8.87e-003 & 7.62e-003 & 29.61\% & 0.096 & 0.082 & 29.32\% & 0.386 & 0.201 & 62.85\% \cr
x = 20$\Lambda$ & 11.1e-003 & 9.66e-003 & 32.05\% & 0.112 & 0.098 & 33.45\% & 0.437 & 0.212 & 57.05\% \cr
$Average$ 		  & 7.46e-003 & 6.21e-003 & 26.62\% & 0.084 & 0.072 & 22.26\% & 0.276 & 0.168 & 58.14\% \cr
\end{tabular}
\caption{Statistical properties of eCF relative to the considered physical quantities of interest in the turbulent region: considered a flow quantity $S$, the deterministic error $eCF_S^d$, the stochastic mean error $\overline{eCF_S}$ and the coefficient of variation $c_{vS}$ are respectively reported.}
\label{tab:turbStatU}
\end{table*} 
\begin{table*}
\centering
\begin{tabular}{l|ccc|ccc|ccc}
  & $eCF_U^d$ &$\overline{eCF_U}$& ${c}_{vU}$& $eCF_{\Theta}^d$ &$\overline{eCF_{\Theta}}$& ${c}_{v \Theta}$ & $eCF_{\tau}^d$ &$\overline{eCF_{\tau}}$& ${c}_{v \tau}$ \cr \hline
x =  2$\Lambda$ & 2.91e-003 & 2.92e-003 & 8.14\% 	& 0.169 & 0.16  & 14.01\% & 0.783 & 0.718 & 31.06\% \cr
x =  4$\Lambda$ & 2.91e-003 & 3.75e-003 & 29.15\% & 0.101 & 0.142 & 40.81\% & 0.618 & 0.312 & 75.70\% \cr
x =  6$\Lambda$ & 1.76e-003 & 3.49e-003 & 54.66\% & 0.052 & 0.102 & 52.43\% & 0.393 & 0.177 & 86.48\% \cr
x =  8$\Lambda$ & 1.60e-003 & 3.16e-003 & 62.81\% & 0.037 & 0.075 & 57.48\% & 0.236 & 0.114 & 59.56\% \cr
x = 10$\Lambda$ & 2.75e-003 & 3.01e-003 & 44.98\% & 0.044 & 0.059 & 44.68\% & 0.103 & 0.071 & 36.33\% \cr
$Average$ 			& 2.39e-003 & 3.27e-003 & 37.92\% & 0.08  & 0.108 & 34.09\% & 0.427 & 0.278 & 38.80\% \cr
\end{tabular}
\caption{Statistical properties of eCF relative to the considered physical quantities of interest in the inlet-dependent region: considered a flow quantity $S$, the deterministic error $eCF_S^d$, the stochastic mean error $\overline{eCF_S}$ and the coefficient of variation $c_{vS}$ are respectively reported.}
\label{tab:transStatU}
\end{table*}
\begin{table*}[h]
\centering
\begin{tabular}{l|cc|cc|cc}
 							  &$\overline{eCF_U}$& ${c}_{vU}$& $\overline{eCF_{\Theta}}$& ${c}_{v \Theta}$ & $\overline{eCF_{\tau}}$& ${c}_{v \tau}$ \cr \hline
x = 12$\Lambda$ & 4.17e-003 			 & 17.49\% 	 &  0.057 									& 14.89\%	 				 & 0.069 									& 33.19\% \cr
x = 14$\Lambda$ & 5.59e-003 			 & 15.04\% 	 &  0.069 									& 15.63\% 				 & 0.108 									& 40.25\% \cr
x = 16$\Lambda$ & 7.28e-003 			 & 11.95\% 	 &  0.086 									& 12.46\% 				 & 0.126 									& 53.37\% \cr
x = 18$\Lambda$ & 9.2e-003	 			 & 13.86\% 	 &  0.099 									& 16.61\% 				 & 0.151 									& 56.5\% \cr
x = 20$\Lambda$ & 11.4e-003 			 & 15.85\% 	 &  0.117 									& 17.69\% 				 & 0.157 									& 49.31\% \cr
$Average$ 		  & 7.53e-003 			 & 12.96\% 	 &  0.086 									& 13.96\% 				 & 0.122 									& 45.59\% \cr
\end{tabular}
\caption{Statistical properties of eCF relative to the considered physical quantities of interest in the turbulent region: considered a flow quantity $S$, the deterministic error $eCF_S^d$, the stochastic mean error $\overline{eCF_S}$ and the coefficient of variation $c_{vS}$ are respectively reported. The database generated through the application of the dynamic Smagorinsky model is considered.}
\label{tab:turbStatU_dyn}
\end{table*} 
\begin{table*}[h]
\centering
\begin{tabular}{l|cc|cc|cc}
 							  &$\overline{eCF_U}$& ${c}_{vU}$& $\overline{eCF_{\Theta}}$& ${c}_{v \Theta}$ & $\overline{eCF_{\tau}}$& ${c}_{v \tau}$ \cr \hline
x =  2$\Lambda$ & 2.95e-003 			 & 5.76\% 	 &  0.17	 									& 5.11\% 				 	& 0.374 									& 32.62\% \cr
x =  4$\Lambda$ & 3.49e-003 			 & 20.32\% 	 &  0.129 									& 27.83\% 				 & 0.22 									& 37.78\% \cr
x =  6$\Lambda$ & 2.52e-003 			 & 38.7\% 	 &  0.073 									& 36.63\% 				 & 0.102 									& 28.58\% \cr
x =  8$\Lambda$ & 2.26e-003 			 & 31.31\% 	 &  0.049 									& 35.31\% 				 & 0.077 									& 21.65\% \cr
x = 10$\Lambda$ & 2.85e-003 			 & 12.82\% 	 &  0.047 									& 14.54\% 				 & 0.058 									& 24.33\% \cr
$Average$ 		  & 2.81e-003 			 & 17.27\% 	 &  0.094 									& 19.07\% 				 & 0.166 									& 30.66\% \cr
\end{tabular}
\caption{Statistical properties of eCF relative to the considered physical quantities of interest in the inlet-dependent region: considered a flow quantity $S$, the deterministic error $eCF_S^d$, the stochastic mean error $\overline{eCF_S}$ and the coefficient of variation $c_{vS}$ are respectively reported. The database generated through the application of the dynamic Smagorinsky model is considered.}
\label{tab:transStatU_dyn}
\end{table*} 
\begin{table*}
\centering
\begin{tabular}{lccc}
 minimum eCF value & $strX$ & $strY$ & $c_S$ \cr \hline
 & \multicolumn{2}{c}{Inlet-dependent region} & \cr \hline
$U_x$ 					 & 5.87 & 12 	 & 0.06 \cr
$\Theta$ 				 & 3.72 & 12 	 & 0 \cr
$Shear$ $stress$ & 4.89 & 7.85 & 0.08 \cr
$Global$ 				 & 4.8	& 12   & 0.06	\cr \hline
 & \multicolumn{2}{c}{Fully developed turbulence region} & \cr \hline
$U_x$ 					 & 10 & 3 & 0 \cr
$\Theta$ 				 & 10 & 12 & 0.2 \cr
$Shear$ $stress$ & 10 & 10.61 & 0.05 \cr
$Global$				 & 3  & 8.77  & 0.072 \cr
\end{tabular}
\caption{Optimum values to minimise eCF in transition regime, turbulent regime and over all the considered domain.}
\label{tab:optPoint}
\end{table*}
\clearpage
\newpage
\section*{Figures}
\begin{figure}[h]
	\centering
	\begin{tabular}{cc}
		\includegraphics[width=0.45\linewidth,height=0.3\linewidth]{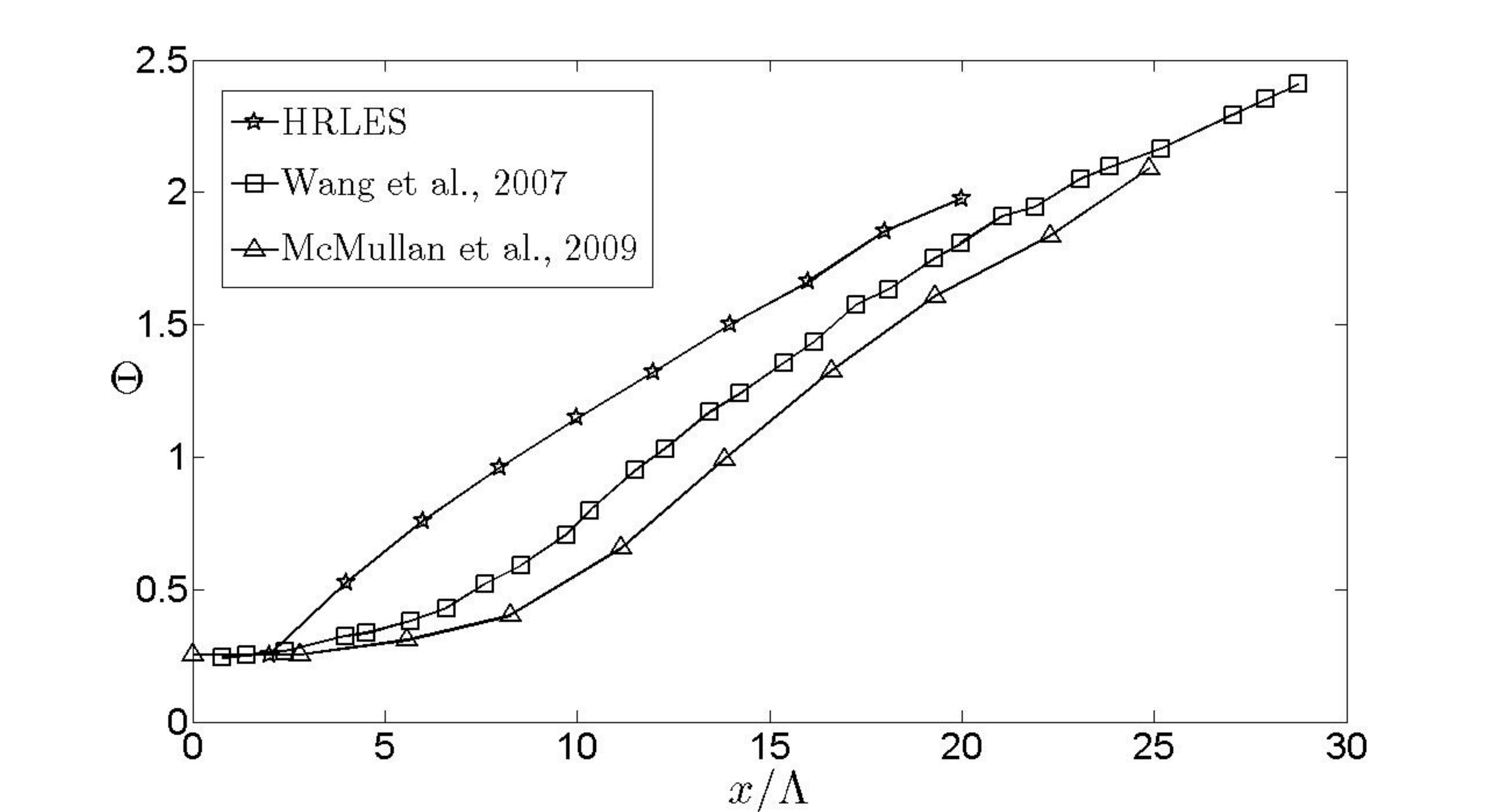} &
		\includegraphics[width=0.45\linewidth,height=0.3\linewidth]{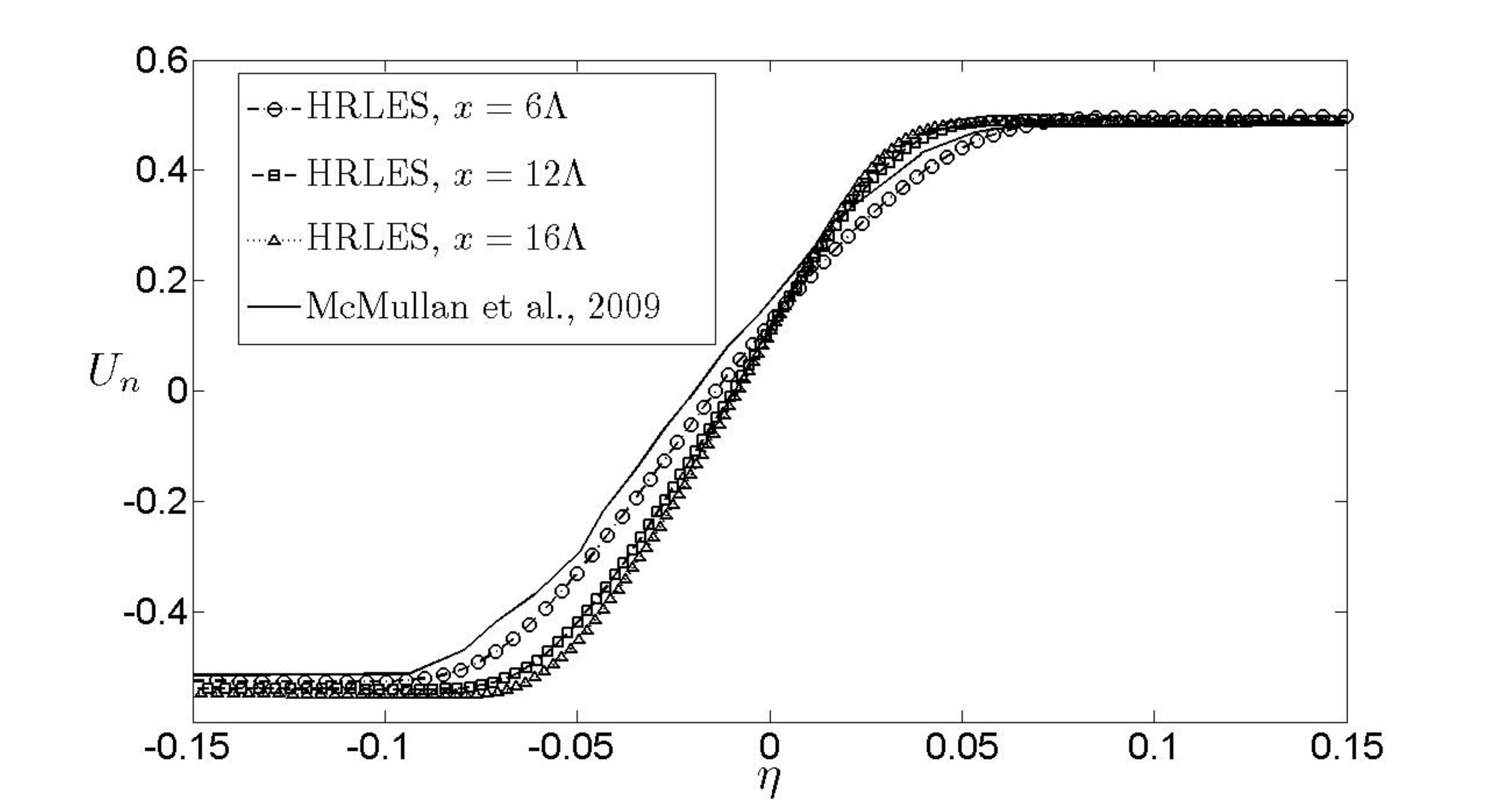}\\
		(a) & (b)
		\end{tabular}
	\caption{Comparison of HRLES results with reference data in the literature; (a) momentum thickness, $\Theta$, (b) normalized average streamwise velocity, $U_n$.}
	\label{fig001}
\end{figure}
\clearpage
\begin{figure*}[h]
	\centering
	\begin{tabular}{cc}
		\includegraphics[width=0.33\linewidth, angle=-90]{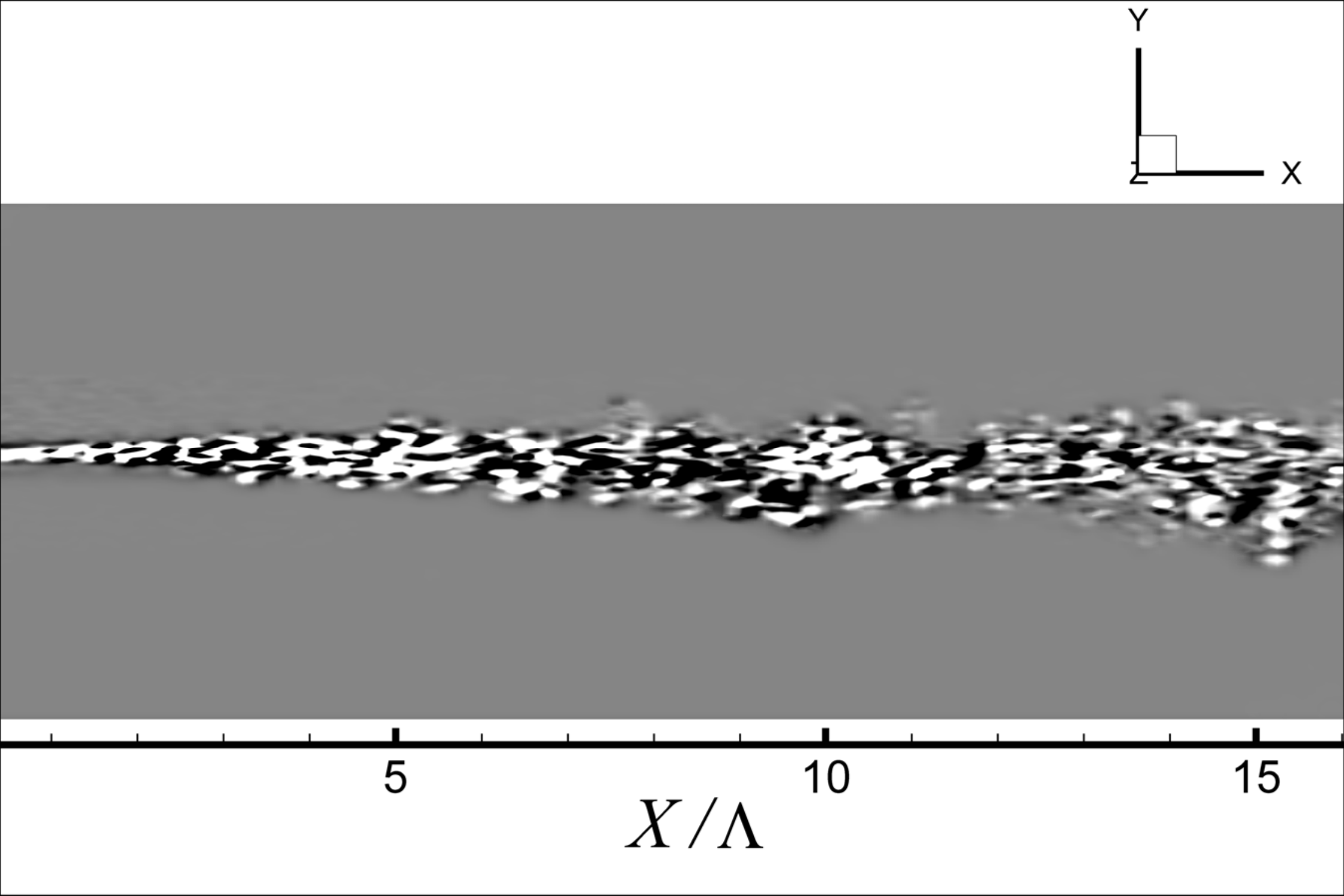}&
		\includegraphics[width=0.33\linewidth, angle=-90]{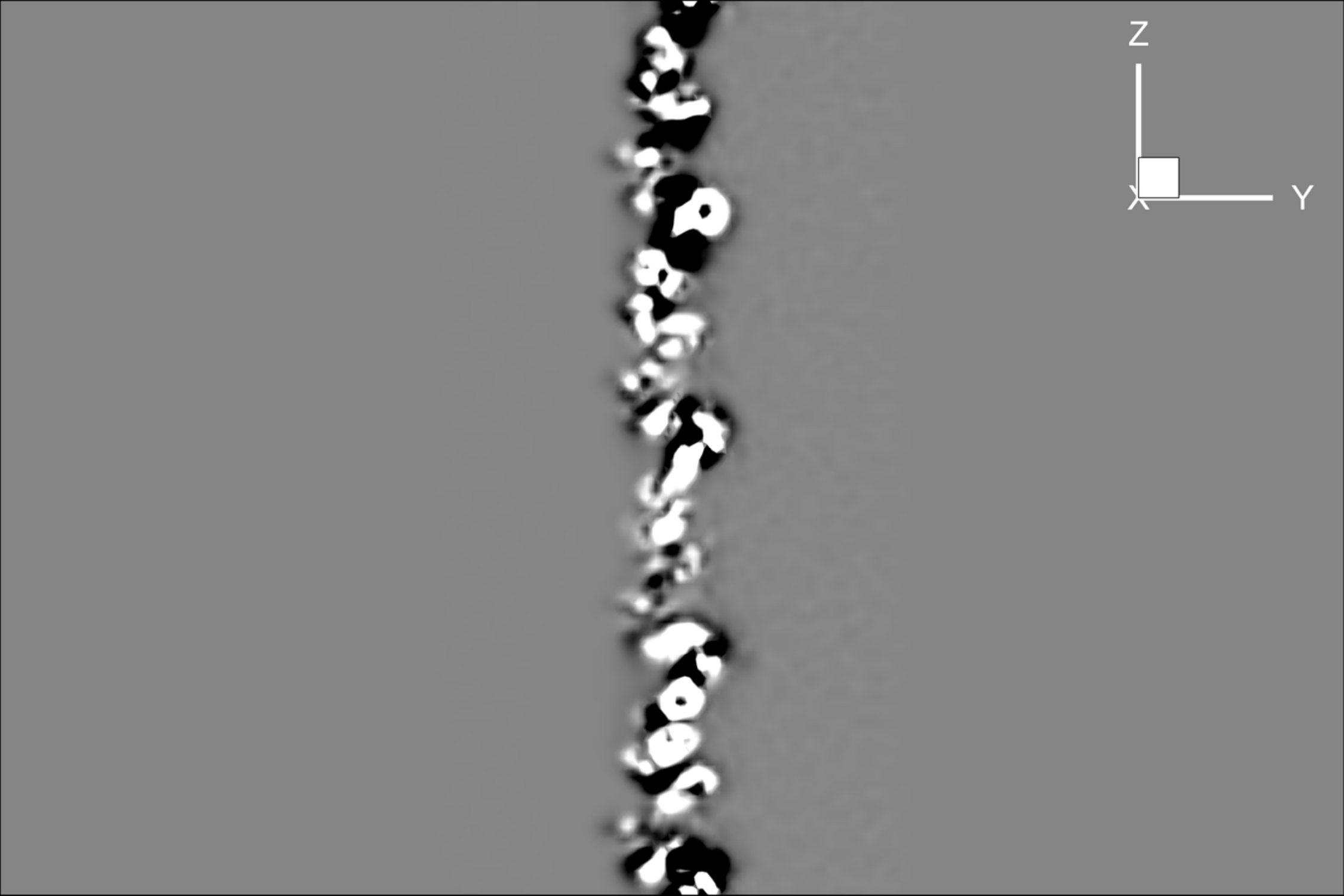}\\
		(a) & (b)\\
		\includegraphics[width=0.33\linewidth, angle=-90]{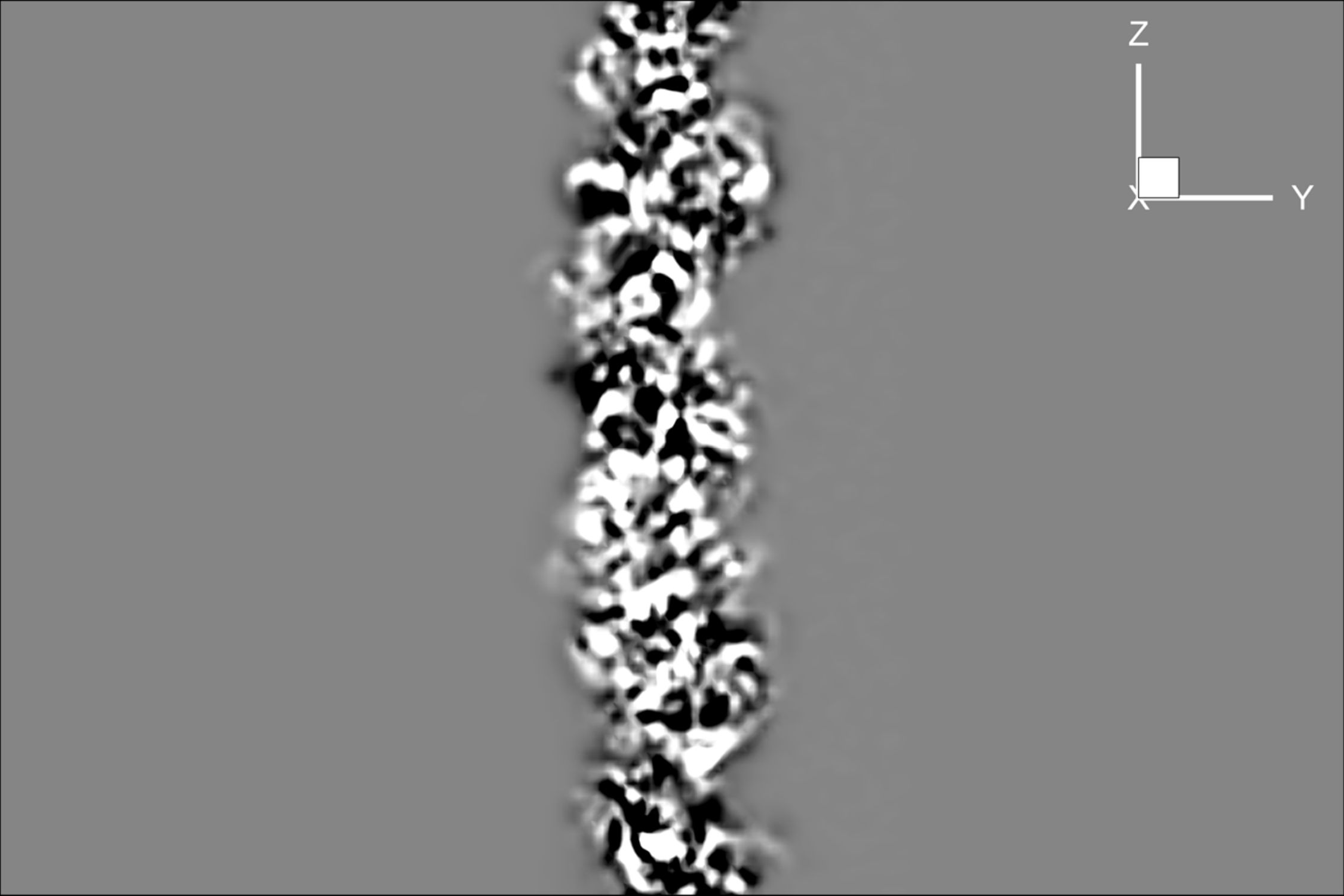}&
 		\includegraphics[width=0.33\linewidth, angle=-90]{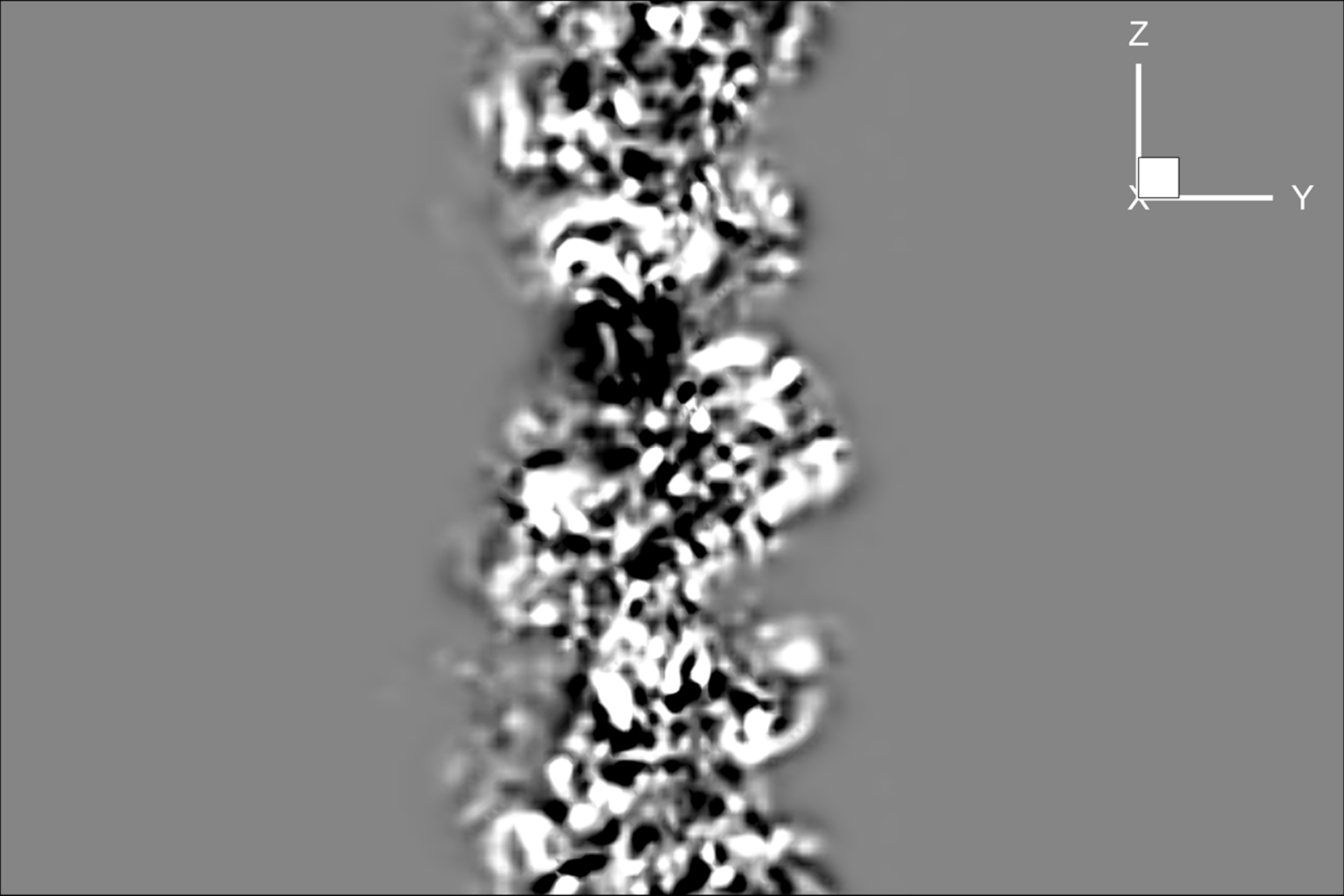}\\
 		(c) & (d)\\
 		\includegraphics[width=0.33\linewidth, angle=-90]{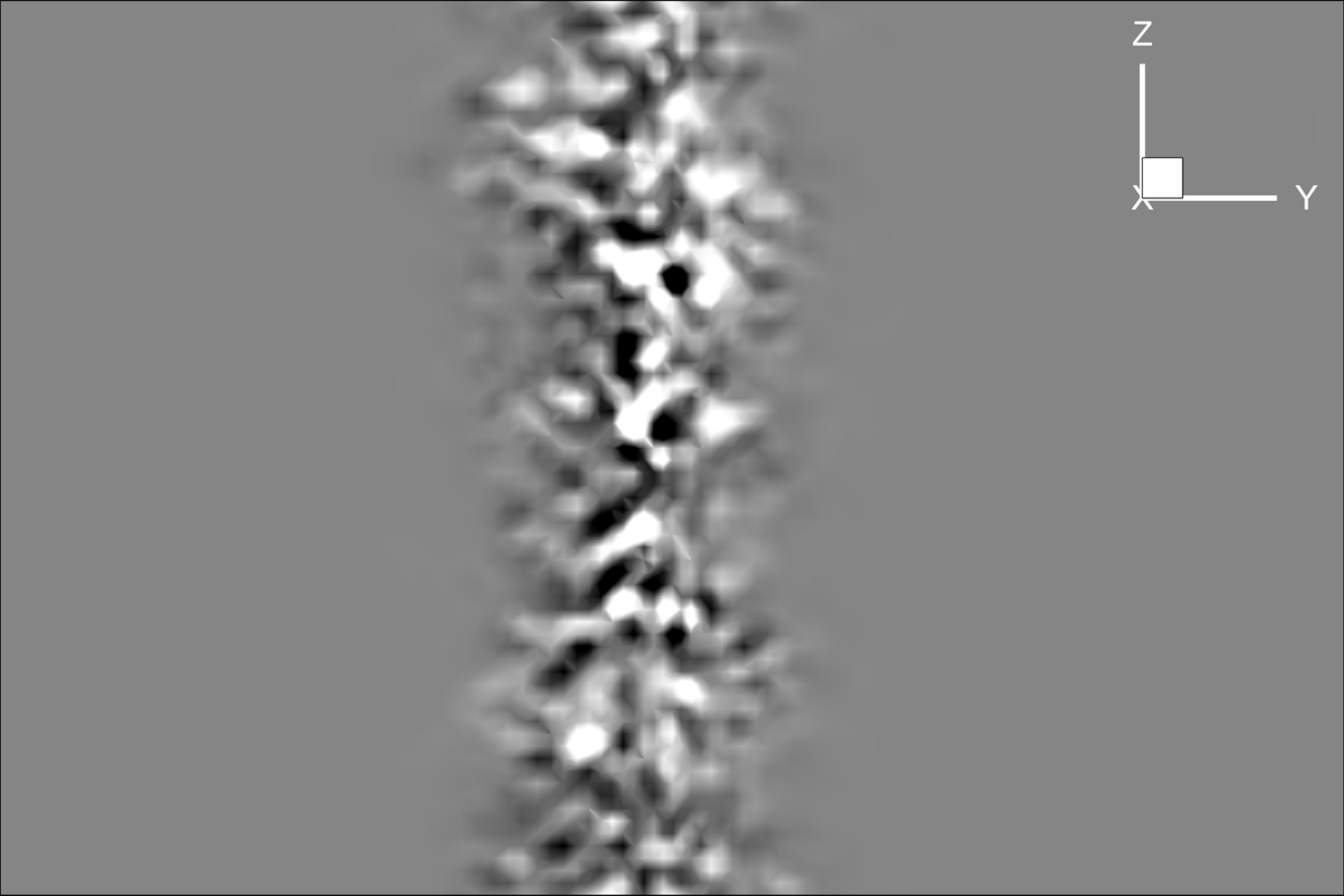}&
 		\includegraphics[width=0.33\linewidth, angle=-90]{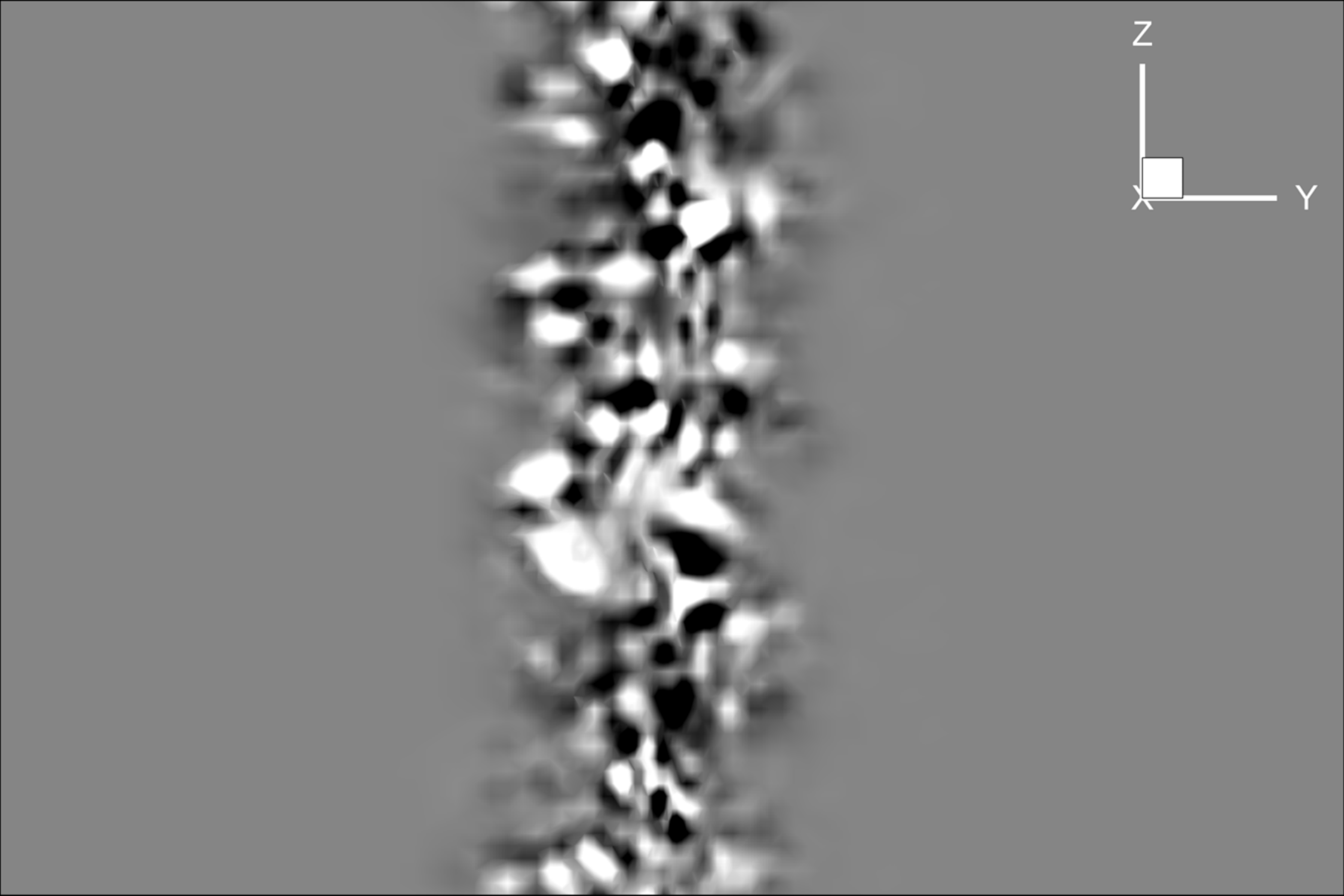}\\
 		(e) & (f)\\
 		\end{tabular}
	\caption{Isocountours of the second invariant of the velocity gradient tensor. HRLES isocontours are shown on a normal plane (a) and at the streamwise sections $x = 2\Lambda$(b), $x = 6\Lambda$(c) and $x = 14\Lambda$(d). Isocontours at the streamwise section $x = 14\Lambda$ are reported also for two LES simulations corresponding to two different Gaussian quadrature points (e)-(f).}
	\label{fig99}
\end{figure*}
\begin{figure*}[h]
	\centering
	\begin{tabular}{cc}
		\includegraphics[width=0.43\linewidth]{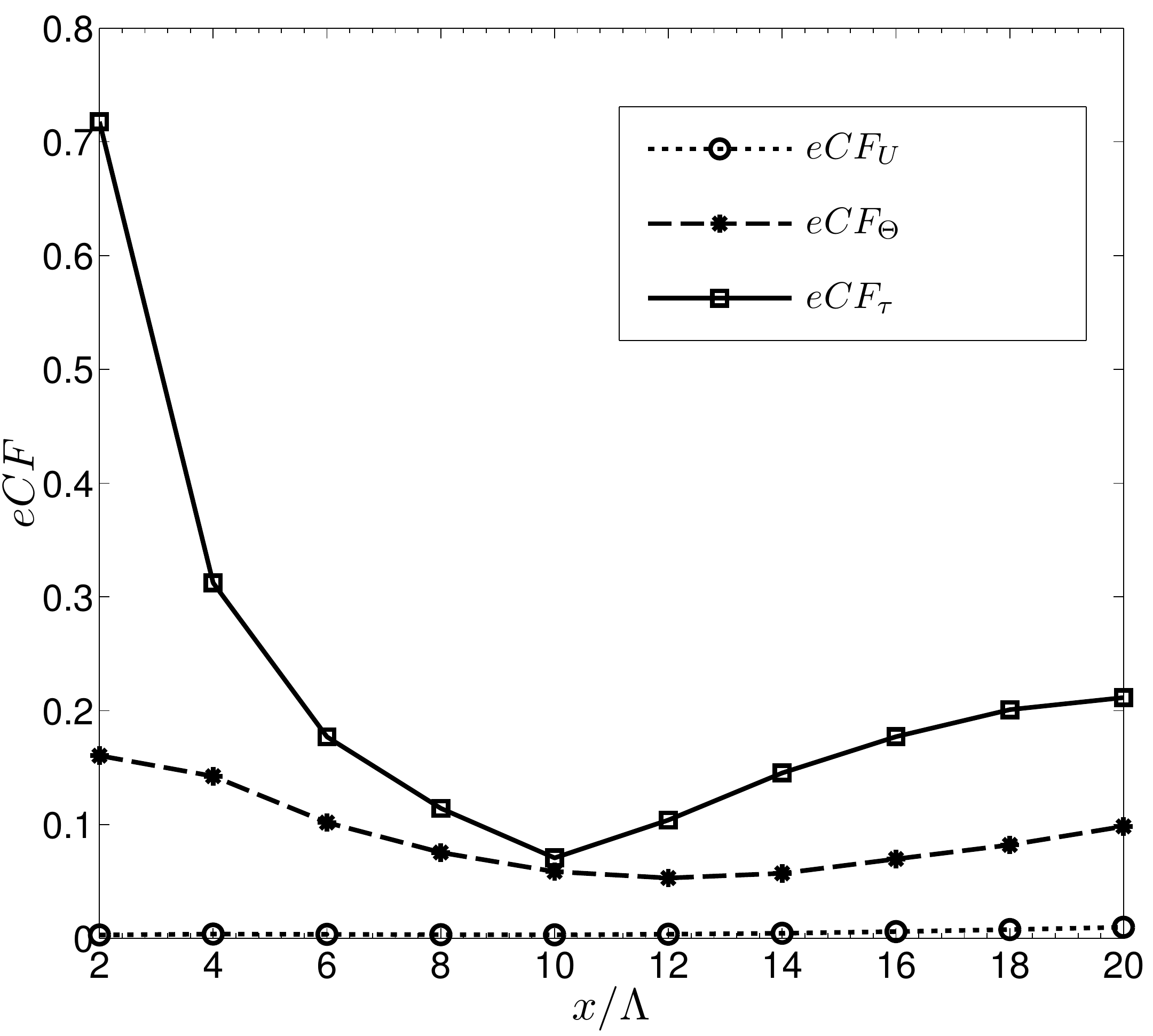}&
		\includegraphics[width=0.43\linewidth]{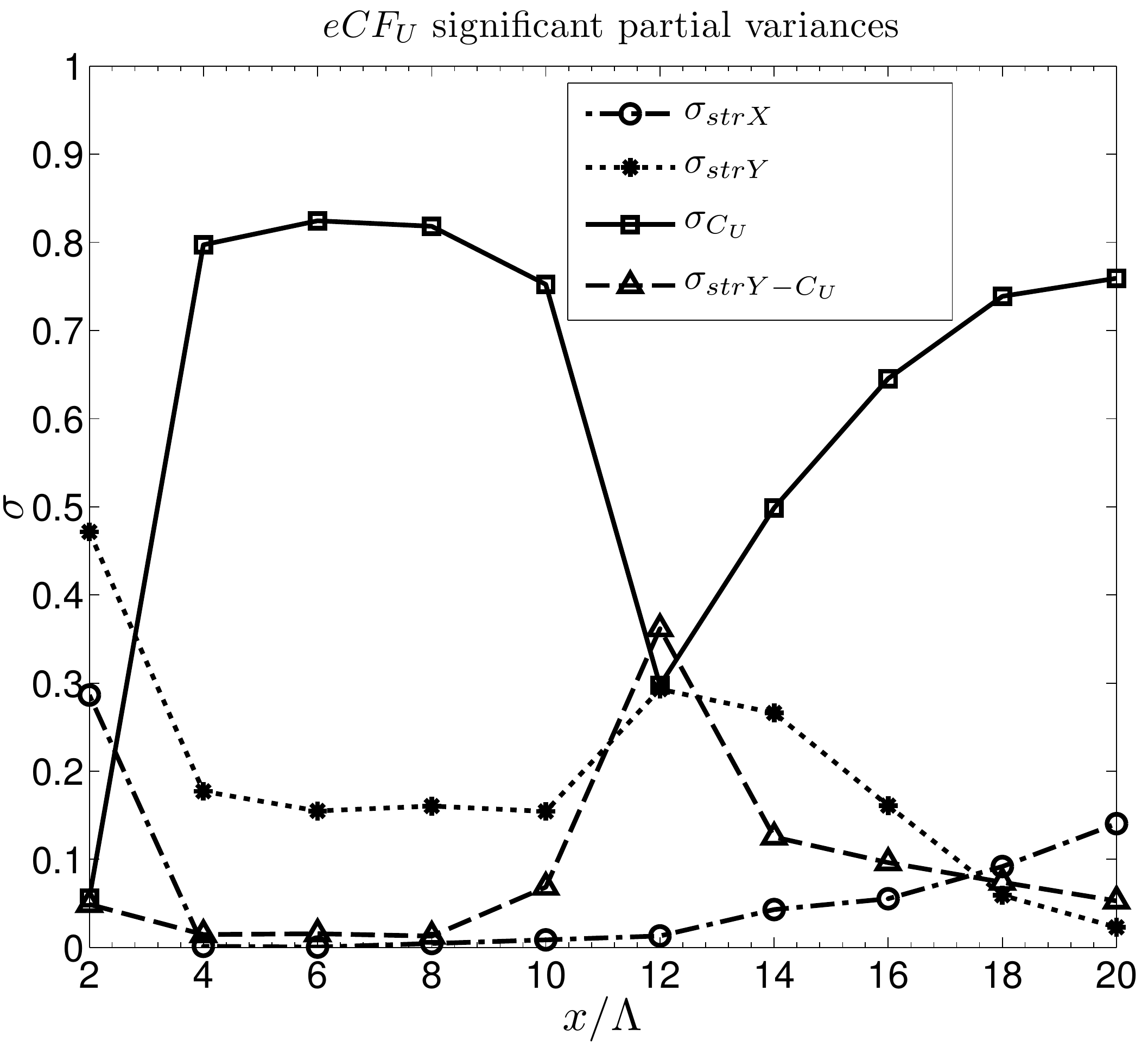}\\
		(a) & (b)\\
		\includegraphics[width=0.43\linewidth]{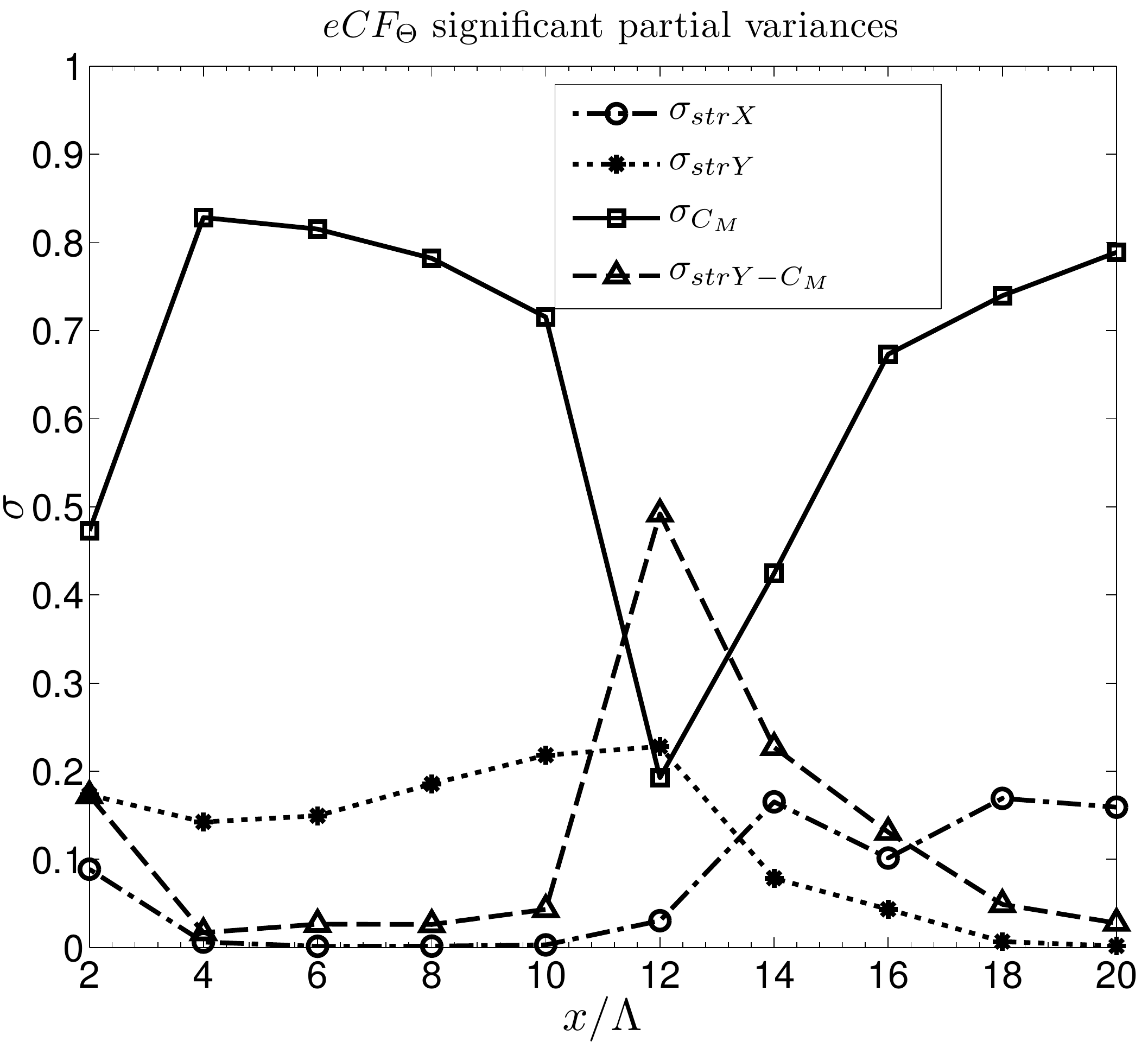}&
 		\includegraphics[width=0.43\linewidth]{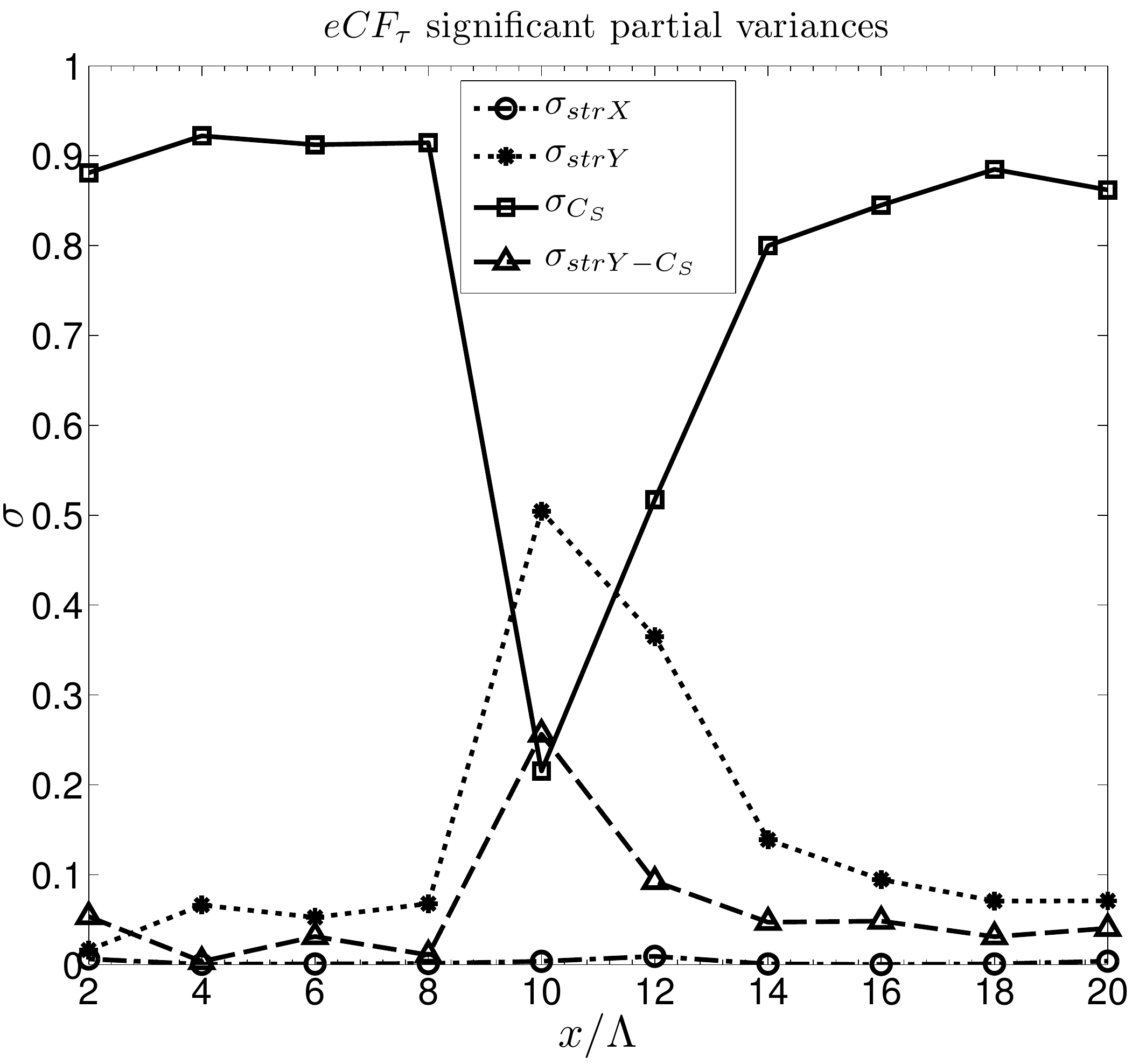}\\
 		(c) & (d)\\
 		\end{tabular}
	\caption{Statistical properties of the error cost function. A comparison between the magnitude of the $eCF$ computed for the different flow quantities is reported (a). The partial variances $\sigma$ of $eCF_U$, $eCF_{\Theta}$ and $eCF_{\tau}$ are shown in pictures (b),(c) and (d) respectively.}
	\label{fig:StatPropECF}
\end{figure*}
\begin{figure*}[h]
	\centering
		\includegraphics[width=0.45\linewidth,height=0.3\linewidth]{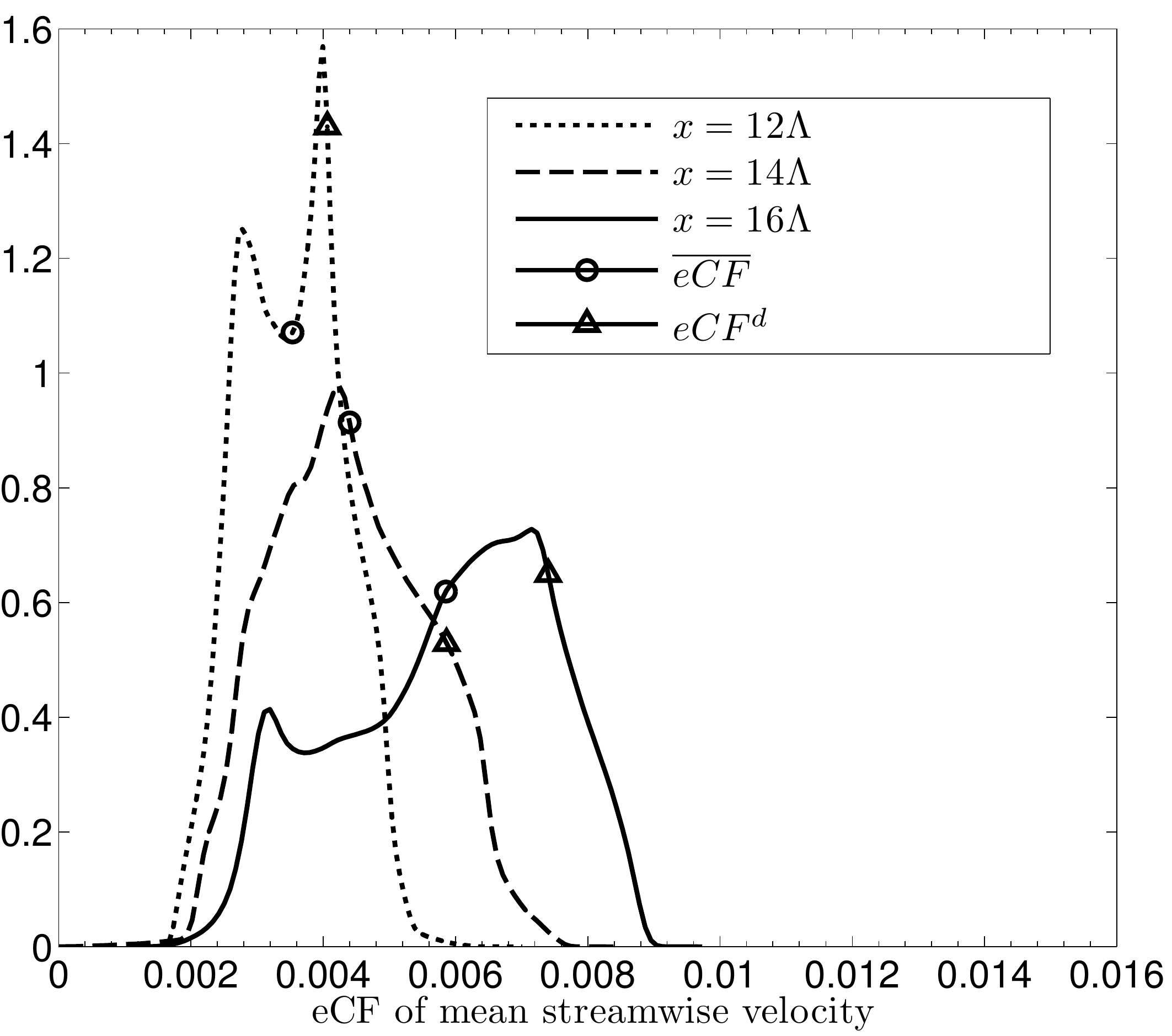}
		\includegraphics[width=0.45\linewidth,height=0.3\linewidth]{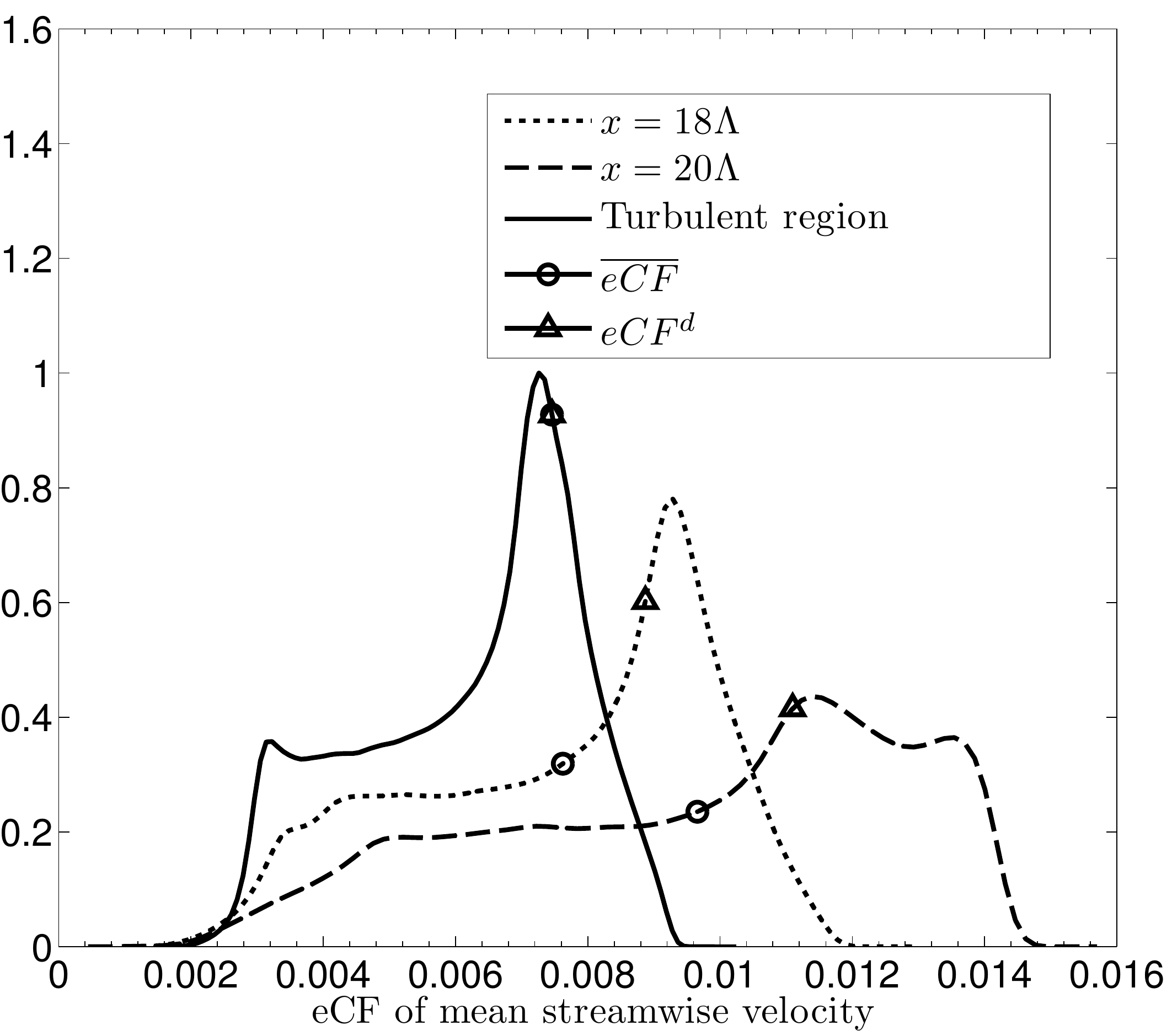}
	\caption{Streamwise velocity $U_x$ eCF pdfs in presence of a fully developed turbulent regime. The stochastic mean and the deterministic values are indicated by a circular mark and a triangular mark, respectively.}
	\label{fig:turbUpdf}
\end{figure*} 
\begin{figure*}[h]
	\centering
		\includegraphics[width=0.45\linewidth,height=0.3\linewidth]{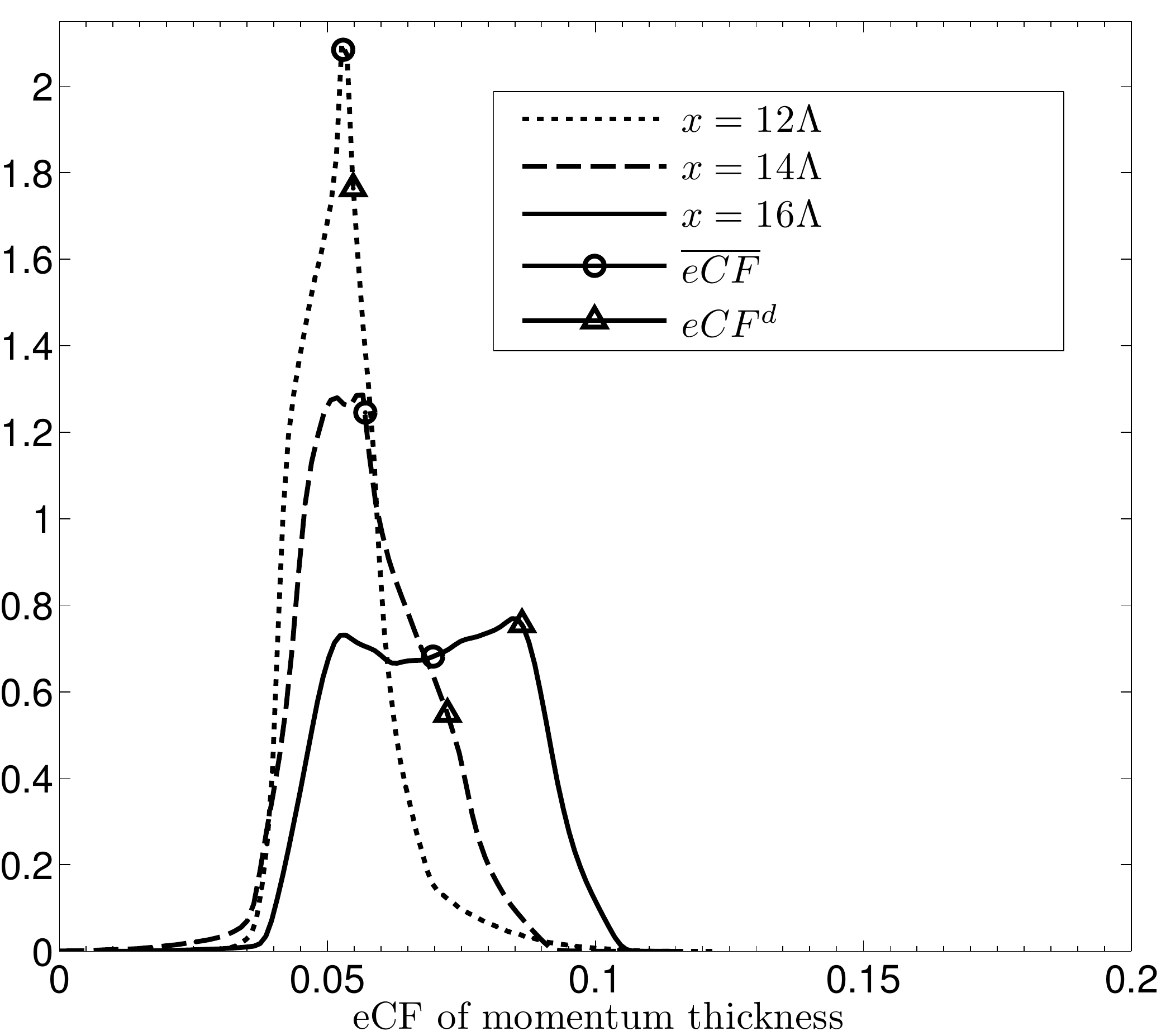}
		\includegraphics[width=0.45\linewidth,height=0.3\linewidth]{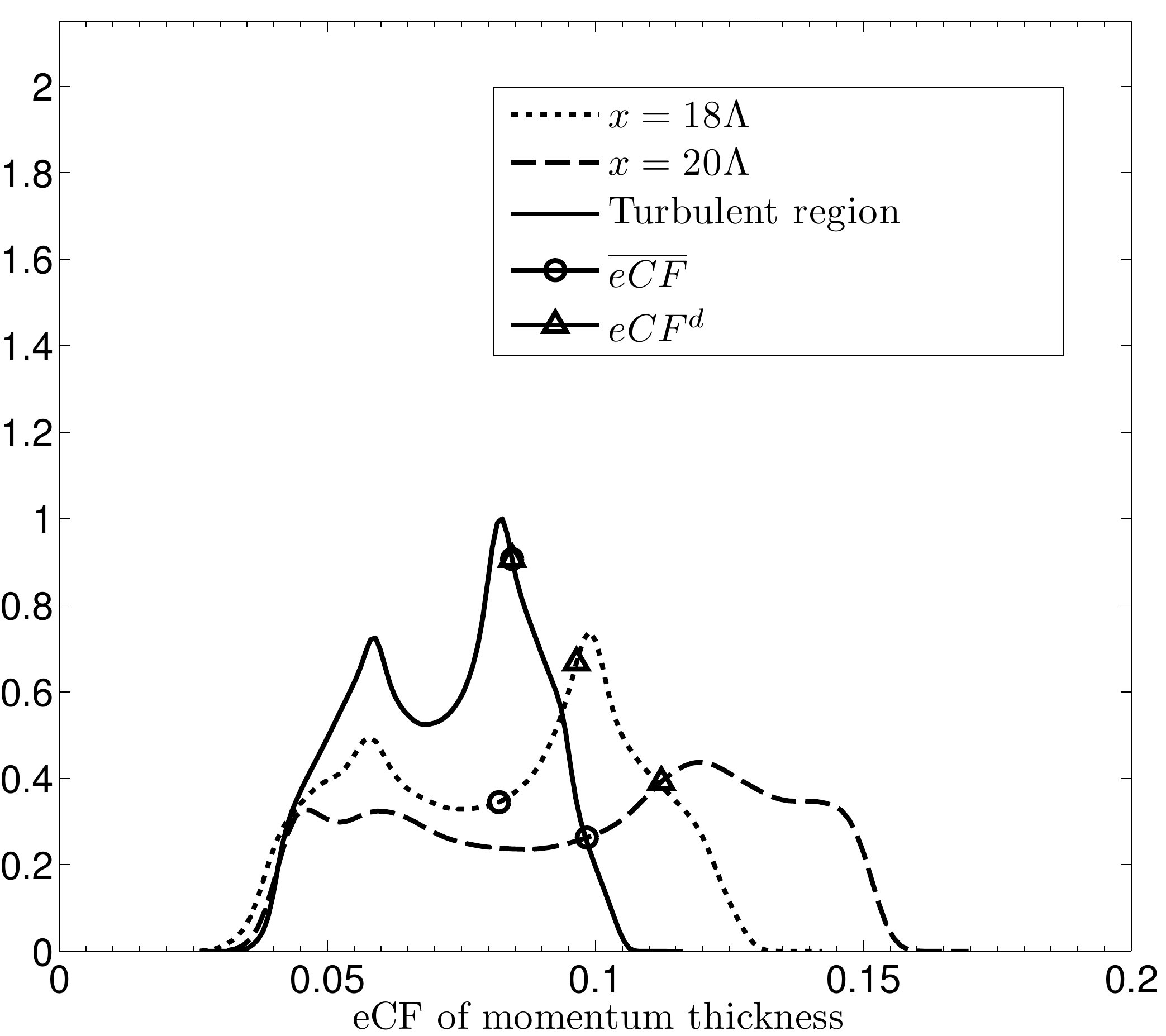}
	\caption{Momentum thickness $\Theta$ eCF pdfs in presence of a fully developed turbulent regime. The stochastic mean and the deterministic values are indicated by a circular mark and a triangular mark, respectively.}
	\label{fig:turbMpdf}
\end{figure*} 
\begin{figure*}[h]
	\centering
		\includegraphics[width=0.45\linewidth,height=0.3\linewidth]{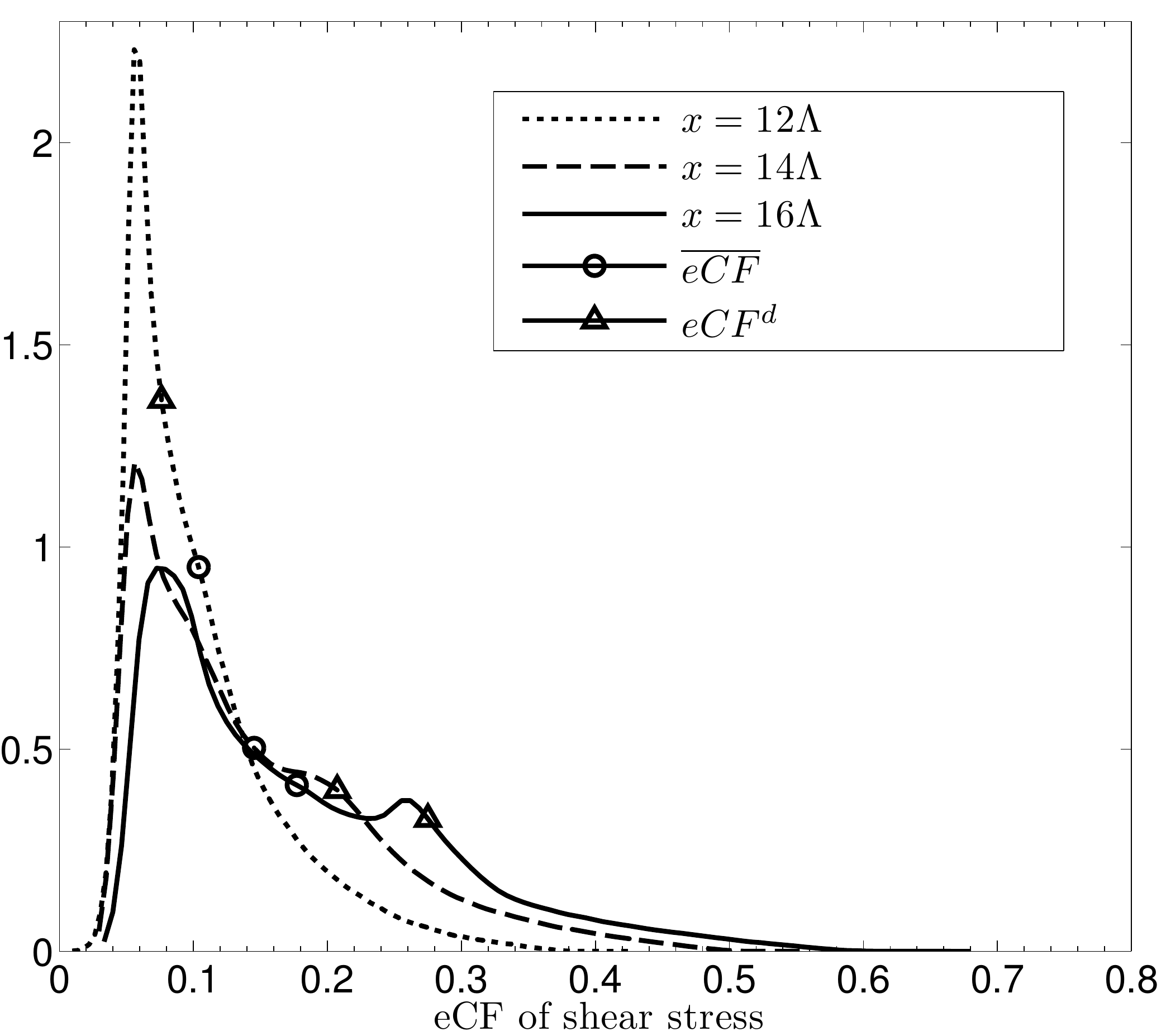}
		\includegraphics[width=0.45\linewidth,height=0.3\linewidth]{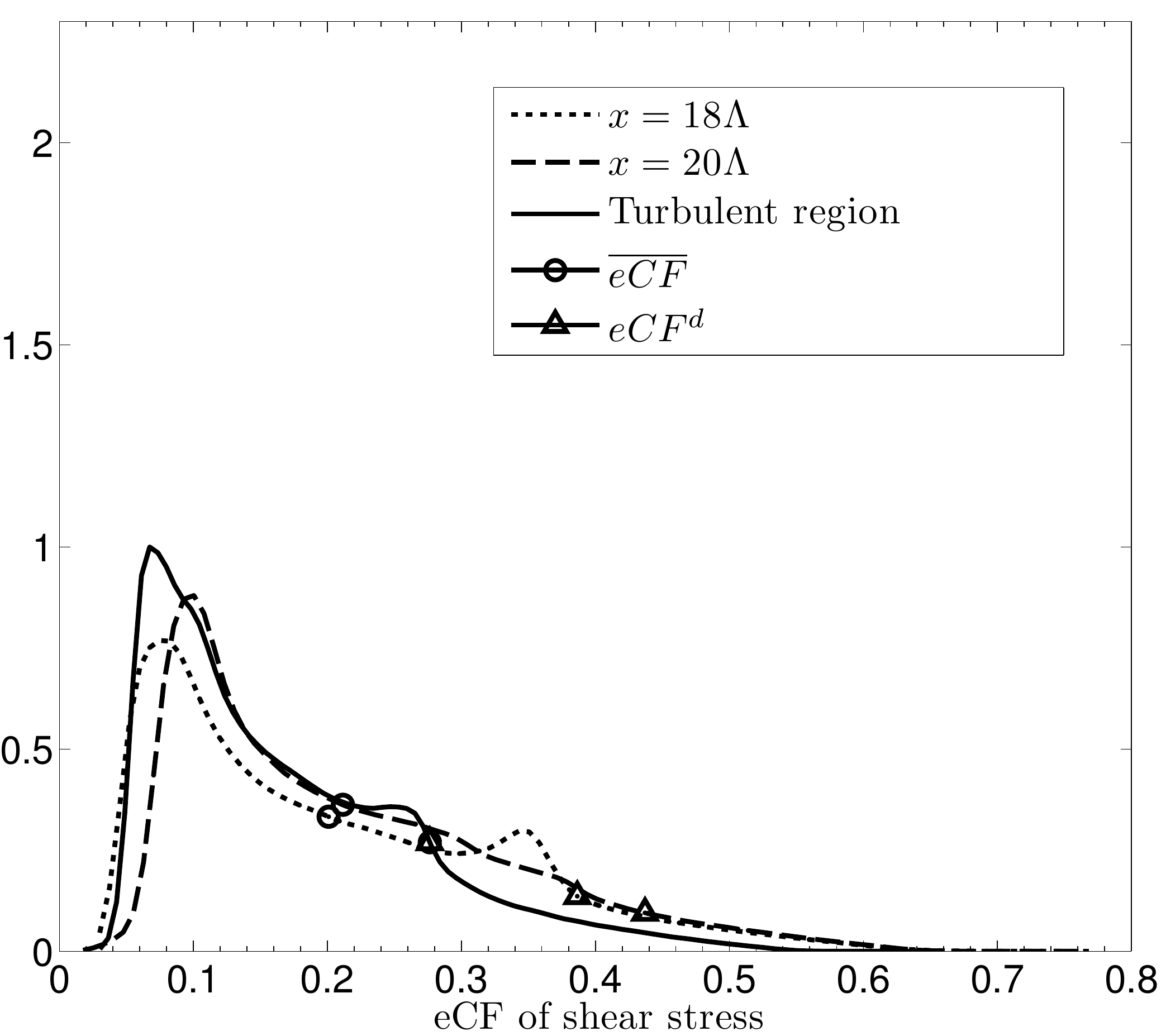}
	\caption{Shear stress $\tau_{uv}$ eCF pdfs in presence of a fully developed turbulent regime. The stochastic mean and the deterministic values are indicated by a circular mark and a triangular mark, respectively.}
	\label{fig:turbSpdf}
\end{figure*}
\begin{figure*}[h]
	\centering
		\includegraphics[width=0.45\linewidth,height=0.3\linewidth]{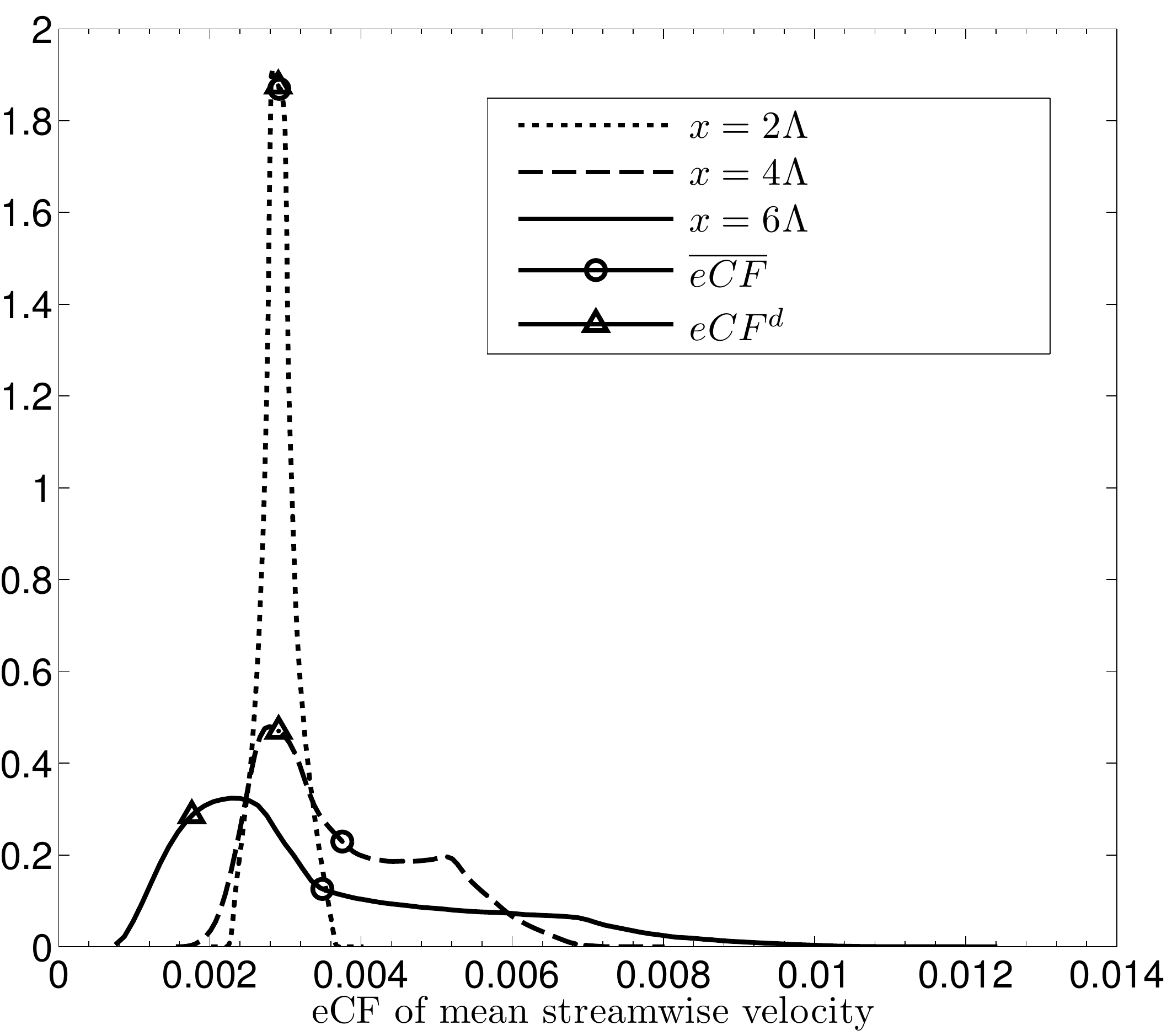}
		\includegraphics[width=0.45\linewidth,height=0.3\linewidth]{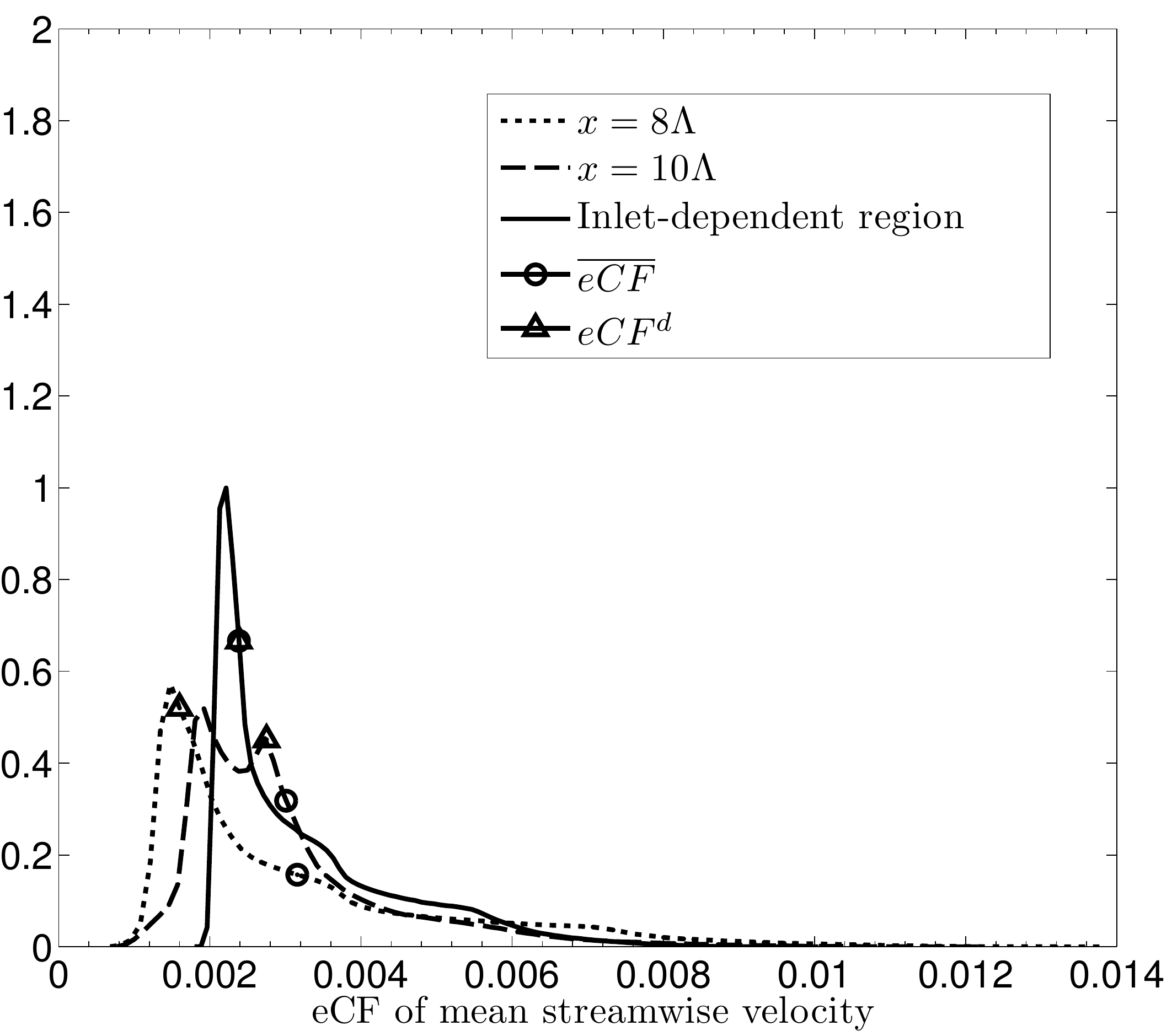}
	\caption{Streamwise velocity $U_x$ eCF pdfs in the transition region. The stochastic mean and the deterministic values are indicated by a circular mark and a triangular mark, respectively.}
	\label{fig:transUpdf}
\end{figure*} 
\begin{figure*}[h]
	\centering
		\includegraphics[width=0.45\linewidth,height=0.3\linewidth]{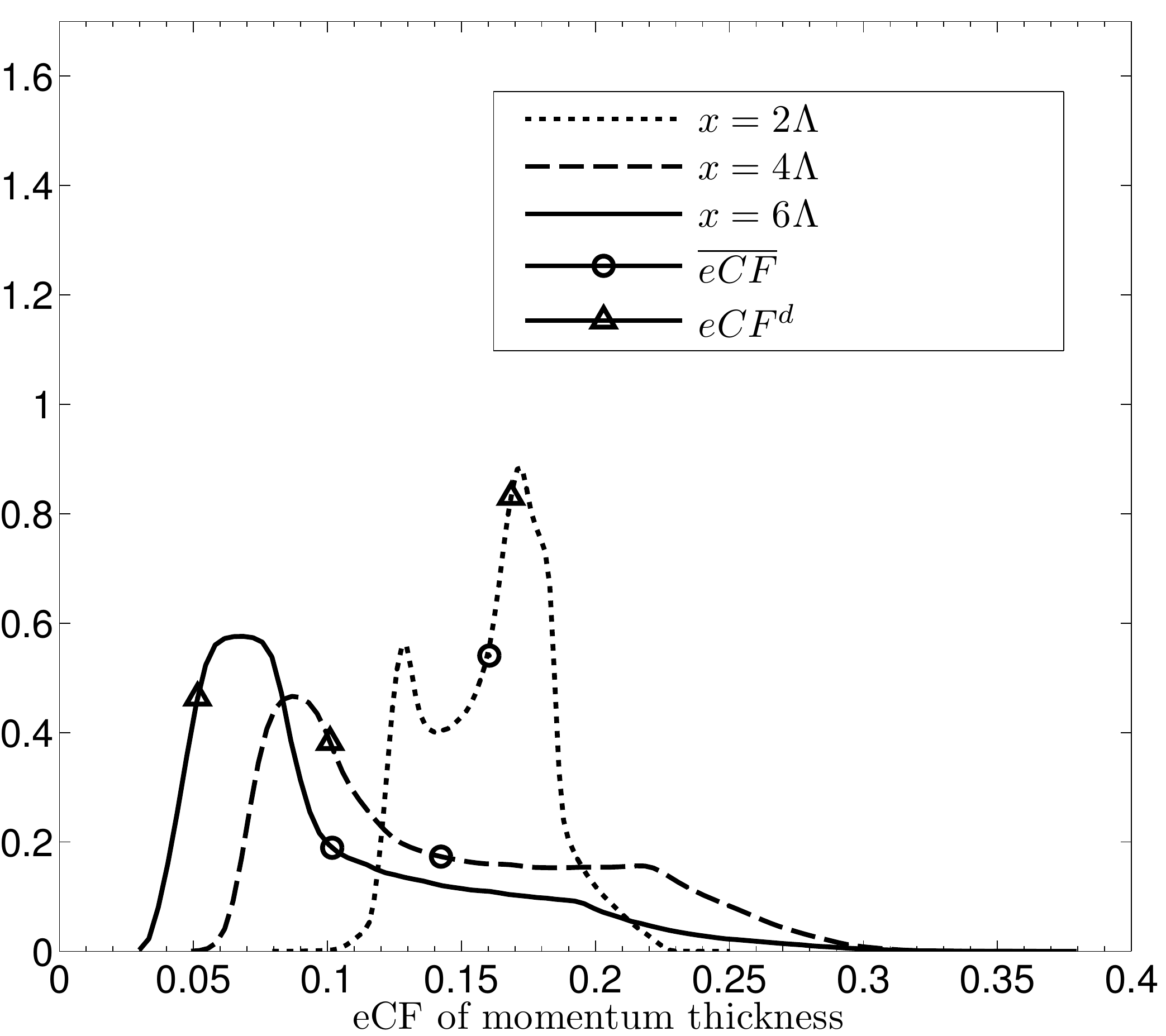}
		\includegraphics[width=0.45\linewidth,height=0.3\linewidth]{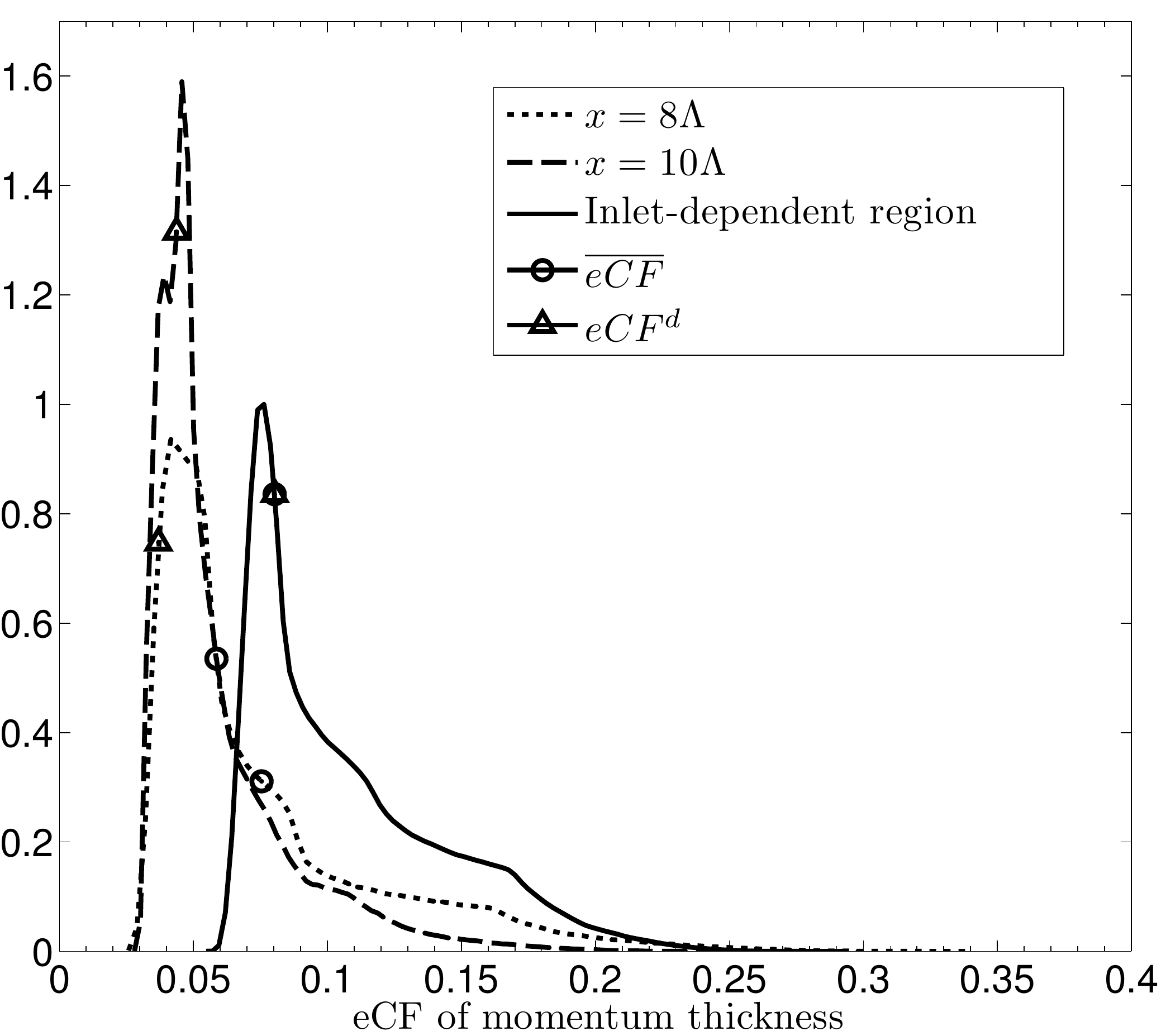}
	\caption{Momentum thickness $\Theta$ eCF pdfs in the transition region. The stochastic mean and the deterministic values are indicated by a circular mark and a triangular mark, respectively.}
	\label{fig:transMpdf}
\end{figure*} 
\begin{figure*}[h]
	\centering
		\includegraphics[width=0.45\linewidth,height=0.3\linewidth]{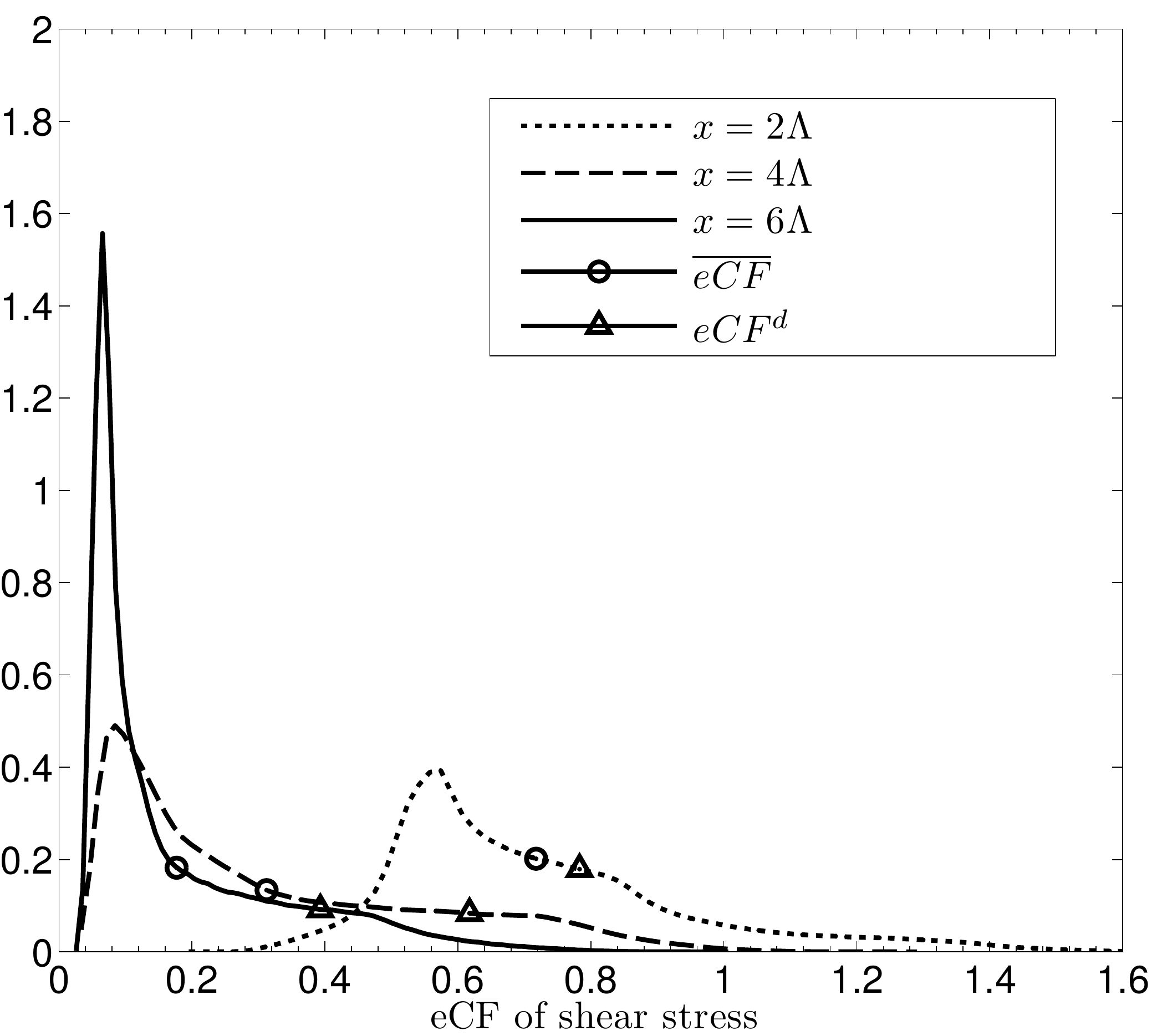}
		\includegraphics[width=0.45\linewidth,height=0.3\linewidth]{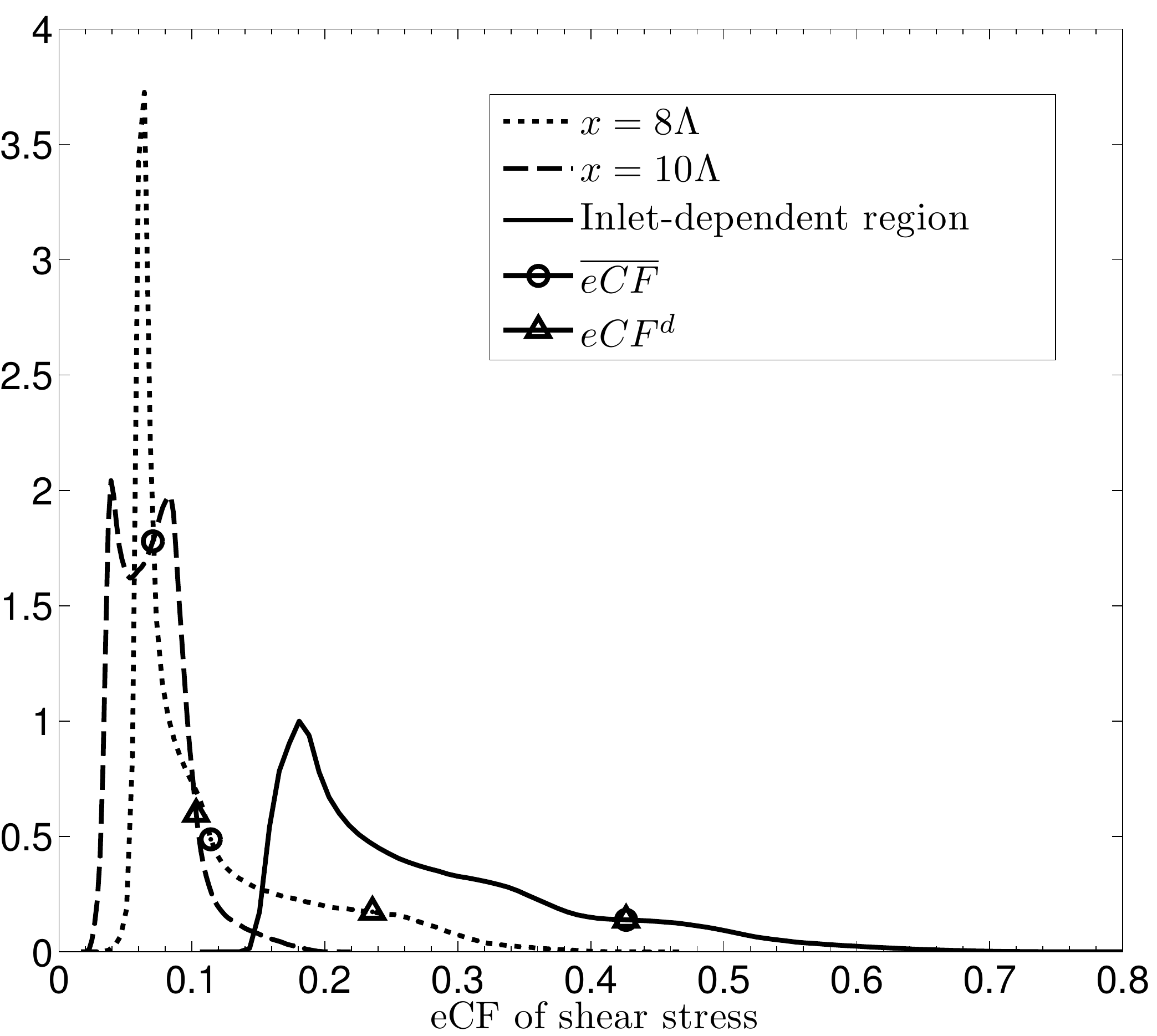}
	\caption{Shear stress $\tau_{uv}$ eCF pdfs in the transition region. The stochastic mean and the deterministic values are indicated by a circular mark and a triangular mark, respectively.}
	\label{fig:transSpdf}
\end{figure*}
\begin{figure*}[h]
	\centering
		\includegraphics[width=0.7\linewidth,height=0.5\linewidth]{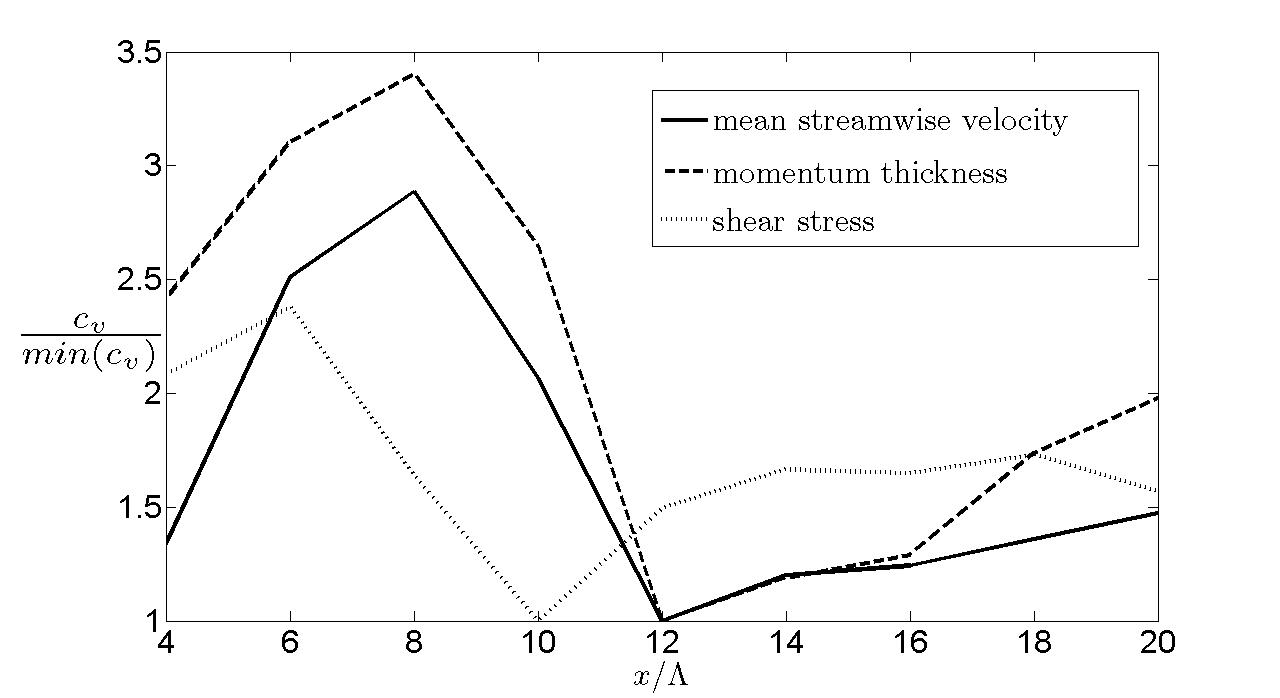}
	\caption{Normalized coefficient of variance for the different physical quantities analyised at sections going from $x = 4\Lambda$ to $x = 20\Lambda$.}
	\label{fig:VarPlot}
\end{figure*}
\begin{figure*}[h]
	\centering
		\includegraphics[width=0.45\linewidth,height=0.3\linewidth]{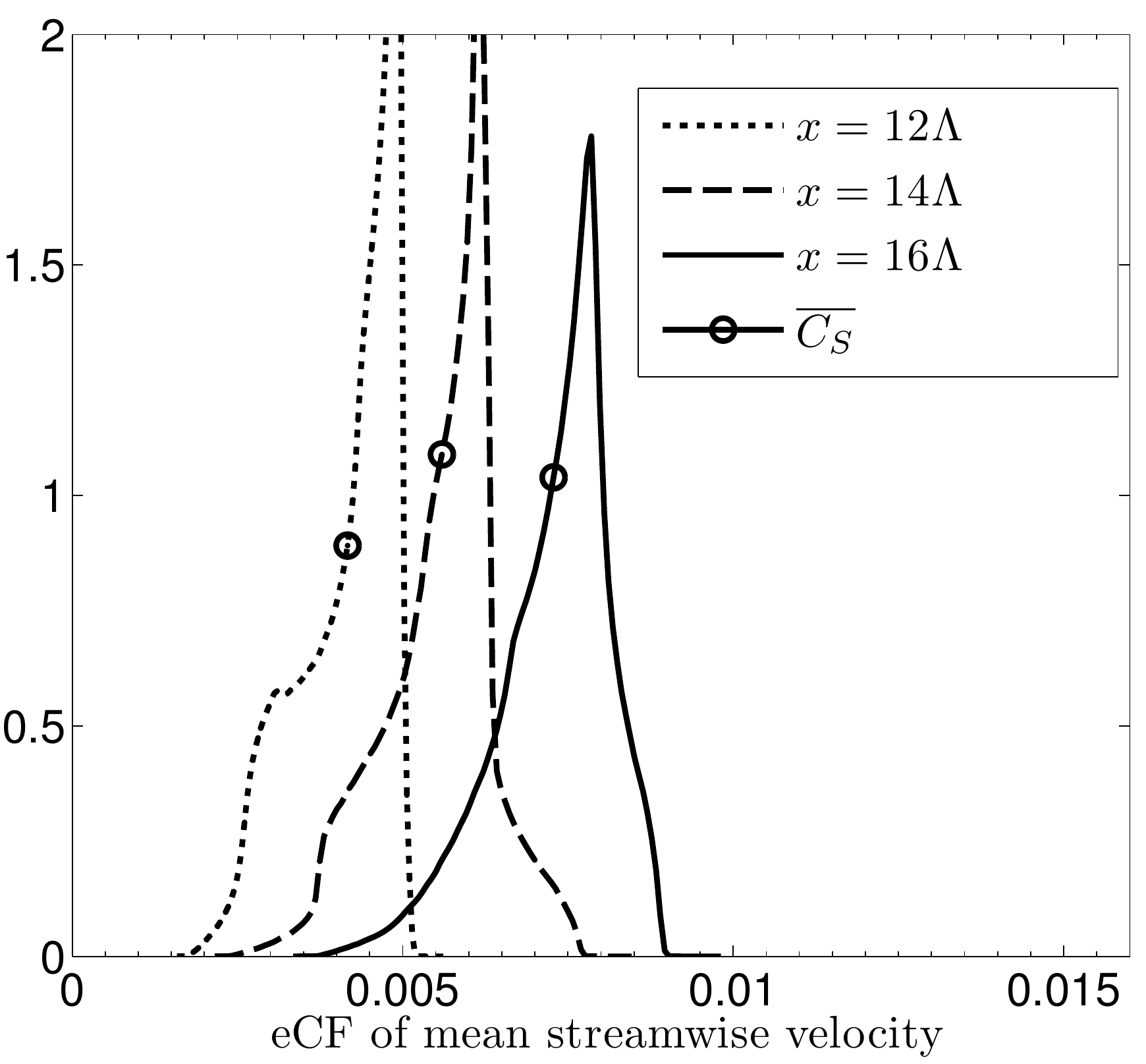}
		\includegraphics[width=0.45\linewidth,height=0.3\linewidth]{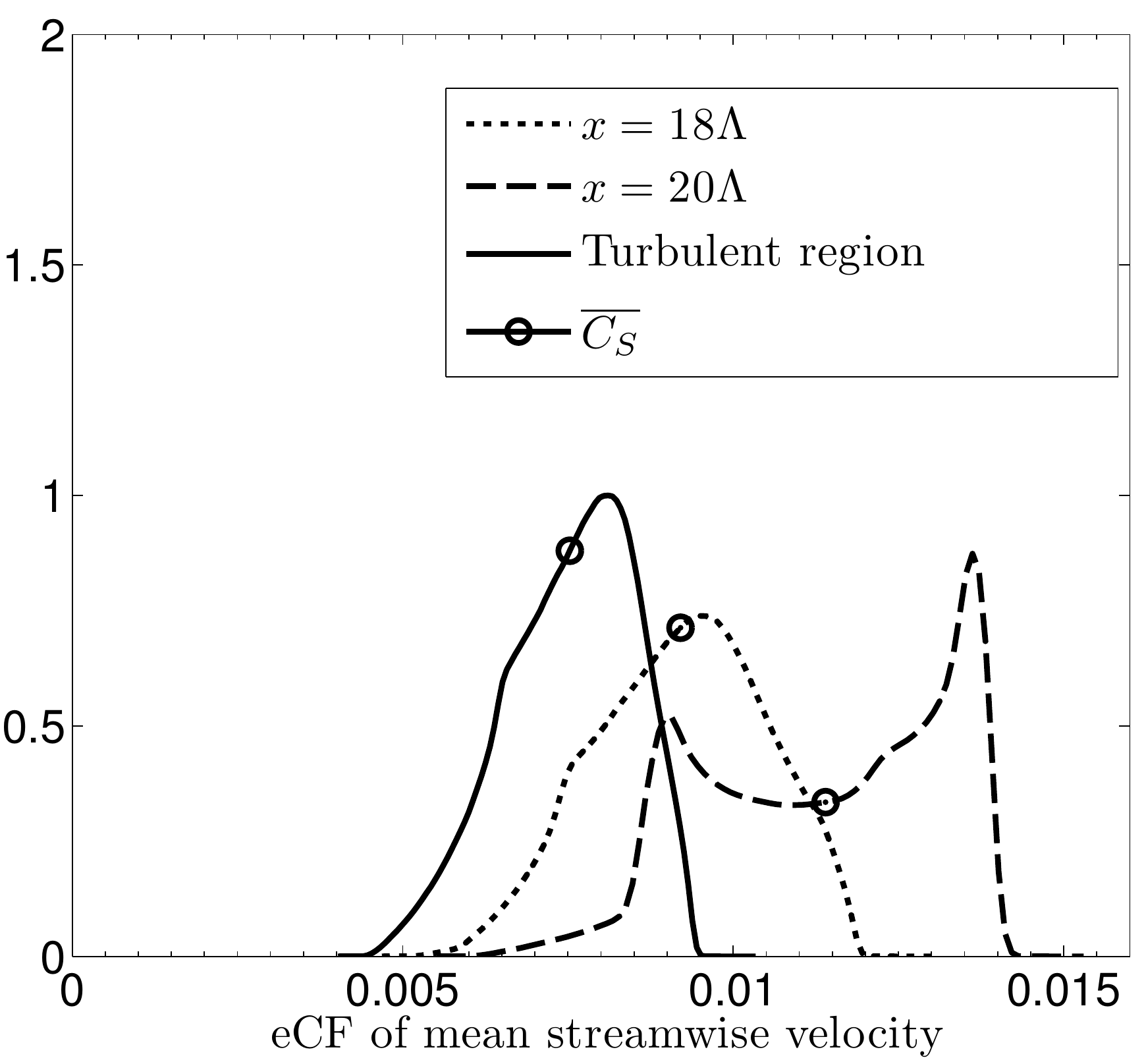}
	\caption{Streamwise velocity $U_x$ eCF pdfs in presence of a fully developed turbulent regime. The stochastic mean value is indicated by a circular mark. The database build using the dynamic Smagorisnky model is considered.}
	\label{fig:turbUpdf_dyn}
\end{figure*} 
\begin{figure*}[h]
	\centering
		\includegraphics[width=0.45\linewidth,height=0.3\linewidth]{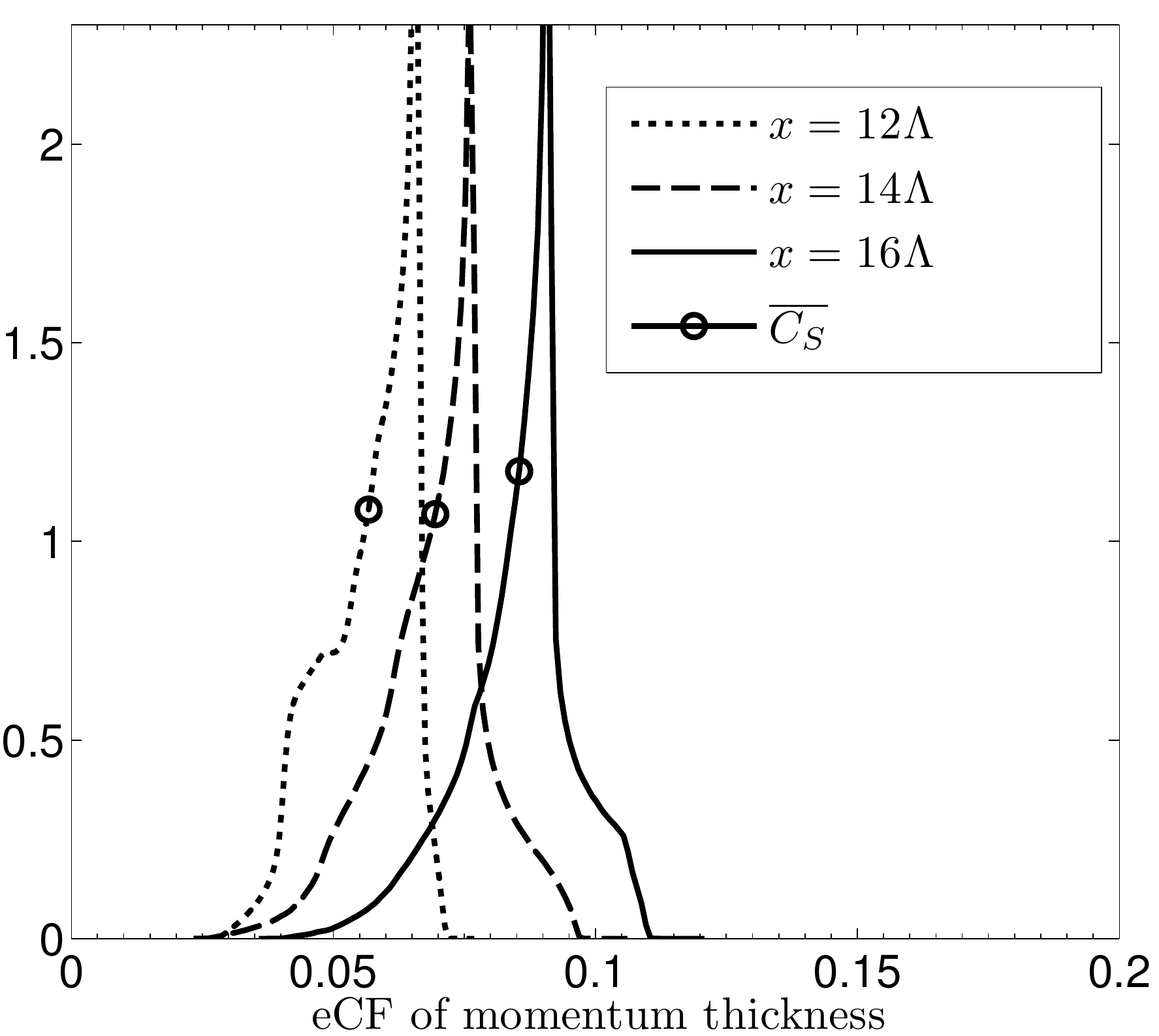}
		\includegraphics[width=0.45\linewidth,height=0.3\linewidth]{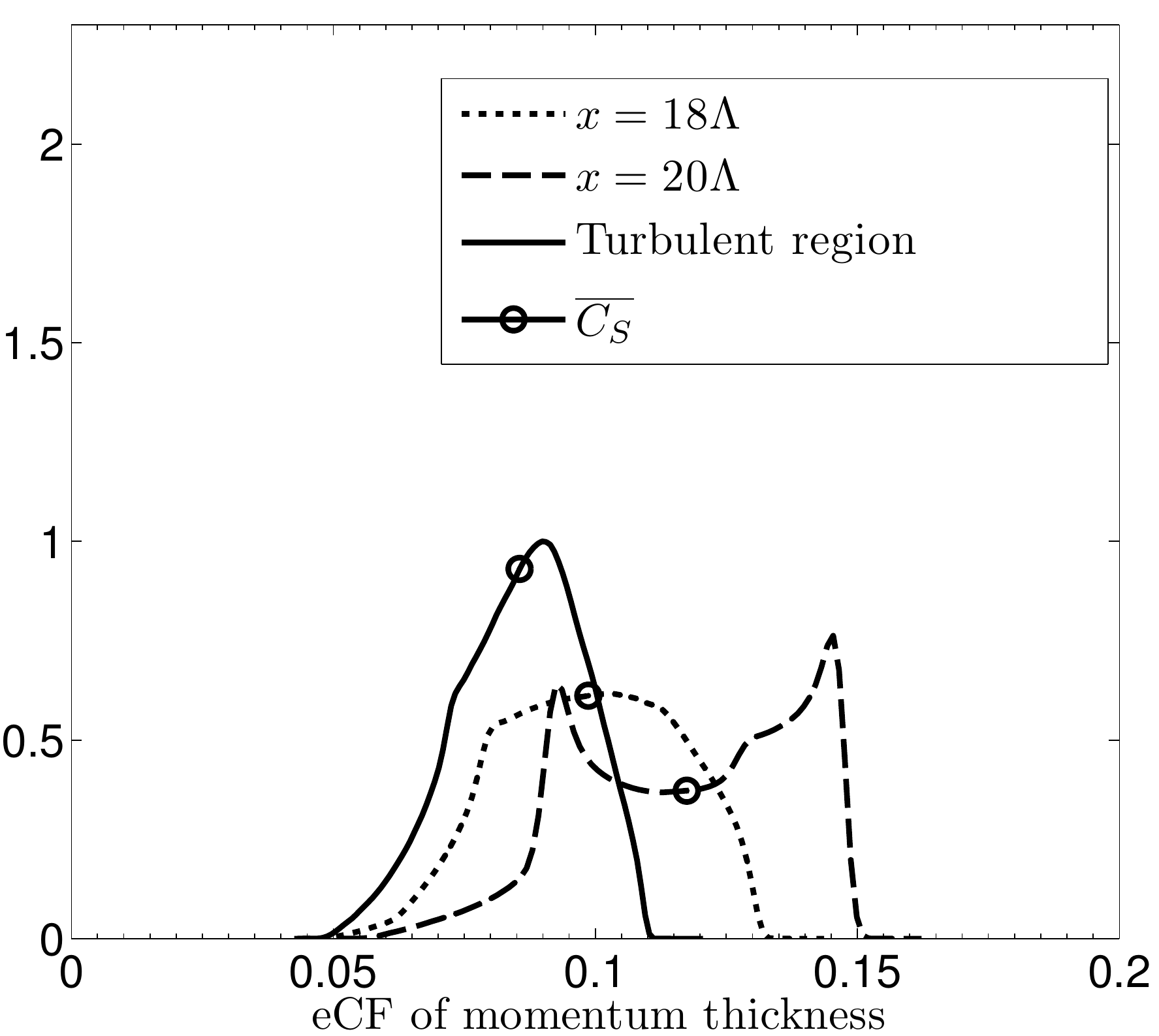}
	\caption{Momentum thickness $\Theta$ eCF pdfs in presence of a fully developed turbulent regime. The stochastic mean value is indicated by a circular mark. The database build using the dynamic Smagorisnky model is considered.}
	\label{fig:turbMpdf_dyn}
\end{figure*} 
\begin{figure*}[h]
	\centering
		\includegraphics[width=0.45\linewidth,height=0.3\linewidth]{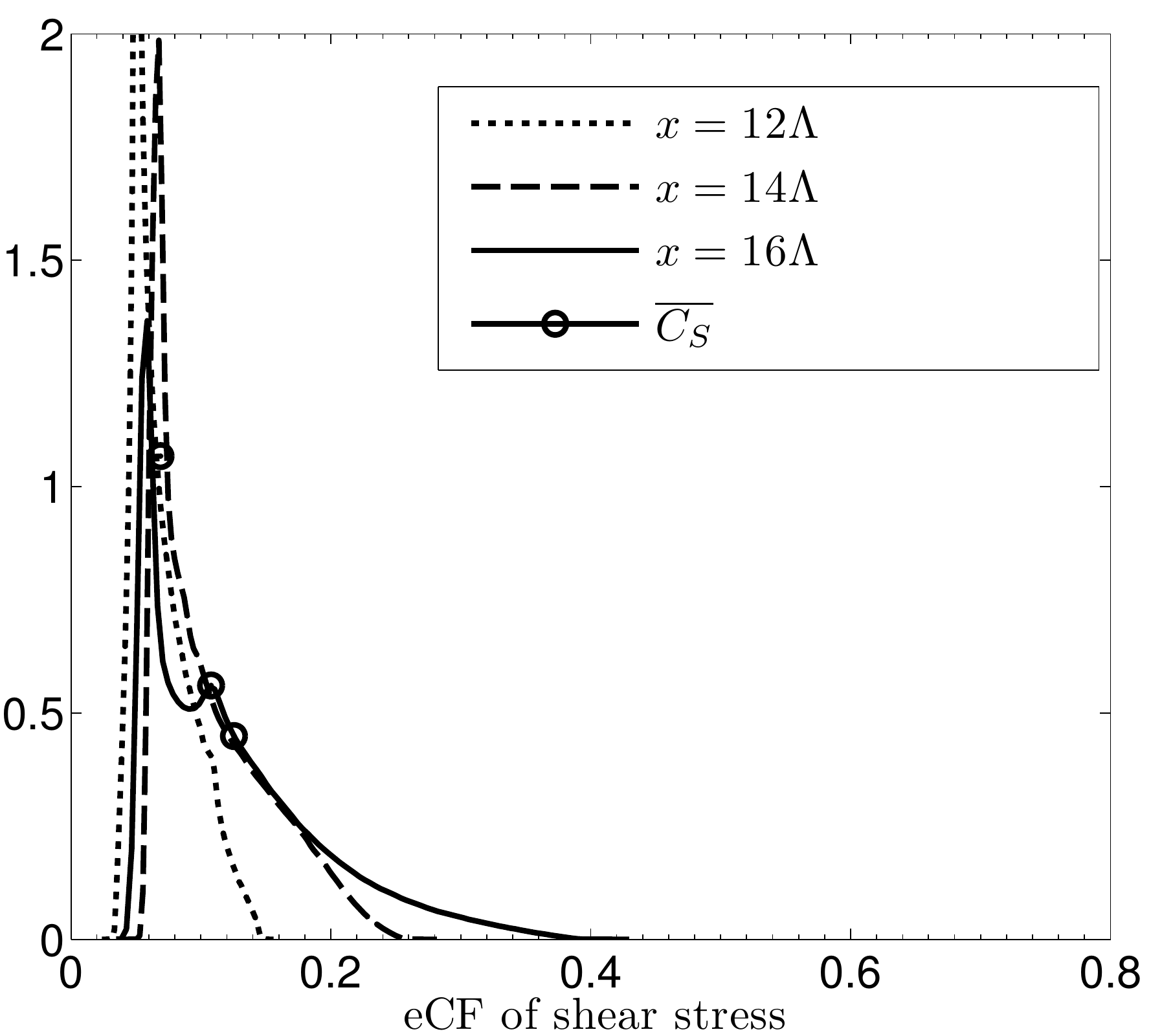}
		\includegraphics[width=0.45\linewidth,height=0.3\linewidth]{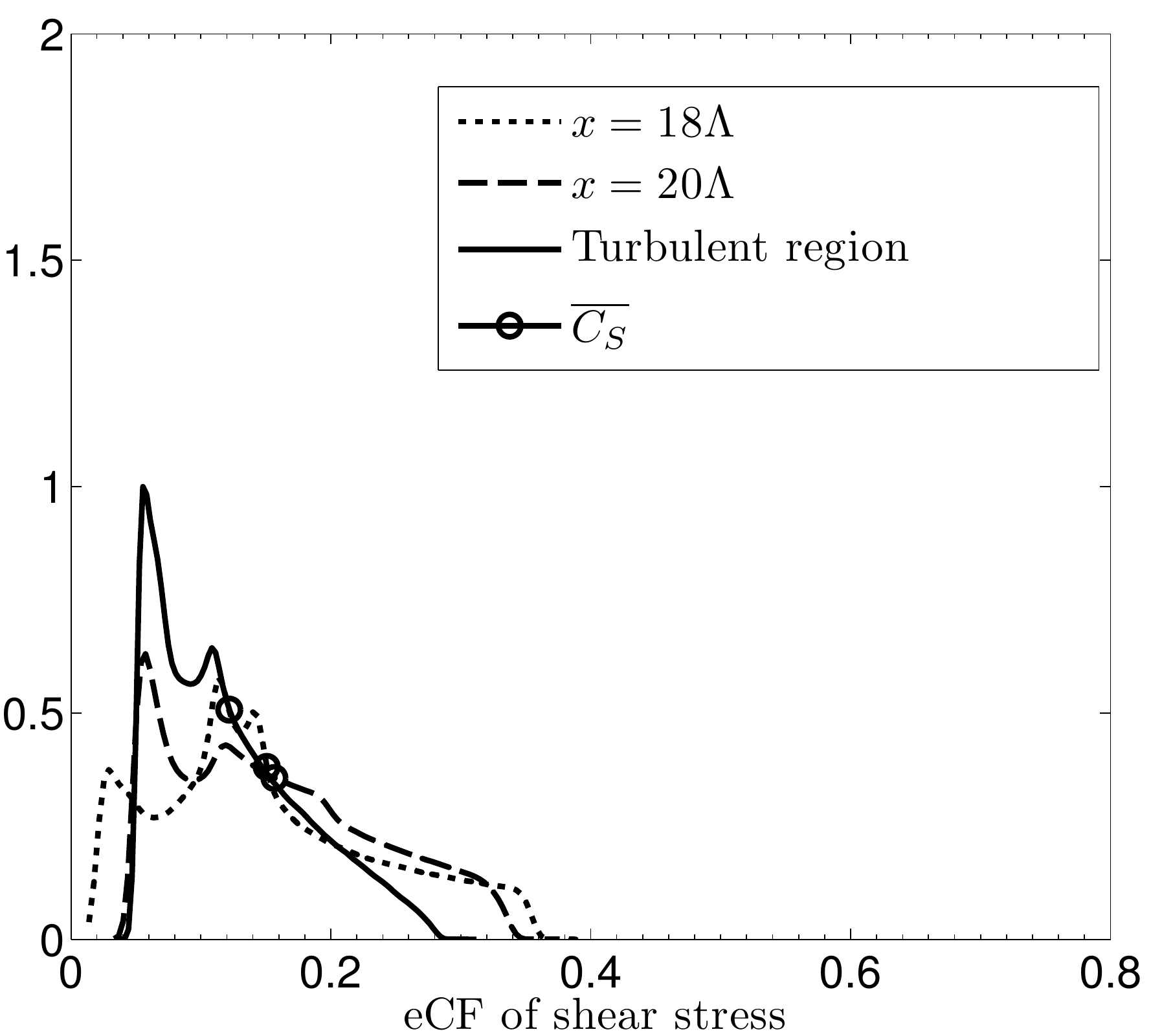}
	\caption{Shear stress $\tau_{uv}$ eCF pdfs in presence of a fully developed turbulent regime. The stochastic mean value is indicated by a circular mark. The database build using the dynamic Smagorisnky model is considered.}
	\label{fig:turbSpdf_dyn}
\end{figure*}
\begin{figure*}[h]
	\centering
		\includegraphics[width=0.45\linewidth,height=0.3\linewidth]{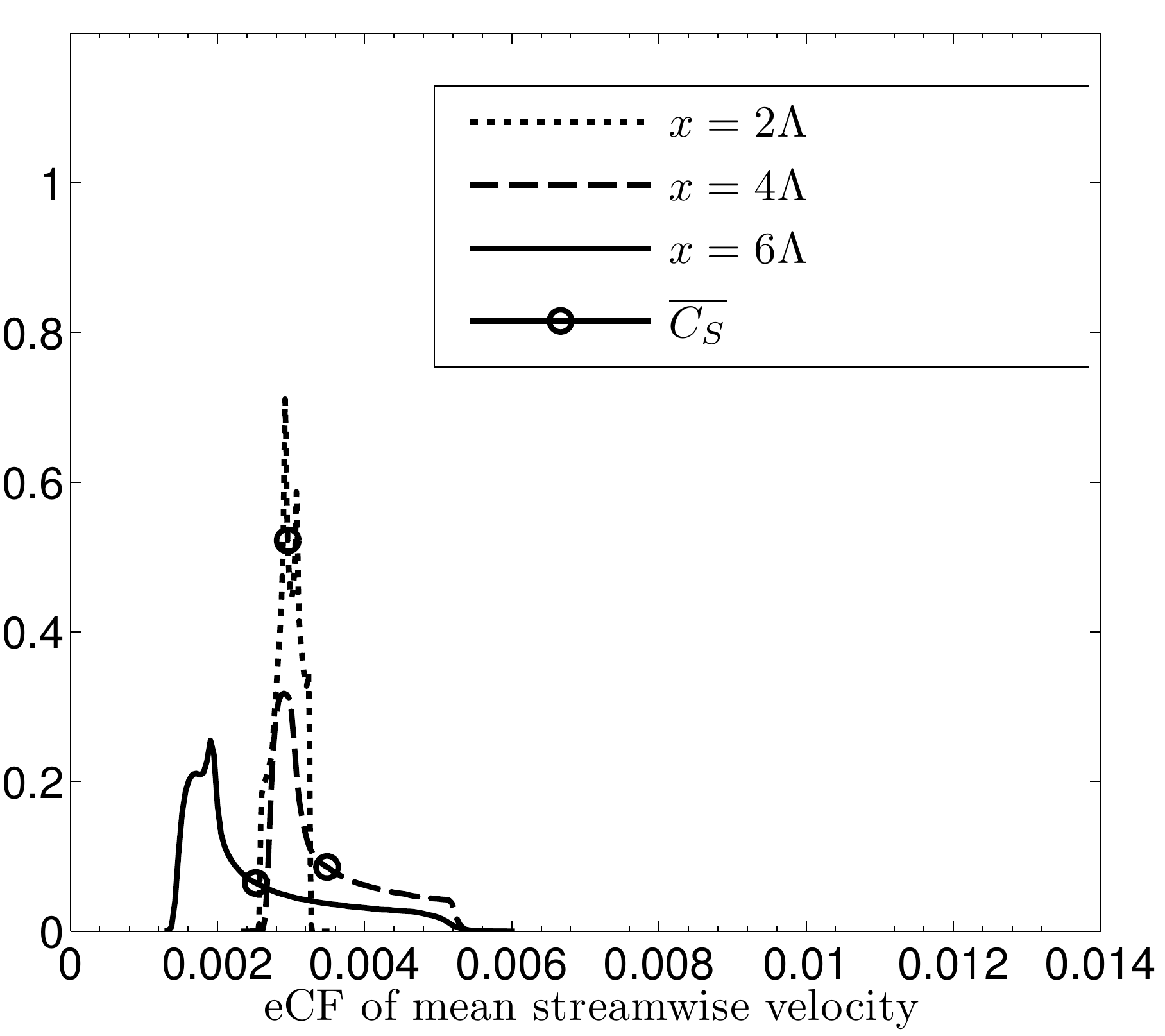}
		\includegraphics[width=0.45\linewidth,height=0.3\linewidth]{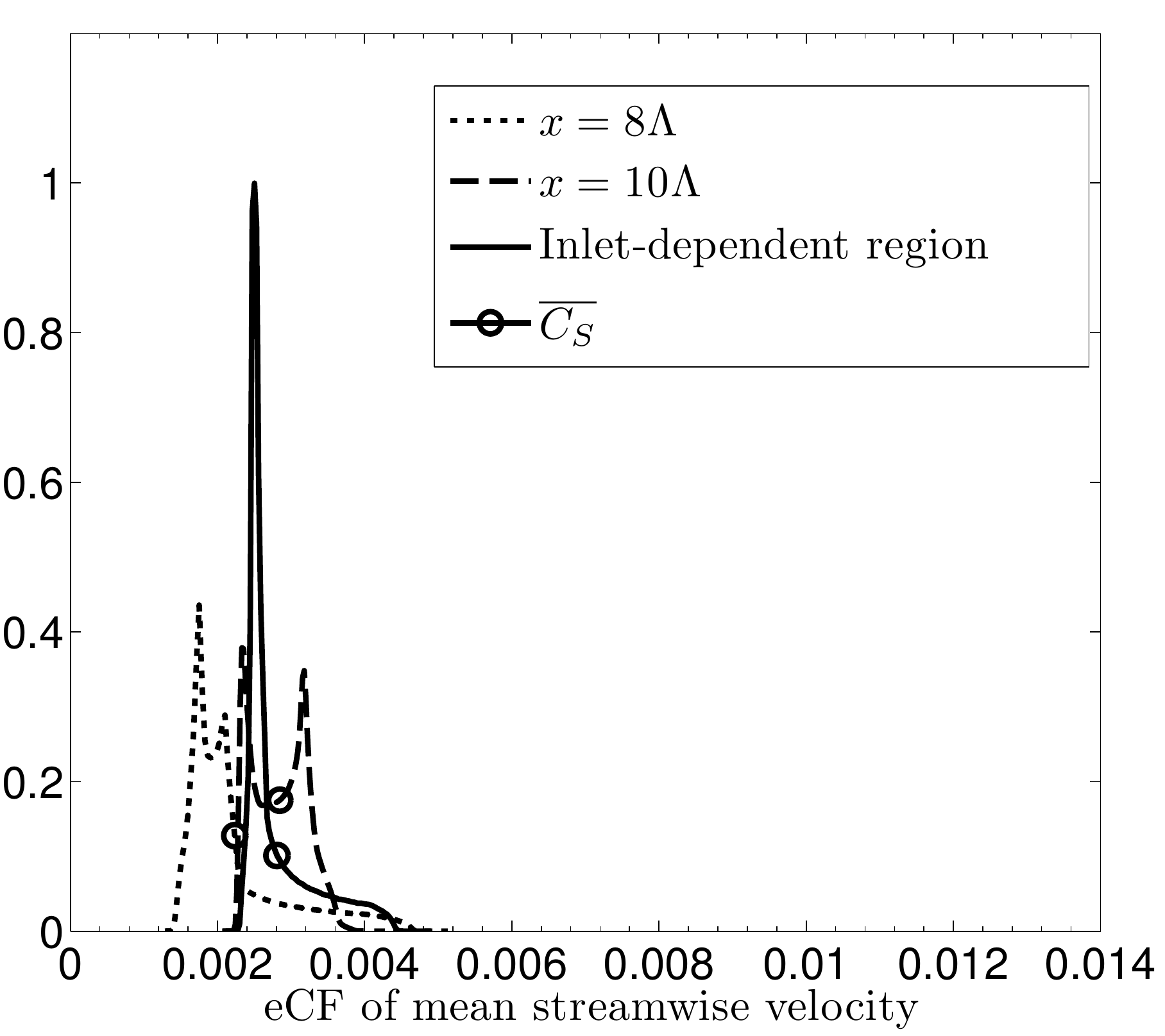}
	\caption{Streamwise velocity $U_x$ eCF pdfs in the transition region. The stochastic mean value is indicated by a circular mark. The database build using the dynamic Smagorisnky model is considered.}
	\label{fig:transUpdf_dyn}
\end{figure*} 
\begin{figure*}[h]
	\centering
		\includegraphics[width=0.45\linewidth,height=0.3\linewidth]{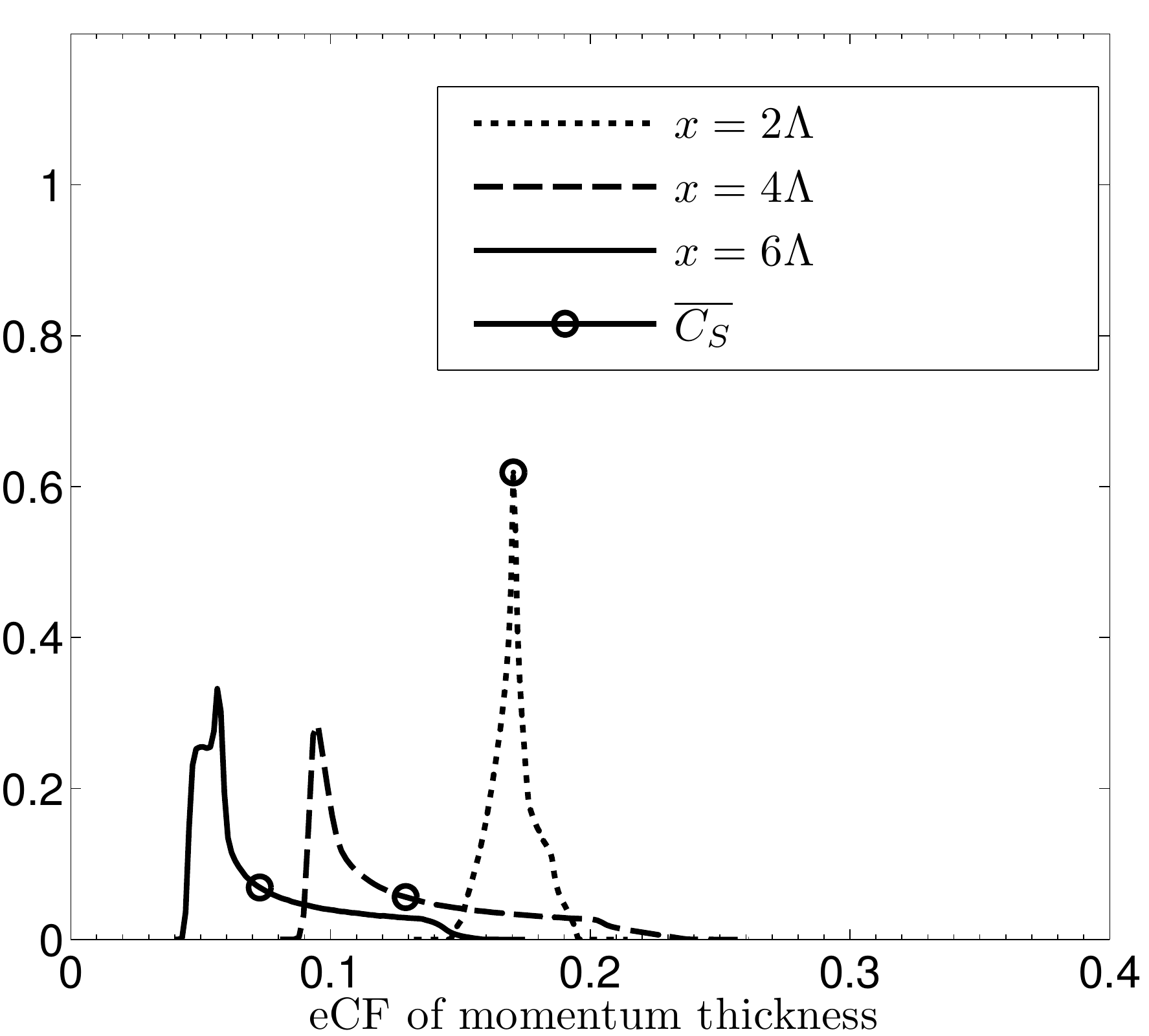}
		\includegraphics[width=0.45\linewidth,height=0.3\linewidth]{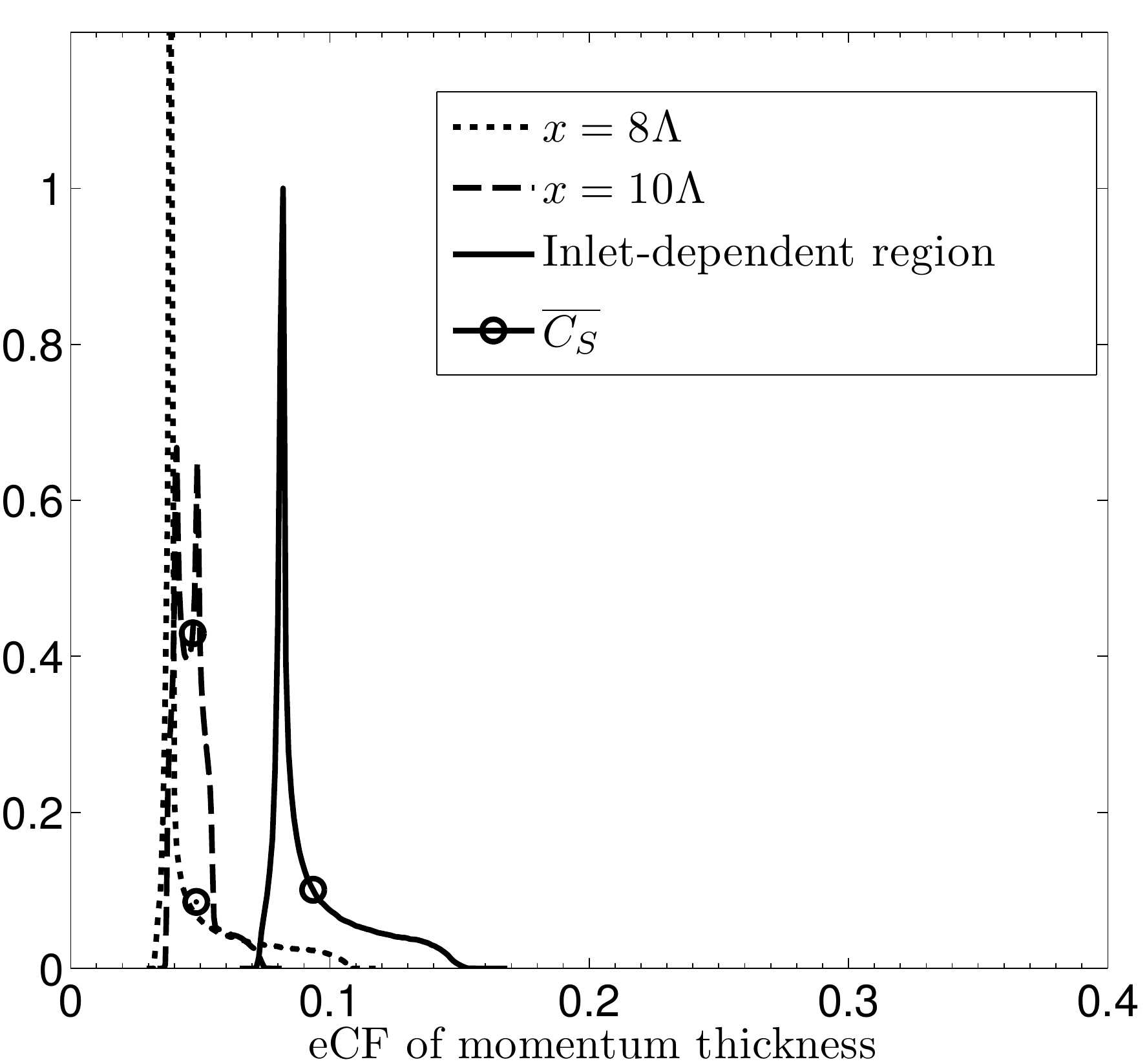}
	\caption{Momentum thickness $\Theta$ eCF pdfs in the transition region. The stochastic mean value is indicated by a circular mark. The database build using the dynamic Smagorisnky model is considered.}
	\label{fig:transMpdf_dyn}
\end{figure*} 
\begin{figure*}[h]
	\centering
		\includegraphics[width=0.45\linewidth,height=0.3\linewidth]{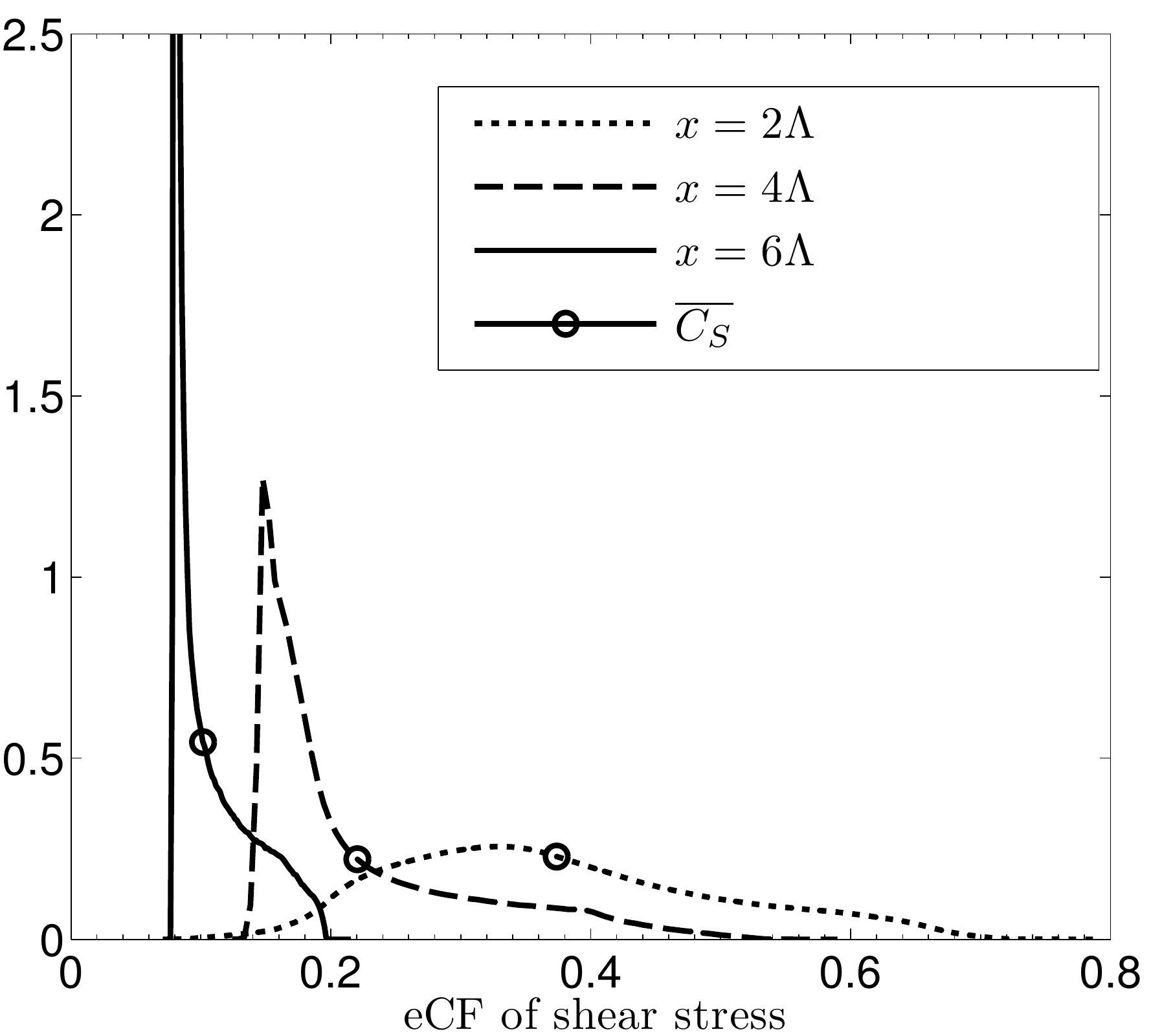}
		\includegraphics[width=0.45\linewidth,height=0.3\linewidth]{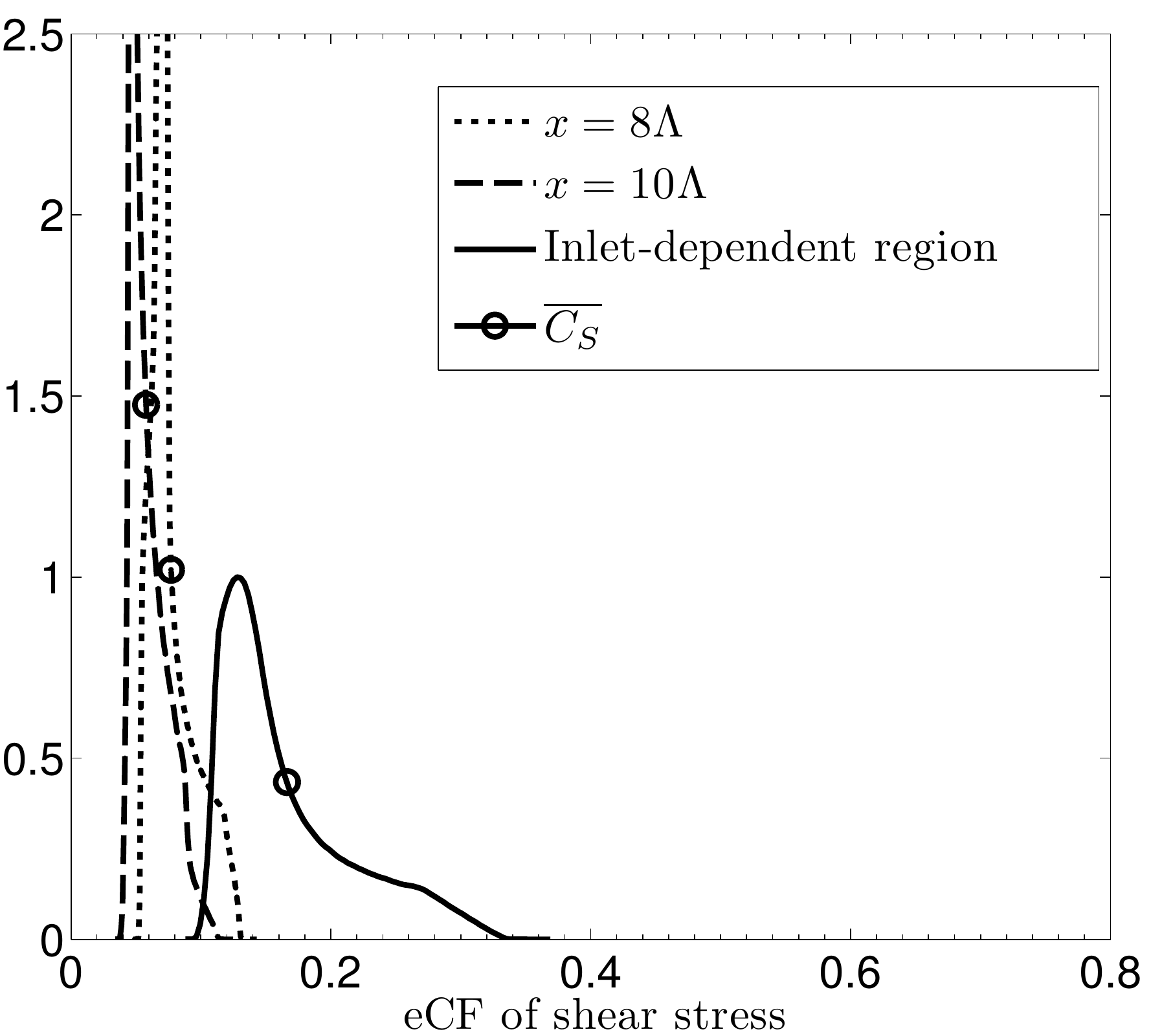}
	\caption{Shear stress $\tau_{uv}$ eCF pdfs in the transition region. The stochastic mean value is indicated by a circular mark. The database build using the dynamic Smagorisnky model is considered.}
	\label{fig:transSpdf_dyn}
\end{figure*}
\clearpage
\begin{figure*}[h]
	\centering
	\begin{tabular}{cc}
		\includegraphics[width=0.42\linewidth]{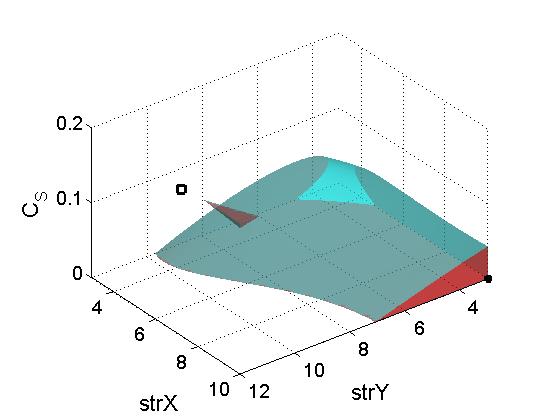} &
		\includegraphics[width=0.42\linewidth]{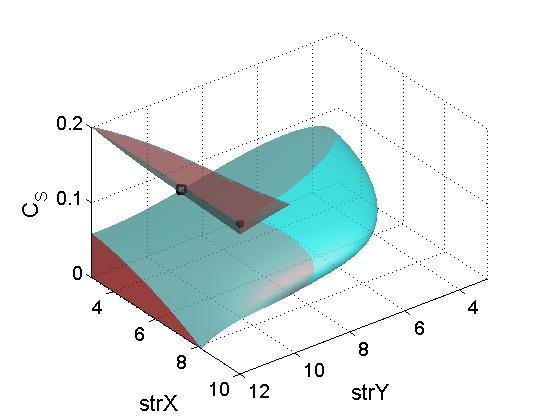}\\
		(a) & (b)\\
		\includegraphics[width=0.42\linewidth]{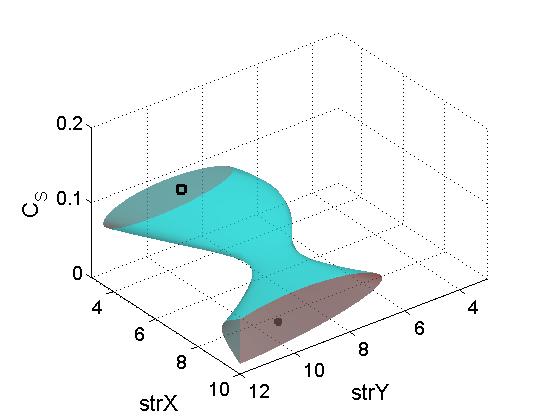} &
		\includegraphics[width=0.42\linewidth]{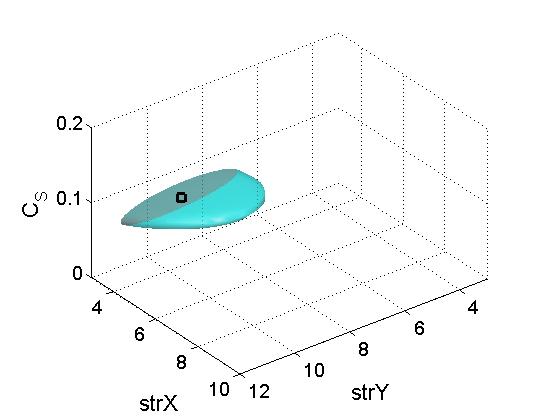}\\
		(c) & (d)\\
	\end{tabular}
	\caption{Isovolumes for eCF $< 1.75$ $*$ min(eCF) in the case of fully developed turbulent flow considering streamwise velocity (a), momentum thickness (b) and shear stress (c). In figure (d) the global optimum zone with a $10\%$ tollerance is shown.}
	\label{fig:optVolumesTurb}
\end{figure*}
\begin{figure*}[h]
	\centering
	\begin{tabular}{cc}
		\includegraphics[width=0.42\linewidth]{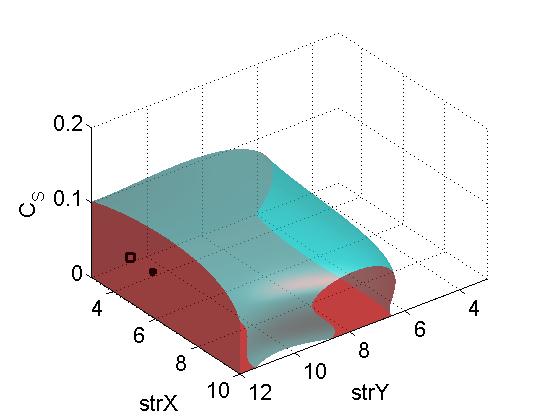} &
		\includegraphics[width=0.42\linewidth]{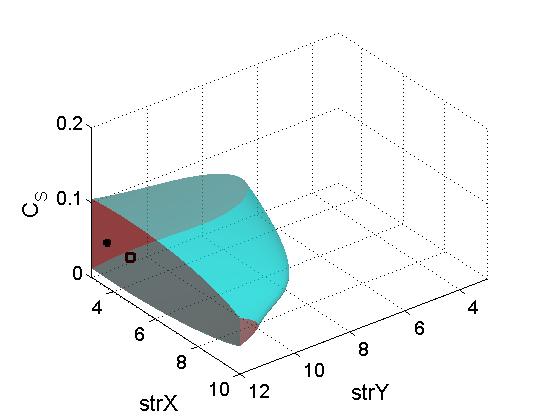}\\
		(a) & (b)\\
		\includegraphics[width=0.42\linewidth]{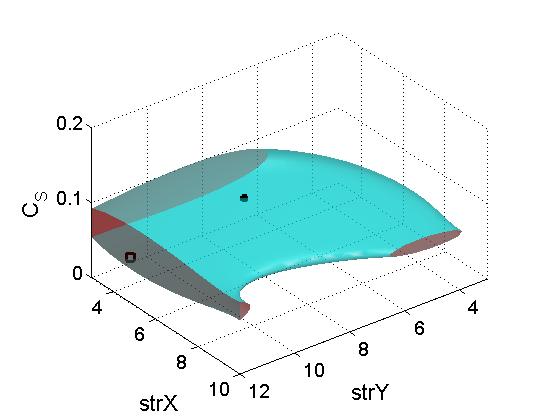} &
		\includegraphics[width=0.42\linewidth]{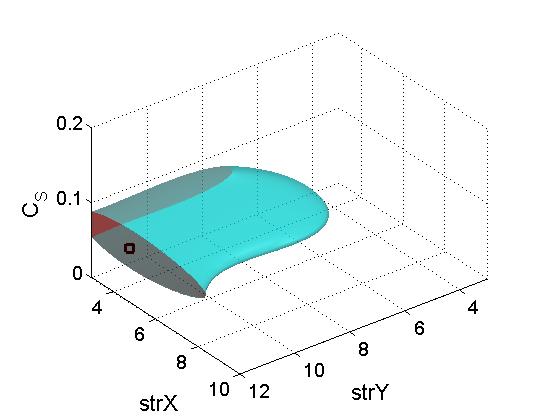}\\
		(c) & (d)\\
	\end{tabular}
	\caption{Isovolumes for eCF $< 1.2$ $*$ min(eCF) in the case of transitioning flow considering streamwise velocity (a), momentum thickness (b) and shear stress (c). In figure (d), global optimization with a $10\%$ tollerance is shown.}
	\label{fig:optVolumesTrans}
\end{figure*}
\begin{figure*}[h]
	\centering
		\includegraphics[width=0.65\linewidth]{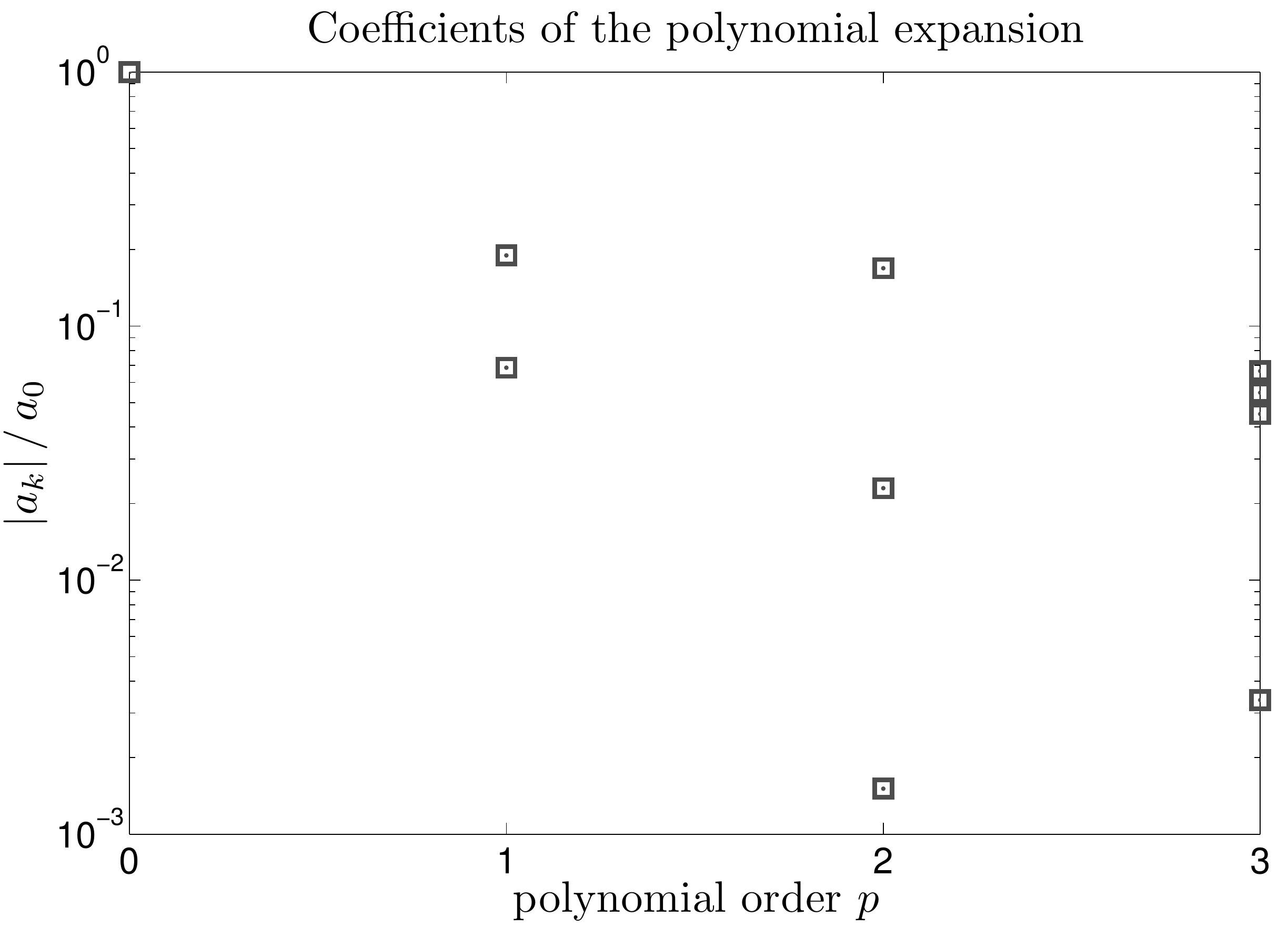}
	\caption{Coefficients $a_k$ of a polynomial expansion, grouped by the polynomial order. The considered case is the expansion of the function $eCF_{\theta}$ at x / $\Lambda$ = 12 starting from a LES database generated using the dynamic version of the Smagorinsky model.}
	\label{fig:gPCaccuracy}
\end{figure*}
\end{document}